\documentclass[oupdraft]{bio}
\usepackage[colorlinks=true, urlcolor=citecolor, linkcolor=citecolor, citecolor=citecolor]{hyperref}
\usepackage{url}

\usepackage[left]{lineno}

\usepackage{algorithm}      
\usepackage{algpseudocode}  

\usepackage{setspace}

\history{}

\begin{document}

\raggedbottom

\title{Pair-based estimators of infection and removal rates for stochastic epidemic models}

\author{SETH D. TEMPLE$^{1,2,\ast}$, JONATHAN TERHORST$^1$\\[4pt]
\textit{$^1$ Department of Statistics, University of Michigan - Ann Arbor, 1085 South University Street, Ann Arbor, Michigan, United States of America}
\\[2pt]
\textit{$^2$ Michigan Institute for Data and AI in Society, University of Michigan - Ann Arbor, 500 Church Street, Ann Arbor, Michigan, United States of America}
\\[2pt]
{Email: sethtem@umich.edu}}

\markboth%
{S. D. Temple and J. Terhorst}
{Pair-based estimators of infection and removal rates}

\maketitle

\footnotetext{To whom correspondence should be addressed.}

\begin{abstract}
{Stochastic epidemic models can estimate infection and removal rates, and derived quantities such as the basic reproductive number ($R_0$), when both infection and removal times are observed. In practice, however, removal times are often available while infection times are not, and existing methods that rely only on removal times can become unstable or biased. We study inference for stochastic SIR/SEIR models in a partial--observation setting. We develop imputation--based estimators that use a small calibration sample of fully observed infectious periods, derive closed--form expressions for the pairwise exposure terms they require, and use a studentized parametric bootstrap for bias correction and uncertainty quantification. In simulations, removal time--only methods performed poorly in moderate to large $R_0$ scenarios, while observing even tens of complete infectious periods substantially improved the estimation of the infection rate. A reanalysis of the 1861 Hagelloch measles outbreak under simulated missingness recovered stable qualitative differences in transmission between school classes. Based on our results, we advocate for the targeted collection of a modest number of complete infectious periods as a means of improving surveillance in the early stages of an epidemic.
}
{Bayesian inference; Compartmental models; Parametric bootstrapping; Renewal processes; Stochastic epidemic.}
\end{abstract}

\nolinenumbers

\newcommand\real{\mathbb{R}}
\newcommand\E{\mathbb{E}}
\newcommand\var{\text{Var}}
\newcommand\cov{\text{Cov}}
\newcommand\hs{\hskip2pt}
\newcommand\expon{\text{Exponential}}
\newcommand\gam{\text{Gamma}}
\newcommand\thh{\text{th}}
\newcommand\Tau{\mathcal{T}}
\newcommand\etal{\textit{et. al.}}
\newcommand\Z{Z}
\newcommand\X{X}
\newcommand\Y{Y}

\newcommand\jkinv{(\delta_k+\delta_j)^{-1}}
\newcommand\bkinv{(\delta_k+\beta)^{-1}}
\newcommand\jk{\delta_j\jkinv}
\newcommand\kj{\delta_k\jkinv}
\newcommand{\exppeir}[3]{\exp(-#1(#2-#3))}
\newcommand{\invpeir}[2]{(#1+#2)^{-1}}
\newcommand{\fracpeir}[2]{#1(#1+#2)^{-1}}

 \newcommand{\beginsupplement}{
    \setcounter{section}{0}
    \renewcommand{\thesection}{S\arabic{section}}
    \setcounter{equation}{0}
    \renewcommand{\theequation}{S\arabic{equation}}
    \setcounter{table}{0}
    \renewcommand{\thetable}{S\arabic{table}}
    \setcounter{tblcap}{0}                 
    \renewcommand{\thetblcap}{S\arabic{tblcap}} 
    \setcounter{figure}{0}
    \renewcommand{\thefigure}{S\arabic{figure}}
    \newcounter{SIfig}
    \renewcommand{\theSIfig}{S\arabic{SIfig}}
    \newcounter{SItab}
    \renewcommand{\theSItab}{S\arabic{SItab}}
    \renewcommand{\thealgorithm}{S\arabic{algorithm}}
    \setcounter{algorithm}{0}
    }

\theoremstyle{lemma} 
\newtheorem{proposition}[theorem]{\sc Proposition}

\section{Introduction}
\label{sec:intro}

Stochastic epidemic models (SEMs) such as the susceptible--infected--recovered (SIR) and susceptible--exposed--infected--recovered (SEIR) models are widely used to quantify transmission dynamics and to estimate quantities such as the basic reproductive number $R_0$ \citep{Blackwood2018-lf,Andersson2012-on}. In the continuous-time formulations we consider, epidemic dynamics is driven by a small set of rate parameters that summarize how quickly infection and removal (e.g. recovery or isolation) occur. Under homogeneous mixing, transmission is governed by an infection rate $\beta$ and removal by a removal rate $\gamma$; in the simplest models, $R_0=\beta/\gamma$.

When all individual infection times and removal times are observed, the estimation of $(\beta,\gamma)$ is routine. However, removal times are often known (e.g., from clinical records), while infection times are completely missing \citep{O-Neill2002-of}. Although infection times can sometimes be learned through contact tracing, this process is time- and labor-intensive. Hence, a number of methods have been developed to infer $\beta$ and $\gamma$ using only removal times. Bayesian approaches to this problem employ data augmentation, sampling missing infection times within a Markov Chain Monte Carlo (MCMC) scheme \citep{gibson1998estimating,Kypraios2007-vq,Jewell2009-gx,Stockdale2017-sr,Stockdale2019-kd,Ball2025-nh}. Frequentist approaches face an intractable marginal likelihood when infection times are latent and therefore typically maximize likelihood approximations \citep{Eichner2003-kv,Stockdale2021-ez}. 

Despite these considerable efforts, it is fair to say that lack of infection time information greatly complicates the accurate estimation of $\beta$ and $\gamma$. Bayesian methods suffer from poor mixing and strong posterior correlations \citep{Kypraios2007-vq,Stockdale2019-kd}; \cite{Ball2025-nh} found that removal time-only likelihoods tended to be multimodal; \cite{Rida1991-vt} pointed out that the maximum likelihood estimate (MLE) of $\beta$ cannot be used when only removal times are known; and \cite{Kypraios2007-vq} indicated that estimating $\beta$ may be impossible in this case.

In this paper, we explore the estimation in a slightly more optimistic, but still plausible, design: removal times are broadly available, and a small subset of cases have fully observed infectious periods. We hypothesized that knowing even a modest number of complete infectious periods could greatly stabilize inference, for the simple reason that $\gamma$ has a statistically efficient maximum-likelihood estimator based only on observed infection durations. Calibrating $\gamma$ in this way reduces the confounding between infection and removal rates and allows us to focus on estimating $\beta$ from partially observed endpoints. 

To that end, we propose an expectation--maximization (EM) approach \citep{Dempster1977-bs,Becker1993-fe} that calibrates the removal rate from a small set of complete infectious periods and then estimates the infection rate by imputing key sufficient statistics, especially the pairwise infective time pressures $\tau_{kj}$. We derive closed-form expressions for the conditional expectations required under partial-observation patterns and quantify uncertainty and correct bias with studentized parametric bootstrap intervals. We also develop a revised MCMC scheme for Bayesian inference with partial endpoints. 

We evaluated these methods in simulations and in a reanalysis of the 1861 Hagelloch measles outbreak under simulated missingness, with the broader goal of quantifying how much complete-case information is needed to stabilize inference. Apart from developing the methodology, our aim here is to argue for limited targeted data collection during emerging epidemics. As our results demonstrate, estimation improves substantially when even a small fraction of cases have completely observed infectious periods.

The remainder of the paper is organized as follows.
Section \ref{sec:prelim} defines the pair-based SIR/SEIR model and summarizes common likelihood-based and Bayesian estimators. Section \ref{sec:methods} presents our imputation-- and EM--based estimators, bootstrap uncertainty quantification, and revised MCMC approach. Empirical results from the simulation and Hagelloch studies are reported in Sections \ref{sec:simulations} and \ref{sec:measles}. We make concluding remarks in Section \ref{sec:discussion}.

\section{Preliminary materials}\label{sec:prelim}

\subsection{Simulating a pair-based epidemic model}

We study a stochastic epidemic model (SEM) represented as a continuous-time Markov chain (CTMC) \citep{Kermack1991-ro,Allen2008-jv}. Let $\mathcal{S}(t), \mathcal{E}(t), \mathcal{I}(t), \mathcal{R}(t) \subseteq \{1,\ldots,N\}$ denote the sets of susceptible, exposed, infectious, and removed individuals at time $t$ in a fixed population of size $N$. These form a partition such that $\mathcal{S}(t) \cap \mathcal{E}(t) \cap \mathcal{I}(t) \cap \mathcal{R}(t) = \emptyset$ and $\mathcal{S}(t) \cup \mathcal{E}(t) \cup \mathcal{I}(t) \cup \mathcal{R}(t) = \{1,\ldots,N\}$. The counts of susceptible, exposed, infectious, and removed individuals are $S(t)$, $E(t)$, $I(t)$, and $R(t)$. The initial state is $S(0)=N-1$, $E(0)=0$, $I(0)=1$, and $R(0)=0$. Each infectious individual $k$ infects a susceptible individual $j$ at rate $\beta_{kj}$ and is removed at rate $\gamma_j$. The duration of the infectious period of individual $j$ is distributed as Exponential$(\gamma_j)$. This setup is analogous to Gillespie--style chemical reaction simulations \citep{Gillespie1976-zs,Gillespie1977-wq,Gillespie2007-jg}.

The stochastic SIR state process is a bivariate CTMC $\{(S(t), I(t)) \hskip2pt | \hskip2pt t \in T \}$ with transition rates
\begin{equation}\label{eq:infect_het_SIR}
\begin{split}
    \lambda_{SI}(t) &= \sum_{k \in \mathcal{I}(t)} \sum_{j \in \mathcal{S}(t)} \beta_{kj},
\end{split}
\end{equation}
\begin{equation}\label{eq:remove_het_SIR}
\begin{split}
    \lambda_{IR}(t) &= \sum_{j \in \mathcal{I}(t)} \gamma_j,
\end{split}
\end{equation}
when there are no incubation periods. The subscripts ${SI}$ and ${IR}$ denote the transitions $S\to I$ (infection) and $I\to R$ (removal), respectively. When rates are homogeneous across individuals, $\beta_{kj}=\beta/N$ and $\gamma_j=\gamma$, these simplify to
\begin{equation}\label{eq:infect_SIR}
\begin{split}
    \lambda_{SI}(t) &= \frac{\beta \cdot S(t) \cdot I(t)}{N},
\end{split}
\end{equation}
\begin{equation}\label{eq:remove_SIR}
\begin{split}
    \lambda_{IR}(t) &= \gamma \cdot I(t).
\end{split}
\end{equation}
We define $R_0 = \beta / \gamma$ in this two-parameter model.

Given the state $(S(t),I(t))$, the waiting time for the next event is distributed as $\text{Exponential}(\lambda_{SI}(t) + \lambda_{IR}(t))$. The event is an infection with probability
\begin{equation}
    \begin{split}
        q(t) = \frac{\lambda_{SI}(t)}{\lambda_{SI}(t) + \lambda_{IR}(t)},
    \end{split}
\end{equation}
and otherwise the event is a removal. Conditional on an infection event, the infected individual is sampled from $\mathcal{S}(t)$ with probabilities
\begin{equation}\label{eq:draw_het_infect}
    \begin{split}
        \Pr(J=j \mid \text{infection at } t + s)= \frac{ \sum_{k \in \mathcal{I}(t)} \beta_{kj}}{\lambda_{SI}(t)},
    \end{split}
\end{equation}
Conditional on a removal event, the removed individual is sampled from $\mathcal{I}(t)$ with probabilities
\begin{equation}\label{eq:draw_het_remove}
    \begin{split}
        \Pr(J=j \mid \text{removal at } t + s)= \frac{\gamma_j}{\lambda_{IR}(t)}.
    \end{split}
\end{equation}
Algorithms \ref{algorithm1} and \ref{algorithm2} simulate the homogeneous--rate (Equations \ref{eq:infect_SIR} and \ref{eq:remove_SIR}) and heterogeneous--rate (Equations \ref{eq:infect_het_SIR} and \ref{eq:remove_het_SIR}) models and output the infection times $\mathbf{i}$ and the removal times $\mathbf{r}$.

To include a fixed incubation period $\delta>0$, we introduce $\mathcal{E}(t)$ and exposure times $e_j$. Exposure at time $t$ moves individual $j$ from $\mathcal{S}(t)$ to $\mathcal{E}(t)$ and schedules the start of the infectious period at $i_j=e_j+\delta$. The next event is the minimum of the infection-- and removal--time clocks and the earliest scheduled infection time. If the earliest scheduled infection time occurs first, we move $\arg\min_{j\in\mathcal{E}(t)} i_j$ to $\mathcal{I}(t)$. We still define the basic reproductive number $R_0=\beta/\gamma$ in this stochastic SEIR model.

To obtain $\mathrm{Erlang}(m,\gamma_j)$ infectious periods, where $m \in \mathbb{N}$, we increment an individual counter $\mu_j$ at each removal--stage event and move $j$ from $\mathcal{I}(t)$ to $\mathcal{R}(t)$ when $\mu_j=m$. Holding the mean fixed, for $m > 1$, the duration $r_j -  i_j$ has a lower variance than that of an exponentially distributed period and the removal time $r_j$ is thus more predictable. Algorithm \ref{algorithm2} includes Erlang--distributed periods and fixed incubation periods as well.

\subsection{The complete likelihood}
\label{sec:completelik}

Following \cite{Stockdale2021-ez}, we first express the complete-data likelihood of infection times $i_1,\dots,i_n$ and removal times $r_1,\dots,r_n$ and the resulting complete-data maximum likelihood estimates (MLEs) of $\beta$ and $\gamma$. These expressions are the starting point for both prior work (which largely relies on removal times only) and our proposed estimators (Section \ref{sec:methods}). Let $\tau_{kj} := r_k \wedge e_j - e_j \wedge i_k$ denote the total time that $k$ can infect $j$, and define $\psi_j$, $\chi_j$, and $\phi_j$ as the terms for the probability that $j$ evades infection until time $i_j$, the infection hazard at $i_j$, and the probability that $j$ fails to infect the never-infected individuals:
\begin{equation}\label{eq:psi}
    \begin{split}
        \psi_j &= \exp\bigg(- \sum_{k \neq j}^n \beta_{kj} \tau_{kj} \bigg),
    \end{split}
\end{equation}
\begin{equation}\label{eq:chi}
    \begin{split}
        \chi_j &= \sum_{k\neq j}^n \beta_{kj} \cdot 1(i_k < i_j < r_k),
    \end{split}
\end{equation}
\begin{equation}\label{eq:phi}
    \begin{split}
        \phi_j &= \exp\bigg(-\sum_{k = n+1}^N \beta_{jk} \tau_{jk} \bigg).
    \end{split}
\end{equation}
The complete-data likelihood is
\begin{equation} \label{eq:likelihood} 
\begin{split}
    \mathcal{L}(\boldsymbol{\beta}, \boldsymbol{\gamma}, m; \mathbf{i}, \mathbf{r} ) &= \left\{ \prod^n_{\substack{j=2}} \psi_j \chi_j \phi_j f_j(r_j - i_j) \right\} \phi_1 f_1(r_1 - i_1).   
\end{split}
\end{equation}
where $f$ is the Erlang infectious period density. The exogenous index case could not have been infected by any other individual, hence the terms $\phi_1$ and $f_1$. The complete-data log--likelihood is
\begin{equation}\label{eq:loglikelihood}
    \begin{split}
        \mathcal{\ell}(\boldsymbol{\beta}, \boldsymbol{\gamma}, m ; \mathbf{r}, \mathbf{i}) &= \log \phi_1 +  \log f_{\gamma_1}(r_1-i_1) + \sum_{j=2}^n \log  \psi_j + \log \chi_j + \log \phi_j + \log f_{\gamma_j}(r_j-i_j) \\
        &= m \cdot \sum_{j=1}^n \log \gamma_j - \gamma_j(r_j-i_j)\\
        &-\sum_{j=1}^n \sum_{k=n+1}^N \beta_{jk} (r_j-i_j) - \sum_{j=2}^n \sum_{k\neq j}^n \beta_{kj} \tau_{kj}\\
        &+ \sum_{j=2}^n \log \bigg( \sum_{k\neq j}^n \beta_{kj} 1(i_k<i_j<r_k) \bigg)\\
        &+ (m-1) \sum_{j=1}^n \log(r_j - i_j) + n \cdot \log ((m-1)!).
    \end{split}
\end{equation}
The $\gamma_j$-- and $\beta_{kj}$--dependent terms are separable in the log--likelihood. Many models assume homogeneous rates across individuals,
\begin{equation}\label{eq:homogeneous_rates}
    \beta_{kj} = \beta/N
    \qquad\text{and}\qquad
    \gamma_j = \gamma,
\end{equation}
and we typically fix the Erlang shape at $m=1$, which simplifies Equation \ref{eq:loglikelihood}.

\subsection{The removal--time only likelihood}

When infection times are unobserved, inference is typically based on the observed removal times $\mathbf{r}$ and treats infection times as latent variables. There are two main strategies for estimation in this setting.

\subsubsection{Pair-based likelihood approximations}

Marginalizing over latent infection times gives
\begin{align}
    \mathcal{L}(\boldsymbol{\beta}, \boldsymbol{\gamma}, m; \mathbf{r})
    &= \int_{\mathcal{I}(\mathbf{r})} \mathcal{L}(\boldsymbol{\beta}, \boldsymbol{\gamma}, m; \mathbf{i}, \mathbf{r}) \, d\mathbf{i} \nonumber\\
    &= \int_{\mathcal{I}(\mathbf{r})} \left\{ \prod_{j=2}^n \psi_j(\mathbf{i},\mathbf{r}) \chi_j(\mathbf{i},\mathbf{r}) \phi_j(\mathbf{i},\mathbf{r}) f_j(r_j-i_j) \right\}\phi_1(\mathbf{i},\mathbf{r}) f_1(r_1-i_1)\, d\mathbf{i}. \label{eq:partial_likelihood_integral}
\end{align}
where $\mathcal{I}(\mathbf{r})$ is the set of infection-time vectors consistent with the observed removal times and the model constraints. The high-dimensional integral \eqref{eq:partial_likelihood_integral} has no closed form and is computationally intractable.
\cite{Stockdale2021-ez} overcome this by approximating expectations of products by products of expectations, e.g.,
\begin{equation}
    \begin{split}
        \E\bigg[ \prod_{j=2}^n \psi_j \chi_j \bigg] &\approx \prod_{j=2}^n \E[\psi_j] \cdot \E[\chi_j].\\
        \E\bigg[ \prod_{j=2}^n \psi_j \bigg] &\approx \prod_{j=2}^n \prod_{k\neq j}^n \E[\psi_{kj}],
    \end{split}
\end{equation}
This so-called pair-based likelihood approximation (PBLA) further uses $\E[\psi_j] \approx \prod_{k\neq j}^n \E[\psi_{kj}]$, where $\psi_{kj}=\exp(-\beta_{kj}\tau_{kj})$. \cite{Stockdale2021-ez} derived key expectations for efficient PBLA computation. They then estimated $\beta$ and $\gamma$ by numerical optimization, either jointly (unconditional PBLA) or for fixed $\gamma$ (conditional PBLA). 

\subsubsection{Markov Chain Monte Carlo samplers}
For Bayesian inference with the complete data likelihood, the priors of the Gamma family $\beta_N \sim \text{Gamma}(\xi_\beta, \zeta_\beta)$ and $\gamma \sim \text{Gamma}(\xi_\gamma, \zeta_\gamma)$ are conjugate priors \citep{Stockdale2019-kd,Kypraios2007-vq}, where $\beta_N = \beta/N$. Let
\begin{equation}\label{eq:A}
    \begin{split}
        A(\mathbf{r},\mathbf{i}) &= \sum_{j=1}^n (r_j - i_j),
    \end{split}
\end{equation}
\begin{equation}\label{eq:B}
    \begin{split}
        B(\mathbf{r},\mathbf{i}) &= \sum_{j=1}^{n} \sum_{k =1}^N \tau_{kj}.
    \end{split}
\end{equation}
The posterior distributions are then
\begin{equation}\label{eq:gammagibbs}
    \begin{split}
        \gamma \hskip2pt | \hskip2pt \mathbf{r}, \mathbf{i} &\sim \text{Gamma}(\xi_\gamma + m \cdot n, \zeta_\gamma  +A(\mathbf{r},\mathbf{i}) ),
    \end{split}
\end{equation}
\begin{equation}\label{eq:betagibbs}
    \begin{split}
        \beta_N \hskip2pt | \hskip2pt \mathbf{r}, \mathbf{i} &\sim \mathrm{Gamma}(\xi_\beta + n - 1, \zeta_\beta  +B(\mathbf{r},\mathbf{i}) ).
    \end{split}
\end{equation}
The posterior means are close to the MLEs (Equations \ref{eq:gammahat} and \ref{eq:betahat}) when the conjugate priors are uninformative.
\begin{equation}
    \begin{split}
        \E[\gamma \hskip2pt  | \hskip2pt \mathbf{r}, \mathbf{i}] &= \frac{\xi_\gamma + m \cdot n}{\zeta_\gamma + \sum_{j=1}^n (r_j-i_j)}, 
    \end{split}
\end{equation}
\begin{equation}
    \begin{split}
        \E[\beta \hskip2pt  | \hskip2pt \mathbf{r}, \mathbf{i}] &= \E[\beta_N \mid \mathbf{r}, \mathbf{i}] \times N = \frac{\xi_\beta + n - 1}{\zeta_\beta + \sum_{j=2}^n \sum_{k\neq j}^n \tau_{kj} + (N-n) \sum_{j=1}^n (r_j - i_j)} \times N, 
    \end{split}
\end{equation}

When we observe $i_j$ but not $r_j$, or vice versa, we can sample the missing observation in a data-augmented MCMC (DAMCMC) sampler. With Equations \ref{eq:betagibbs} and \ref{eq:gammagibbs}, we use Gibbs steps to obtain posterior samples of $\beta_N$ (and thus $\beta$) and $\gamma$, and for augmented variables $\tilde{i}_j$ or $\tilde{r}_j$, we use a Metropolis step. The most straightforward approach to data augmentation is to sample a time endpoint from the Gamma family with the current posterior sample of $\gamma$, and then devise a Hastings ratio to accept or reject the sampled endpoint. Variations of this MCMC scheme appear in \cite{Neal2005-lo}, \cite{Kypraios2007-vq}, and \cite{Stockdale2019-kd}. There can be high dependence between the augmented variables $\tilde{i}_j$ or $\tilde{r}_j$ and the posterior samples of $\beta$ and $\gamma$. \cite{Neal2005-lo} and \cite{Kypraios2007-vq} provided algorithms to decouple the dependence between the missing data and the parameters to some success, particularly for when the infection times $\mathbf{i}$ are missing. In Section \ref{sec:methods}, we propose another DAMCMC sampler for when $i_j$ or $r_j$ are missing but not both.

\section{Imputation-based estimators of infection and removal rates}
\label{sec:methods}

As noted in Section \ref{sec:intro},
the removal rate $\gamma$ is in principle easy to estimate once even a modest number of infectious periods are fully observed: with $r_j-i_j\sim\text{Gamma}(m,\gamma)$, the usual MLE is unbiased, consistent, and efficient, and should be accurate with tens of complete cases \citep{Casella1993-tf}. This suggests a practical design for outbreak investigations: collect removal times broadly, but also collect a small calibration sample of fully observed infectious periods (or close proxies) to estimate $\gamma$ largely independently of $\beta$.

\subsection{Complete--data maximum likelihood estimation}

For our EM estimators, we use the complete--data MLEs under homogeneous rates $\beta_{kj}=\beta/N$ and $\gamma_j=\gamma$. The removal--rate estimator is the usual MLE of an exponential decay parameter:
\begin{equation}\label{eq:gammahat}
    \begin{split}
        \hat{\gamma} &= \frac{m \cdot n}{\sum_{j=1}^n (r_j - i_j)}.
    \end{split}
\end{equation}
The score for $\beta$ is
\begin{equation}\label{eq:partial}
    \begin{split}
        \frac{\partial{\ell(\cdot)}}{\partial{\beta}} &= \beta^{-1} (n-1) - N^{-1} \cdot \bigg(\sum_{j=1}^n \sum_{k=n+1}^N (r_j - i_j) + \sum_{j=2}^n \sum_{k\neq j}^n \tau_{kj} \bigg).
    \end{split}
\end{equation}
Setting this score to zero gives
\begin{equation}\label{eq:betahat}
    \begin{split}
        \hat{\beta} &= \frac{n-1}{\sum_{j=2}^n \sum_{k\neq j}^n \tau_{kj} + (N-n) \sum_{j=1}^n (r_j - i_j)} \times N
    \end{split}
\end{equation}
The numerator counts post-index infections, and the denominator is the total infectious exposure over infected and never-infected individuals. This estimator is consistent and asymptotically normally distributed \citep{Rida1991-vt}. In Appendix \ref{sec:hsem}, we describe models with group--specific infection rates $\beta_g$ and/or a baseline infection rate $\beta_0$ scaled by a kernel function $h(\mathbf{x}_k,\mathbf{x}_j)$. The MLEs of  $\hat{\beta}_g$ and $\hat{\beta}_0$ are defined similarly.

\subsection{Expectation--maximization estimators}

Our starting point is the complete--data MLE $\hat{\beta}$ (Equation \ref{eq:betahat}), whose denominator is a sum of two kinds of terms: infectious period lengths $(r_j-i_j)$ and pairwise infective time pressures $\tau_{kj}$. We focus on partial endpoint observation, where each infected individual $j$ has at least one of $(i_j,r_j)$ observed, and a subset of individuals have both endpoints observed (i.e. complete infectious periods). When some infection or removal times are missing, the terms in the $\hat{\beta}$ denominator cannot be computed directly. We therefore adopt an EM--style plug-in strategy: estimate $\gamma$ from the complete infectious periods, and then impute the missing terms in the $\hat{\beta}$ denominator using their conditional expectations under this calibrated $\hat{\gamma}$. Specifically, if $\mathcal{C}$ is the set of individuals with fully observed infectious periods and $n_{\mathcal{C}}:=|\mathcal{C}|$, we take $\hat{\gamma}=m \cdot  n_{\mathcal{C}}/\sum_{j\in\mathcal{C}}(r_j-i_j)$.

\paragraph{A baseline endpoint--imputation estimator.}
We define $\bar{\beta}$ as the complete--data MLE (Equation \ref{eq:betahat}) computed after replacing $(\mathbf{i},\mathbf{r})$ with $(\bar{\mathbf{i}},\bar{\mathbf{r}})$, where
\[
\bar{i}_j :=
\begin{cases}
i_j, & i_j \text{ observed},\\
r_j - m/\hat{\gamma}, & i_j \text{ missing},
\end{cases}
\qquad
\bar{r}_j :=
\begin{cases}
r_j, & r_j \text{ observed},\\
i_j + m/\hat{\gamma}, & r_j \text{ missing}.
\end{cases}
\]
This estimator uses only the mean infectious period length to fill missing endpoints and does not condition on the relative order of observed endpoints across individuals, which is what drives the pairwise exposure terms $\tau_{kj}$. We include $\bar{\beta}$ primarily as a simple baseline for comparison in the simulation study (Section \ref{sec:simulations}) to isolate the benefit of imputing the exposure terms $\tau_{kj}$ using their conditional expectations rather than imputing endpoints by a mean--duration rule.

\paragraph{A $\tau_{kj}$--imputation estimator.}
Our main estimator imputes the exposure terms directly by replacing $\tau_{kj}$ with $\E_{\hat{\gamma}}[\tau_{kj}]$ when the endpoints needed to compute $\tau_{kj}$ are missing, and similarly replacing $(r_j-i_j)$ by $\E_{\hat{\gamma}}[r_j-i_j]$ when one of its endpoints is missing. Here, $\E_{\hat{\gamma}}[\cdot]$ denotes a conditional expectation given the observed endpoints for the individuals involved; in particular, if the relevant endpoints are observed then the expectation reduces to the observed value. The resulting estimator is
\begin{equation}\label{eq:betahattau}
    \begin{split}
        \tilde{\beta} &= \frac{n-1}{\sum_{j=2}^n \sum_{k\neq j}^n \E_{\hat{\gamma}}[\tau_{kj}] + (N-n) \sum_{j=1}^n \E_{\hat{\gamma}}[r_j - i_j]} \times N.
    \end{split}
\end{equation}
The $\tau_{kj}$--based group--specific and baseline infection rate estimators $\tilde{\beta}_g$ and $\tilde{\beta}_0$ are defined similarly (Appendix \ref{sec:hsem}). 

To implement $\tilde{\beta}$ we need closed-form expressions for $\E_{\hat{\gamma}}[\tau_{kj}]$ under the different patterns of which endpoints among $\{i_k,r_k,i_j,r_j\}$ are observed. Since $\tau_{kj}$ is known exactly when $r_k,i_k,$ and $i_j$ are observed, there are seven distinct partial observation patterns for which $\E[\tau_{kj}]$ is needed:
\[
(r_k,r_j),\ (i_k,i_j),\ (r_k,i_j),\ (r_j,i_k),\ (r_k,r_j,i_k),\ (r_k,r_j,i_j),\ (r_j,i_k,i_j).
\]
(The PBLA method requires only the $(r_k,r_j)$ case.) The expected value $\E[\tau_{kj}]$ is the same for the partial observation patterns $(r_j,i_k,i_j)$ and $(i_k,i_j)$ and $(r_k,r_j,i_j)$ and $(r_k,i_j)$, respectively, because $r_j$ does not provide additional information when $i_j$ is observed. We derive each expectation in Appendix \ref{sec:appendix}.

For incubation periods of fixed length $\delta > 0$, the infective time pressure is $\tau_{kj} = r_k \wedge e_j - e_j \wedge i_k$ instead of $\tau_{kj} = r_k \wedge i_j - i_j \wedge i_k$. If $i_j$ is observed, then $e_j=i_j-\delta$ is observed, and if $i_j$ is missing but $r_j$ is observed, then $e_j=r_j-\delta-(r_j-i_j)$ is determined by the duration of the latent infectious period. The same expressions (Lemmas \ref{lemma2}, \ref{lemma4}, \ref{lemma3}, \ref{lemma1}, and \ref{hardlemma} in Appendix \ref{sec:appendix}) apply after shifting times as in \cite[p.~591]{Stockdale2021-ez}.

\paragraph{A pathology under removal time--only observation}
We suggest in the following that the imputed value $\E[\tau_{kj}]$ may poorly estimate the underlying term $\tau_{kj}$ when only removal times $r_k$ and $r_j$ are observed. Consider the homogeneous SIR model and assume that only removal times are observed. Let $\gamma^\dagger$ be whatever estimate of $\gamma$ is plugged into the imputation formulas.
\begin{proposition}\label{pathological}For any estimate $\gamma^\dagger$ of the removal rate $\gamma$, the basic reproductive number estimate $\tilde{R}_0 = \tilde{\beta} / \gamma^\dagger$ is 2 when the outbreak infects the entire population.

    \begin{proof} The estimate $\tilde{\beta}$ simplies when $n = N$ to
    \begin{equation}
        \tilde{\beta} = \frac{N-1}{\sum_{j=2}^N \sum_{k\neq j}^N \E_{\gamma^\dagger}[\tau_{kj}]} \times N.
    \end{equation}
    By Lemma 2 of \cite[p. 585]{Stockdale2021-ez}, the denominator is distributed as
    \begin{equation}
        \sum_{j=2}^N \sum_{k\neq j}^N \tau_{kj} \sim \sum_{j=1}^{N-1} j \cdot W_j, 
    \end{equation}
    where $W_1,\dots,W_{N-1}$ are independent, identically distributed $\mathrm{Exponential}(\gamma)$ random variables. Next, the imputed sum is
    \begin{equation}
        \sum_{j=2}^N \sum_{k\neq j}^N \E_{\gamma^\dagger}[\tau_{kj}] = \gamma_\dagger^{-1} \sum_{j=1}^{N-1} j = \gamma_\dagger^{-1} N(N-1)/2.
    \end{equation}
    Finally, the basic reproductive number estimate is
    \begin{equation}
        \tilde{R}_0 = \gamma_\dagger^{-1} \times \frac{N-1}{\gamma^{-1}_\dagger (N(N-1)/2)} \times N = 2.
    \end{equation}
    \end{proof}
\end{proposition}

We claim that this behavior is pathological: in the deterministic SIR model under homogeneous mixing, the final epidemic size in the infinite-population limit satisfies the final-size relation \citep{Miller2012-dq} 
\begin{equation}
    1-n/N = \exp(-R_0(1-(1- n/N))) = \exp(-R_0 \cdot n/N)
\end{equation}
Hence, $R_0$ has to diverge as $n/N\to 1$. An estimator that returns $\tilde{R}_0=2$ even when the entire population is infected is therefore ignoring precisely the feature of the data that should signal very rapid transmission.

More generally, let $\mathcal{K}_{\mathbf{r}}$ be the subset of infected individuals for whom only removal times are observed. Then, the contribution of the removal time--only pairs within $\mathcal{K}_{\mathbf{r}}$ to the imputed denominator is
\begin{equation}
    \sum_{j \in \mathcal{K}_\mathbf{r}} \sum_{\substack{k \neq j \\ k \in \mathcal{K}_\mathbf{r}}} \E_{\hat{\gamma}}[\tau_{kj}]
    = \frac{|\mathcal{K}_\mathbf{r}|(|\mathcal{K}_\mathbf{r}|-1)}{2\hat{\gamma}}.
\end{equation}
As the number of removal--only cases grows, a larger fraction of the denominator in $\tilde{\beta}$ is driven by this mechanically averaged term rather than by the observed temporal ordering of endpoints. This helps explain the downward bias we see in simulations when many infection times are missing. The PBLA method uses a different approximation, but it faces the same broader problem: when too many pairwise contributions are replaced by removal time--only expectations, the data carry too little information to distinguish transmission from removal.

\subsection{Uncertainty Quantification}

To quantify uncertainty in our estimates we studied two approaches. 
The first was bootstrapping. Note that the dependency structure of the data preclude naive nonparametric bootstrapping: resampling removal and infection times $\{\{(r_j,i_j)^*\}_{1:n}\}_{1:B}$ from the observed data $\{(r_j,i_j)\}_{1:n}$ (nonparametric bootstrapping) could result e.g.~$I(t_{j})=0$ and $E(t_{j})=0$ but $I(t_{j+1}) > 0$ for some $t_j < t_{j+1}$. As a result, we relied on parametric bootstrapping.

Given infection and removal rate estimates $\tilde{\beta}$ and $\hat{\gamma}$, we used Algorithms \ref{algorithm1} and \ref{algorithm2} to simulate another SEM and estimate bootstrap samples $\{\tilde{\beta}^*\}_{1:B}$ and $\{\tilde{R}_0^*\}_{1:B}$, where $R_{0}^* = \tilde{\beta}^* / \hat{\gamma}^*$. We performed conditional simulations until the bootstrap epidemic size $n^*$ is within 10\% of the initial epidemic size $n$. These conditional simulations might require a couple of runs when $n/N$ is close to 1 but many more when $n/N$ is far from 1 (consistent with lower $R_0$).

We can construct a variety of confidence intervals \citep{Carpenter2000-pe,DiCiccio1996-jv,Efron1994-hw} with different asymptotic behaviors on the accuracy and correctness of confidence intervals \citep{Hall1988-rp}. From our simulation study (Section \ref{sec:simulations}), we explored the basic and percentile bootstrap procedures with recentering (bias-correction), as well as the bootstrap$-t$ procedure (otherwise known as the studentized bootstrap). We found that the bootstrap$-t$ approach offered better empirical coverage, so we focus on it in our results. We also considered the midpoint of the bootstrap$-t$ interval as a bias-corrected estimate of the infection rate $\beta$. This approach involved double bootstrapping \citep{Carpenter2000-pe}, which further reduced the bias of $\tilde{\beta}$ for $R_0 \geq 3$ simulations. The precise details of our bootstrap$-t$ procedure are in Algorithm \ref{algorithm3}. 

For the removal rate $\gamma$, we can form a Wald-type confidence interval due to the asymptotic normality of the MLE $\hat{\gamma}$ (Equation \ref{eq:gammahat}). The ideal performance of such an interval estimator is well-established when there are enough fully observed infectious periods \citep{Casella1993-tf}, so we do not discuss $\hat{\gamma}-$confidence intervals.

Lastly, we implemented a full Bayesian approach based on DAMCMC.
We sampled the SEM parameters $\beta_N$ and $\gamma$ and the missing data from a Metropolis-within-Gibbs algorithm. The Gibbs steps involved updating $\beta_N$ and $\gamma$ with Equations \ref{eq:betagibbs} and \ref{eq:gammagibbs}. The missing data was augmented and updated with a marginal Metropolis--Hastings step \citep{Metropolis1953-bq,Hastings1970-ch}. To update the times, we sampled $\tilde{u} \sim \text{Erlang}(m,1)$, and then we proposed either the infection time $\tilde{i}_j = r_j - \gamma \cdot \tilde{u}$ or the removal time $\tilde{r}_j = i_j + \gamma \cdot \tilde{u}$. We accepted the augmented infection and removal times based on the Hastings ratios in Equations \ref{eq:hastinginfection} and \ref{eq:hastingremoval}. The full details are in Algorithm \ref{algorithm4}. In our code implementation, we generalized the sampling algorithm to include multitype and spatial models as in Appendix \ref{sec:hsem} (details omitted).

\section{Simulation study}
\label{sec:simulations}

We conducted a simulation study to evaluate the estimators under different epidemic dynamics. The point estimators we evaluated were the unconditional and conditional PBLA estimates, the $\tilde{\beta}$ and $\bar{\beta}$ estimators, the midpoint of the bootstrap$-t$ interval, and the Bayesian posterior mean. For each parameter setting, we generated epidemic datasets using the event-driven simulators in Algorithms \ref{algorithm1} and \ref{algorithm2}. Unless otherwise noted, the simulations were run in a closed population of size $N=100$ with the removal rate $\gamma=1$, the infection rate $\beta \in \{1,2,3,4,5\}$, and the Erlang shape $m=1$. We retained only outbreaks with at least $n\geq 20$ infected individuals and performed 1000 Monte Carlo replicates for each combination of parameters. Minor deviations from this baseline design are described in the corresponding results, which include increasing the population size $N$, changing the Erlang shape to $m$, and estimating smaller and larger infection rates $\beta$.

To measure the empirical coverage of the interval estimates, we studied only 200 of the 1000 Monte Carlo replicates. We calculated 95\% bootstrap$-t$ and Bayesian credible intervals (the 2.5th and 97.5th percentiles), respectively. For each bootstrap$-t$ interval, we used 20 inner samples $\{\tilde{\beta}^{**}\}_{1:20}$ to normalize the bootstrap estimates $\{\tilde{\beta}^*\}_{1:200}$ and then the 200 outer samples $\{t^*\}_{1:200}$ to calculate $t^*_{0.025}$ and $t^*_{0.975}$. For each credible interval, we used 300 posterior samples without thinning or burn-in. We used a modest number of bootstrap and MCMC samples to limit computational runtime. Calculating the 2.5th and 97.5th percentiles from this few samples could result in noisy estimates of coverage. 

\subsection{Estimation without infection times}

\begin{figure}[!htp]
\centering\includegraphics[scale=0.85]{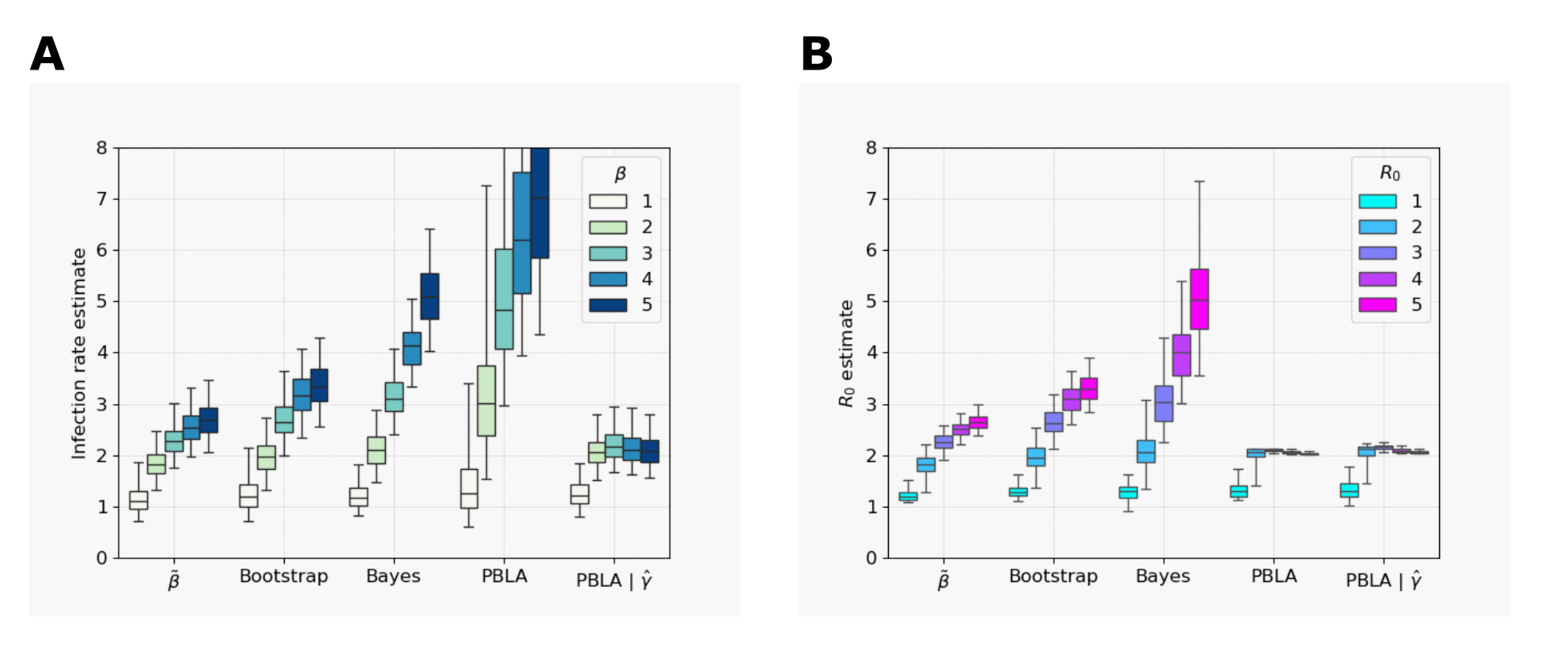}
\caption{SIR estimators that use some versus zero fully observed infectious periods. Box plots show the 2.5th, 25th, 50th, 75th, and 97.5th percentiles of  A) infection rate $\beta$ and B) basic reproductive number $R_0$ estimates (y-axis) from different methods (x-axis). The uncorrected EM method $\tilde{\beta}$, the midpoint of a studentized bootstrap interval, and the Bayesian posterior mean came from simulations with $p_1=0.4$ and $p_2=0.8$. The unconditional and conditional PBLA estimates came from simulations with $p_1=0$ and $p_1=0.5$, respectively. The fixed incubation period was 0.}
\label{fig:pbla}
\end{figure}

The PBLA and DAMCMC methods can estimate $\beta$ and $\gamma$ without infection times $\mathbf{i}$. We found that the unconditional PBLA estimates of $\beta$ were heavily biased and varied widely when $\beta \geq 2$ (Figure \ref{fig:pbla}A). On the other hand, the unconditional PBLA estimates of $R_0$ were almost always close to 2 when $R_0 \geq 2$ (Figure \ref{fig:pbla}B). The Pearson correlation coefficient of the unconditional PBLA estimates of $\beta$ and $\gamma$ was nearly 1 (Figure \ref{fig:pearson}A), as was the Pearson correlation coefficient of the unconditional PBLA estimates of $R_0$ and the prevalences $n/N$. 

Next, we studied whether the PBLA method could accurately estimate $\beta$ conditional on the MLE $\hat{\gamma}$ from 50\% of infectious periods $\{r_j-i_j\}_{1:n}$. The conditional PBLA estimates of $\beta$, and also of $R_0$, were less variable than the unconditional estimates, but still biased when $\beta > 2$ (Figure \ref{fig:pbla}). The Pearson correlation coefficient of the conditional PBLA estimates of $\beta$ and the MLEs $\hat{\gamma}$ was nearly 0.70 (Figure \ref{fig:pearson}A). \cite{Stockdale2021-ez} remarked that their likelihood approximation performed poorly when the prevalence exceeded 0.70, which was almost always the case for the $\beta \geq 2$ ($R_0 \geq 2$) simulations. Regarding Proposition \ref{pathological}, the unconditional and conditional PBLA methods are very close to 2 in all such epidemics. 

The performance of the PBLA estimators compared unfavorably to that of the $\tilde{\beta}$ estimator, the bootstrap midpoint, and the Bayesian posterior mean. Those methods that observed entire infectious periods for 40\% of infections provided less biased estimates of $\beta$ and $R_0$ for $R_0 \geq 2$ (Figure \ref{fig:pbla}).  The Pearson correlation coefficients of their estimates of $\beta$ and $\gamma$ were between 0.18 and 0.30, and the Pearson correlation coefficients between the prevalences and the estimates of $\beta \geq 2$ were less than 0.50 (Figure \ref{fig:pearson}A-B). Finally, the Pearson correlation coefficients of their estimates of $\beta$ and $\gamma$ decreased as more infectious periods were fully observed (Figure \ref{fig:pearson}C-D).

\begin{figure}[!htp]
\centering\includegraphics[scale=0.85]{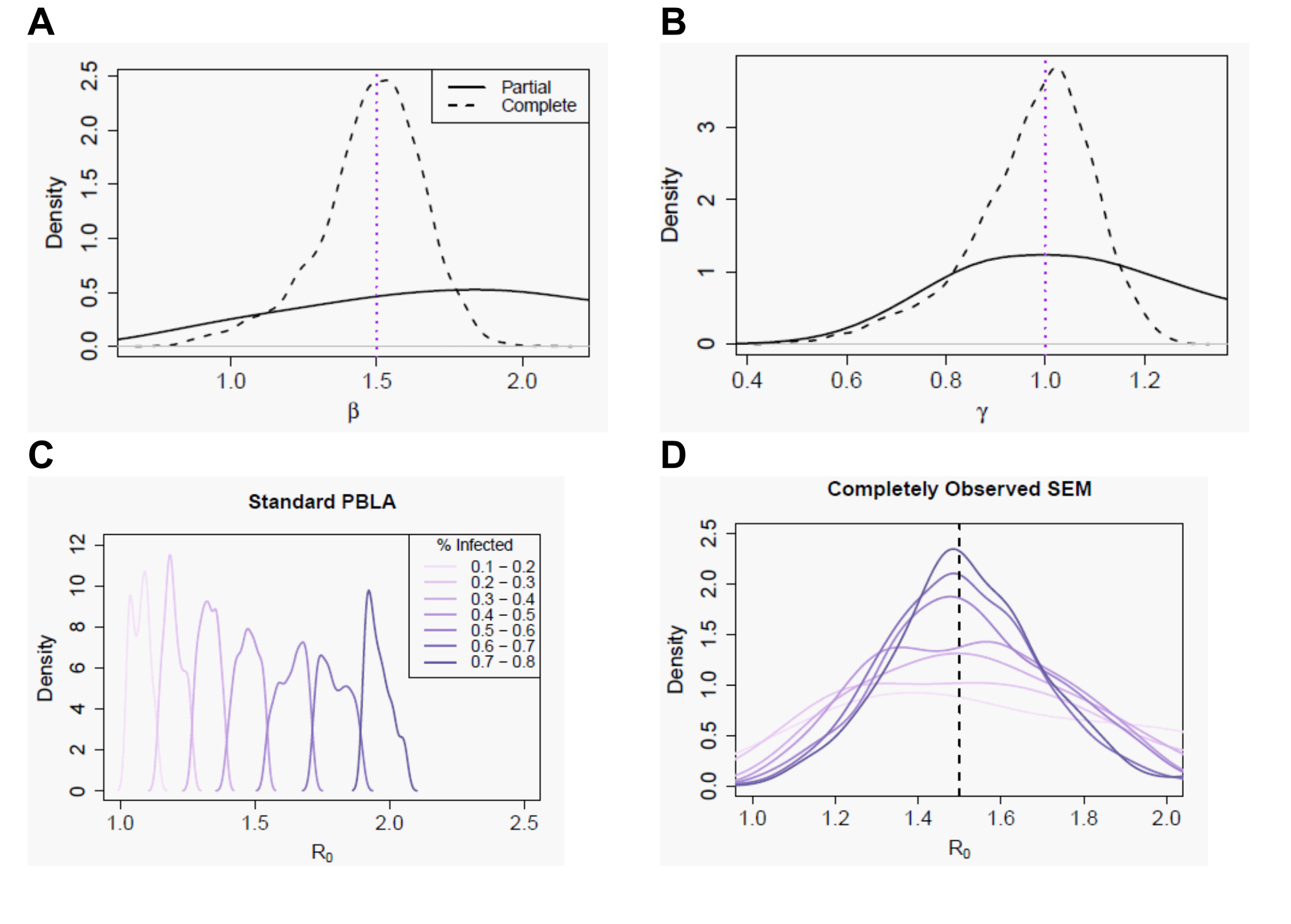}
\caption{The pair-based likelihood approximation is flat and heavily affected by the epidemic size. A-B) Kernel density estimates (KDEs) of the unconditional PBLA estimates (solid black) and the MLEs from complete data (dashed black) are juxtaposed for the infection rate $\beta=1.5$ and the removal rate $\gamma=1$. C) KDEs of the basic reproductive number $R_0=\beta/\gamma$ from the unconditional PBLA method changed as we increased the percent of susceptible individuals infected (purple gradients in legend), whereas D) KDEs of the MLEs from complete data remained centered around the true $R_0=1.5$. The population size $N$ is 200, the Erlang shape $m$ is 1, and there is no fixed delay (SIR model). For each decile in C,D), we performed enough simulations so that the KDEs are based on at least 50 simulations.}
\label{fig:prelim}
\end{figure}

In a small experiment, the unconditional PBLA method performed poorly in relation to the MLEs $\hat{\gamma}$ and $\hat{\beta}$ (Equations \ref{eq:gammahat} and \ref{eq:betahat}). Although \cite{Stockdale2021-ez} show that the central tendencies of the unconditional PBLA estimates of $\beta$ and $\gamma$ were close to their true values for  stochastic SIR models with $R_0 < 2$, Figure \ref{fig:prelim}A-B shows that the variability of the PBLA estimates of $\beta=1.5$ and $\gamma=1$ were much greater than the variability of the MLEs. Moreover, Figure \ref{fig:prelim}C shows that the PBLA estimates of $R_0$ were highly correlated with the percentage of infected individuals $n/N$. Figure \ref{fig:prelim}D shows that the MLEs of $R_0$ were centered on the true value of 1.5 for all $n/N$. This result suggests that the unconditional PBLA estimates of $\beta$ and $\gamma$ were also unreliable when $R_0 < 2$, which is a regime where the prevalences were less than 0.70 in many simulations. \cite{Becker1993-fe} would contest that comparing the results from complete data with partial data is unfair, which we agree with. We are not interested in which removal time-only estimators perform better, but rather if any removal time-only estimators are accurate and have low variance.

\begin{figure}[!htp]
\centering\includegraphics[scale=0.85]{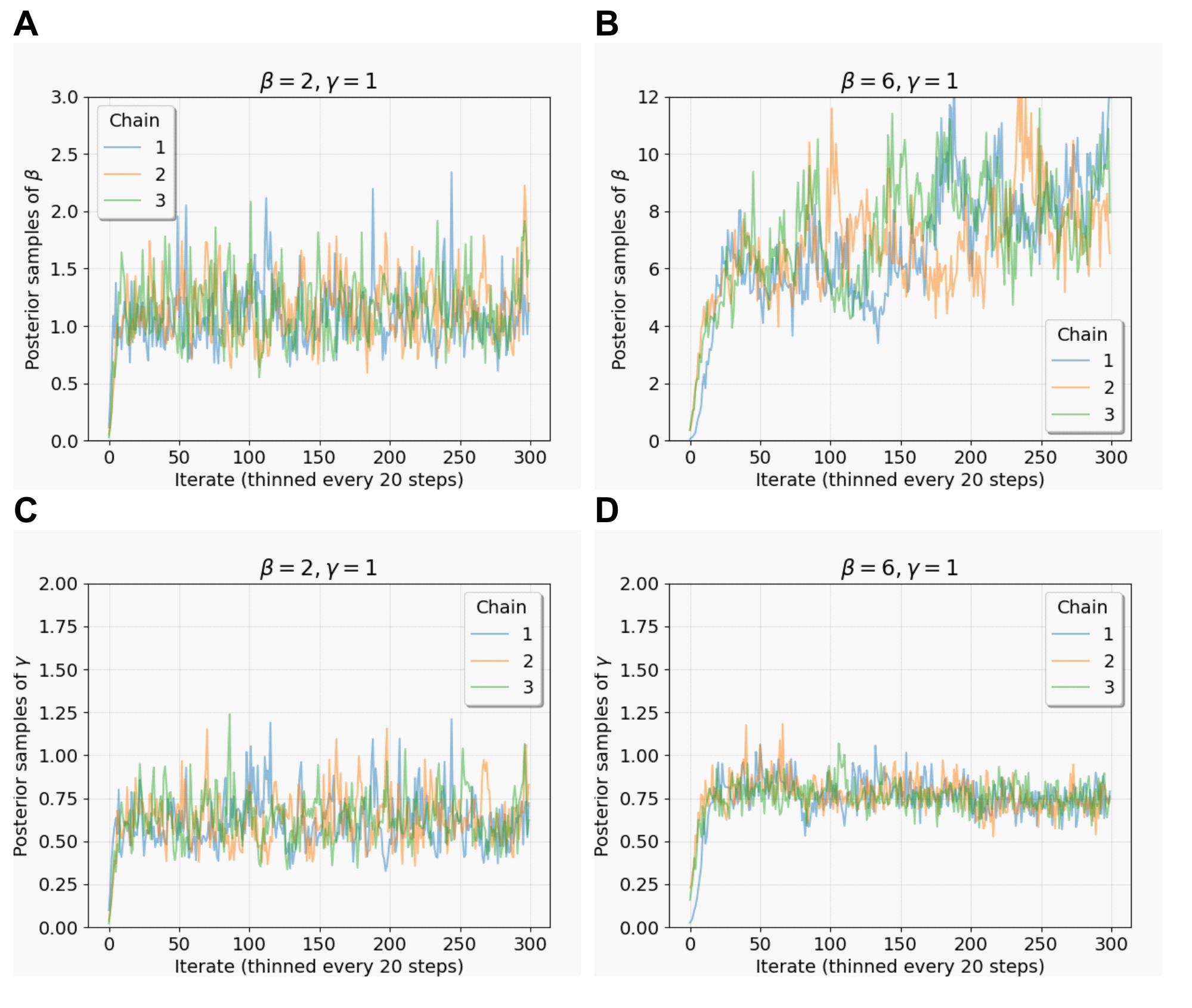}
\caption{Example trace plots for Bayesian SIR models without infection times. Line plots show the posterior samples of A,C) $\beta$ and B,D) $\gamma$ (y-axes) over iterations thinned every 20 steps (x-axis) for three chains (legend). The true simulation values of $\beta$ and $\gamma$ are specified in the subplot titles.}
\label{fig:nonetrace}
\end{figure}

\begin{figure}[!htp]
\centering\includegraphics[scale=0.85]{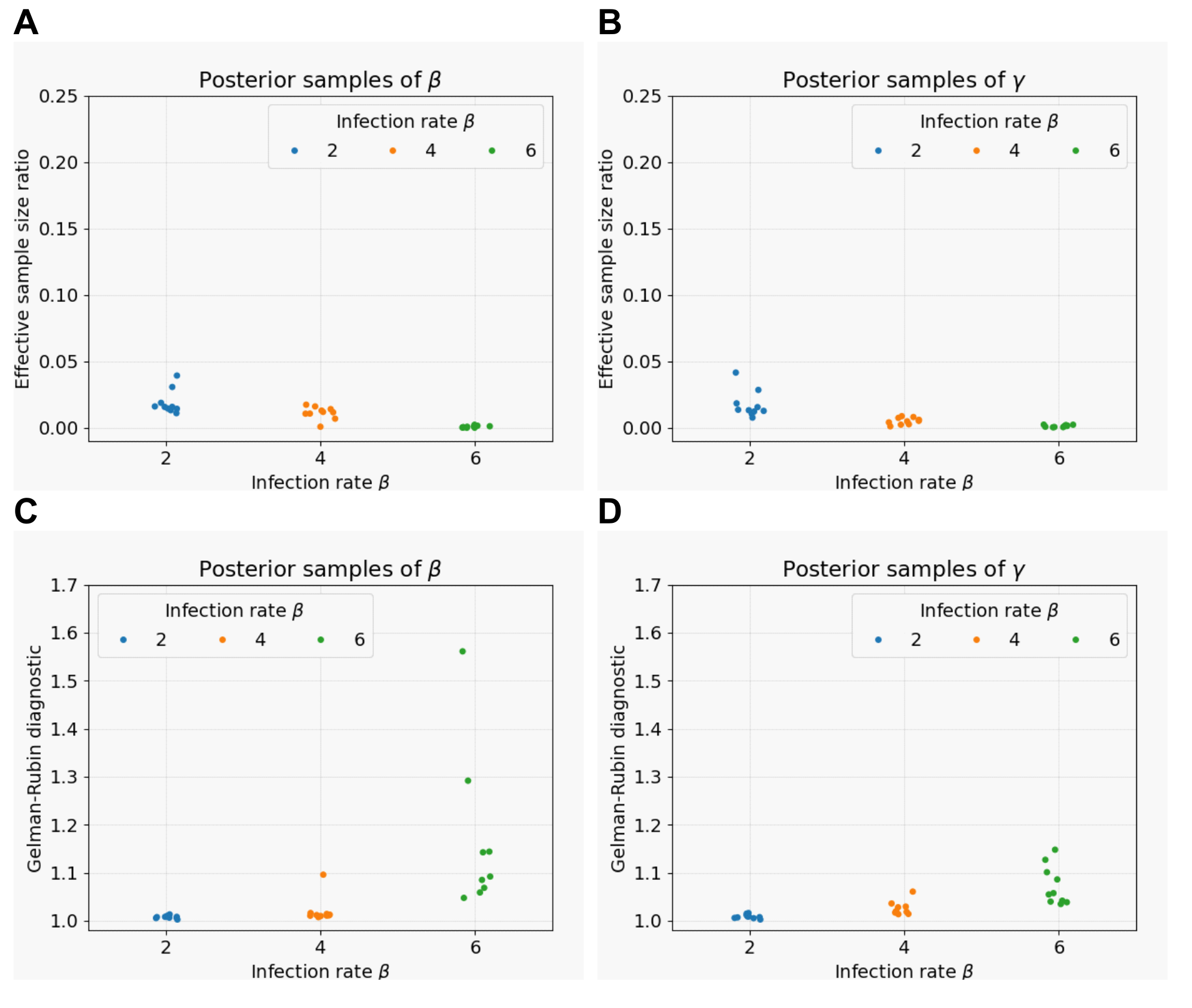}
\caption{Markov Chain Monte Carlo diagnostics for Bayesian SIR models without infection times. Strip plots show the A-B) effective sample size ratios and C-D) Gelman-Rubin statistic (y-axes) from posterior samples of A,C) $\beta$ and B,D) $\gamma$ (subplot titles). There are 10 simulations for each infection rate $\beta \in \{2,4,6\}$ (x-axis, legend). We used 10 mixing chains. We discarded the first 2000 posterior samples as burn-in and then calculated diagnostics from the remaining 8000 posterior samples. The diagnostics are computed with the Python functions \texttt{ess} and \texttt{rhat} in the \texttt{arviz} package.}
\label{fig:nonemcmc}
\end{figure}

\begin{figure}[!htp]
\centering\includegraphics[scale=0.85]{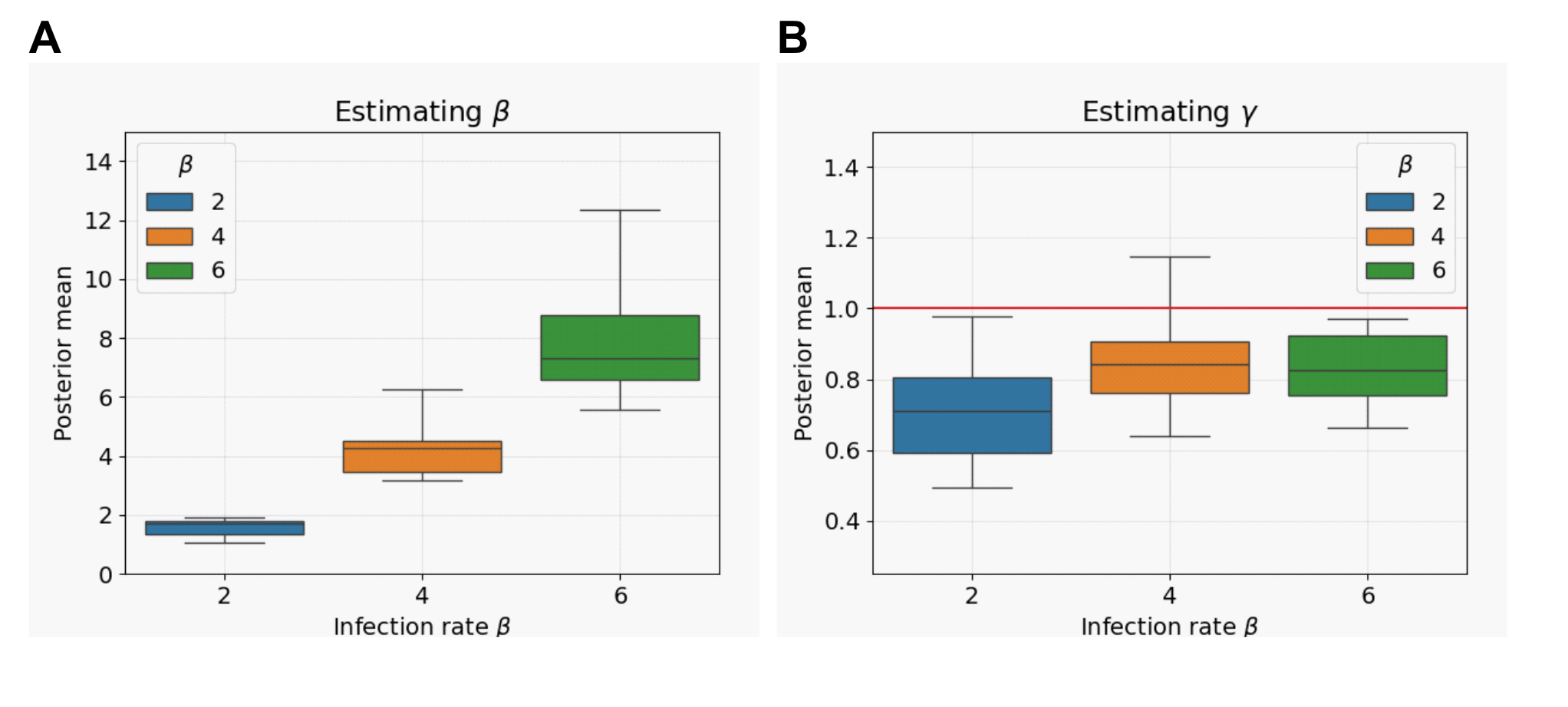}
\caption{Posterior mean infection and removal rates in SEIR epidemic models when no infection times are observed. Box plots show the 2.5th, 25th, 50th, 75th, and 97.5th percentiles of posterior means (y-axis) of A) $\beta$ and B) $\gamma$ as we increased the infection rate $\beta$ (x-axis). The true removal rate $\gamma=1$ (horizontal red line). The posterior means are calculated from 400 samples (MCMC chains of 10000 samples followed by a burn-in period of 2000 samples and thinning every 20 samples). Each box is based on 100 posterior means. The fixed incubation period $\delta$ is 0, the total susceptible population size $N$ is 100, the minimum epidemic size $n$ is 20, the removal rate $\gamma$ is 1, and the Erlang shape parameter $m$ is 1.}
\label{fig:nonemean}
\end{figure}

In another small experiment, we found that the DAMCMC method was biased and inefficient when no infection times were observed. We generated ten SIR datasets each for $\beta \in \{2,4,6\}$ and $\gamma=1$. For each dataset, we ran 10 MCMC chains to obtain 10,000 posterior samples of $\beta$ and $\gamma$ with uninformative priors and initial values far from the true values. From the trace plots in Figure \ref{fig:nonetrace}, we observed that the posterior samples approached the true values even when poorly initialized. However, we measured that the effective sample size ratios were less than 0.05 in all simulations and that the Gelman-Rubin diagnostics were greater than 1.05 and sometimes as high as 1.55 in the $\beta=6$ simulations (Figure \ref{fig:nonemcmc}) \citep{Gelman1995-ru}. We observed that the posterior correlations between $\beta$ and $\gamma$ were greater than 0.5 when $\beta=2$ and decreased towards 0 for $\beta>2$. The posterior means of each chain were also biased, especially for the removal rate $\gamma=1$ (Figure \ref{fig:nonemean}). The removal rate estimate informs how long an individual is infectious for, which is an important factor in public health decision making. With expertise and careful monitoring of the DAMCMC chains, it might be possible to attain more accurate results for this simple SIR model.

\subsection{Estimation with some infection times}

For our primary analysis, we randomly set some infection and removal times to be missing, and then we estimated $\beta$ with the EM estimators $\tilde{\beta}$ and $\bar{\beta}$, the midpoint of the bootstrap$-t$ interval (Algorithm \ref{algorithm3}), and the posterior mean of Algorithm \ref{algorithm4}. We investigated 1) how many infectious periods were enough to reasonably estimate $\beta$ and 2) whether infection or removal times were more important in the estimation. We designed most of the following computer experiments to demonstrate the qualitative trends of $\tilde{\beta}$ under different missing data scenarios, not to highlight the best possible estimation accuracy, which was often achieved by the midpoint of the bootstrap$-t$ interval or the posterior mean. Let $p_1,p_2\in \{0.4,0.6,0.8\}$ be the expected proportion of infectious periods with missing values and the expected proportion of those for which the missing value is an infection time.

\subsubsection{Behavior of the imputation-based estimators}\label{sec:initial}

First, we fixed $p_2=0.8$ and varied $p_1 \in \{0.4,0.6,0.8\}$ to measure how the magnitude of missingness impacts inference. The frequentist $\tilde{\beta}$ estimates were centered around the true values $\beta=1$ and $ \beta= 2$ and approached the true $\beta \geq 3$ as $p_1$ increased (Figure \ref{fig:eminfection}A). In the sum $\sum_{j=2}^n \sum_{k\neq j}^n \tau_{kj}$, around $100(p_1^2 + p_1(1-p_1)(1-p_2)/2)$ percent of the $\tau_{kj}$ terms are known exactly and do not require imputation. For example, around 18.4\% versus 65.6\% of the terms are known exactly when $p_1=0.4$ versus $p_1=0.8$.

The EM estimator $\bar{\beta}$ that sets missing $i_j \gets \E[i_j]$ and $r_j \gets \E[r_j]$ was marginally more accurate than $\tilde{\beta}$ but the bias still increased as $\beta$ increased (Figure \ref{fig:eminfection}D). In the EM approaches, we identified that the term $\sum_{j=2}^n \sum_{k\neq j}^n \E[\tau_{kj}]$ was typically an overestimate of the term $\sum_{j=2}^n \sum_{k\neq j}^n \tau_{kj}$ in the denominator of Equation \ref{eq:betahat}, leading to the downward biases of $\tilde{\beta}$ and $\bar{\beta}$.

The basic reproductive number $R_0$ was the main factor that affected the biases of $\tilde{\beta}$ and $\bar{\beta}$. Since $\gamma=1$, the box plots of the $\tilde{R}_0$ and $\bar{R}_0$ estimates look similar to the box plots of the $\tilde{\beta}$ and $\bar{\beta}$ estimates (Figures \ref{fig:embasicnumber}A,D and \ref{fig:eminfection}A,D). For $\gamma=2$ and fixed $R_0 \in \{1,2,3,4,5\}$, Figure \ref{fig:bigremovalrate}D shows that $\tilde{R}_0$ estimates were similar to those in Figure \ref{fig:embasicnumber}A, whilst the $\tilde{\beta}$ estimates are closer to the true values $\beta \in \{2,4,6,8,10\}$ (Figure \ref{fig:bigremovalrate}A).

The infectious period model $r_j-i_j \sim$ Erlang($m,\gamma$) and the population size $N$ affected the variance of the estimates. Recall that we assumed exponentially distributed infectious periods in the EM estimator $\tilde{\beta}$. The $\tilde{\beta}$ estimates were more biased and had lower variance when we increased the Erlang shape $m$ from 1 to 2 (Figure \ref{fig:erlang}). And, as anticipated, the variance of the $\tilde{\beta}$ estimates decreased and their bias remained unchanged as we increased the population size $N$ from 100 to 200 (Figure \ref{fig:populationsize}).

Figure \ref{fig:smallinfectionrate}A shows that the $\tilde{\beta}$ estimates were reasonably accurate for $\beta \in \{1,1.33,1.66,2\}$ despite the method being biased for $\beta>2$. The $\tilde{\beta}$ estimates for $\beta \in \{0.66,1,1.33\}$ were slightly inflated, but recall that we simulated datasets conditional on prevalence $n/N > 0.2$. Thus, the datasets were more reflective of $\beta > 1$ than their true values because many epidemics with $R_0 \leq 1$ do not exceed the prevalence threshold of 20\%.

Second, we fixed $p_1=0.4$ and varied $p_2 \in \{0.4,0.6,0.8\}$ to evaluate whether infection times were more or less useful than removal times for estimation (Figures \ref{fig:eminfection}B,E, \ref{fig:embasicnumber}B,E, \ref{fig:bigremovalrate}B,E, and \ref{fig:smallinfectionrate}B). For $R_0 \leq 3$, the $\tilde{\beta}$ and $\bar{\beta}$ estimates did not change much when we increased $p_2$. For $R_0 > 3$, we noticed that all estimates decreased slightly as we increased $p_2$. The variance of the estimates did not change significantly as we increased $p_2$. In the sum $\sum_{j=2}^n \sum_{k\neq j}^n \tau_{kj}$, around $100\cdot p_2^2(1-p_1)^2$ percent of the terms are imputed with only $r_k$ and $r_j$ observed, which we have suggested could be poor replacements (Section \ref{sec:methods}). For example, around 23.4\% versus 5.8\% of the terms involve such imputations when $p_2=0.8$ versus $p_2=0.4$.

Third, we fixed $p_1=0.4$ and $p_2=0.8$ and varied the fixed incubation period $\delta \in \{0,1,2\}$. In these SEIR models with $R_0 \in \{3,4,5\}$, the biases of the $\tilde{\beta}$ and $\bar{\beta}$ and $\tilde{R}_0$ and $\bar{R}_0$ estimates were often between 0.25 and 1 point less than those of the SIR model (Figures \ref{fig:eminfection}C,F and \ref{fig:embasicnumber}C,F). Nevertheless, when $\delta\geq 1$, the 2.5th and 97.5th percentiles of estimates mostly contained the true parameter value. When $\delta > 0$, the infections spread more slowly, such that even from incomplete $\{(r_j,i_j)\}_{1:n}$ we might know that certain individuals could not have infected others (Lemmas \ref{lemma2} and \ref{lemma4}). The impact of $\delta$ on estimation is relative to the expected length $\gamma^{-1}$ of an infectious period. From this analysis, we say that the estimator may not be as biased in practice because incubation periods are almost certainly nonzero.

\begin{figure}[!thp]
\centering\includegraphics[scale=0.85]{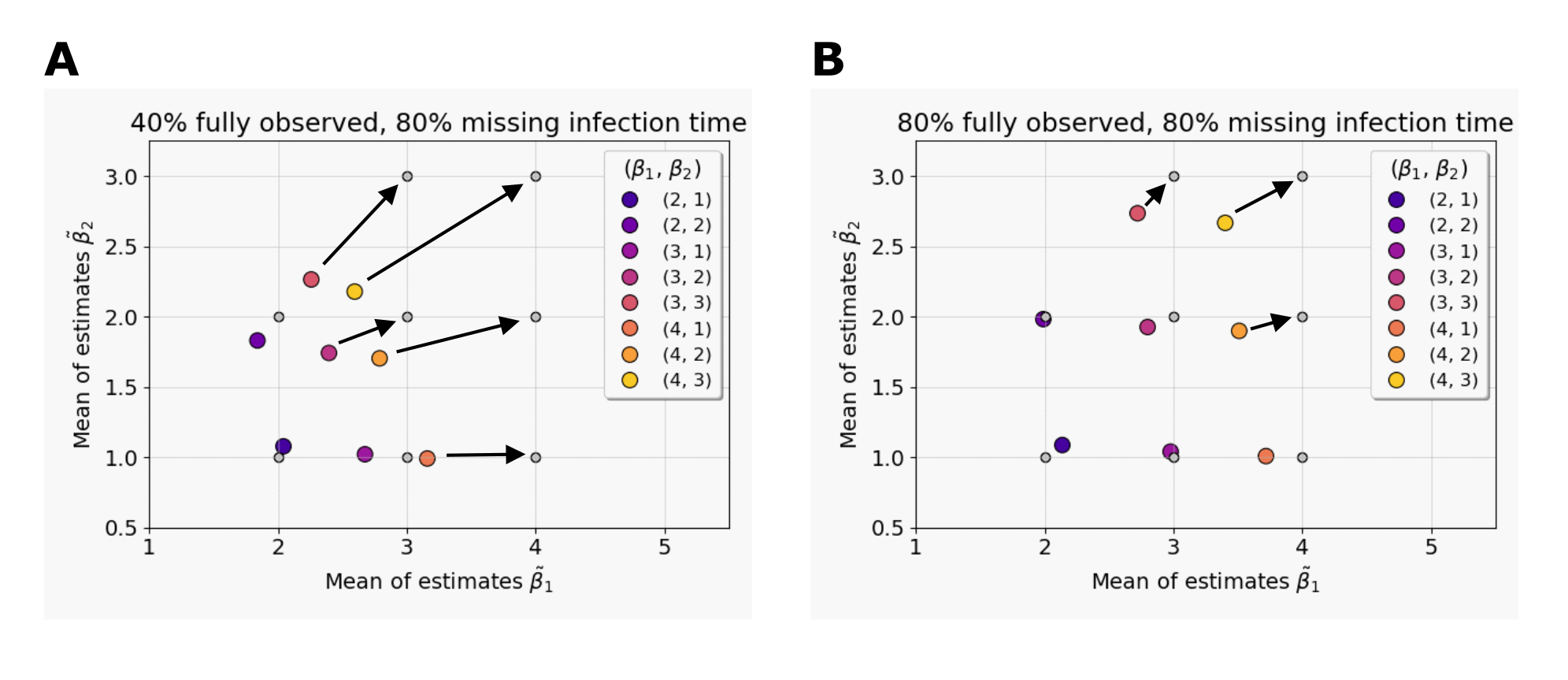}
\caption{Infection rate estimates in multitype SIR epidemic models. Scatter plots show the means of class-specific infection rate estimates $\tilde{\beta}_1$ (x-axis) by $\tilde{\beta}_2$ using the new conditional expectation method. Each combination of  $(\beta_1, \beta_2)$ is denoted in the legend. The true values are represented as small gray dots. When not clear, we draw black arrows to distinguish which (under-)estimates correspond to which true values. The class-specific susceptible population sizes $N_{\beta_1}$ and $N_{\beta_2}$ are both 50, and the minimum epidemic size is 20. The removal rates $\gamma_1$ and $\gamma_2$ are both 1, and the class-specific susceptible population sizes $N_{\gamma_1}$ and $N_{\gamma_2}$ are both 50. In the plot titles A-B), we denote the expected proportion of completely observed infectious periods $p_1$ and the expected proportion of missing infection times $p_2$.}
\label{fig:multitype}
\end{figure}

Fourth, we verified that the biases and trends of the $\tilde{\beta}$ estimator applied to group-specific and spatial models as well. We simulated datasets 1) with group-specific infection rates $\beta_1$ and $\beta_2$ and 2) with a baseline infection rate $\beta \in \{1,2,3,4,5\}$ times an exponential decay kernel $h(\mathbf{x}_j, \mathbf{x}_k) = \exp(-0.05 \cdot ||\mathbf{x}_j- \mathbf{x}_k||_2)$. ``Latitude'' and ``longitude'' values were drawn independently from the Uniform(0,100) random variable, and then the locations were transformed so that the mean and standard deviation of the distances $\{||\mathbf{x}_j- \mathbf{x}_k||_2\}_{(j,k)}$ were 0.9 and 0.01. Figure \ref{fig:multitype} shows the means of the $\tilde{\beta}_1$ and $\tilde{\beta}_2$ estimates of $\beta_1 \in \{2,3,4\}$ and $\beta_2 \in \{1,2,3\}$ for $p_1 \in \{0.4,0.8\}$, $p_2=0.8$, and $\gamma=1$. As before, the group-specific infection rate estimates were more biased the smaller $p_1$ was. Figure \ref{fig:spatial} shows the same overall behavior and accuracy of the baseline $\tilde{\beta}$ infection rates of the spatial model as in the standard SIR models (Figure \ref{fig:eminfection}).

\subsubsection{Parametric bootstrap bias-correction}

\begin{figure}[!htp]
\centering\includegraphics[scale=0.85]{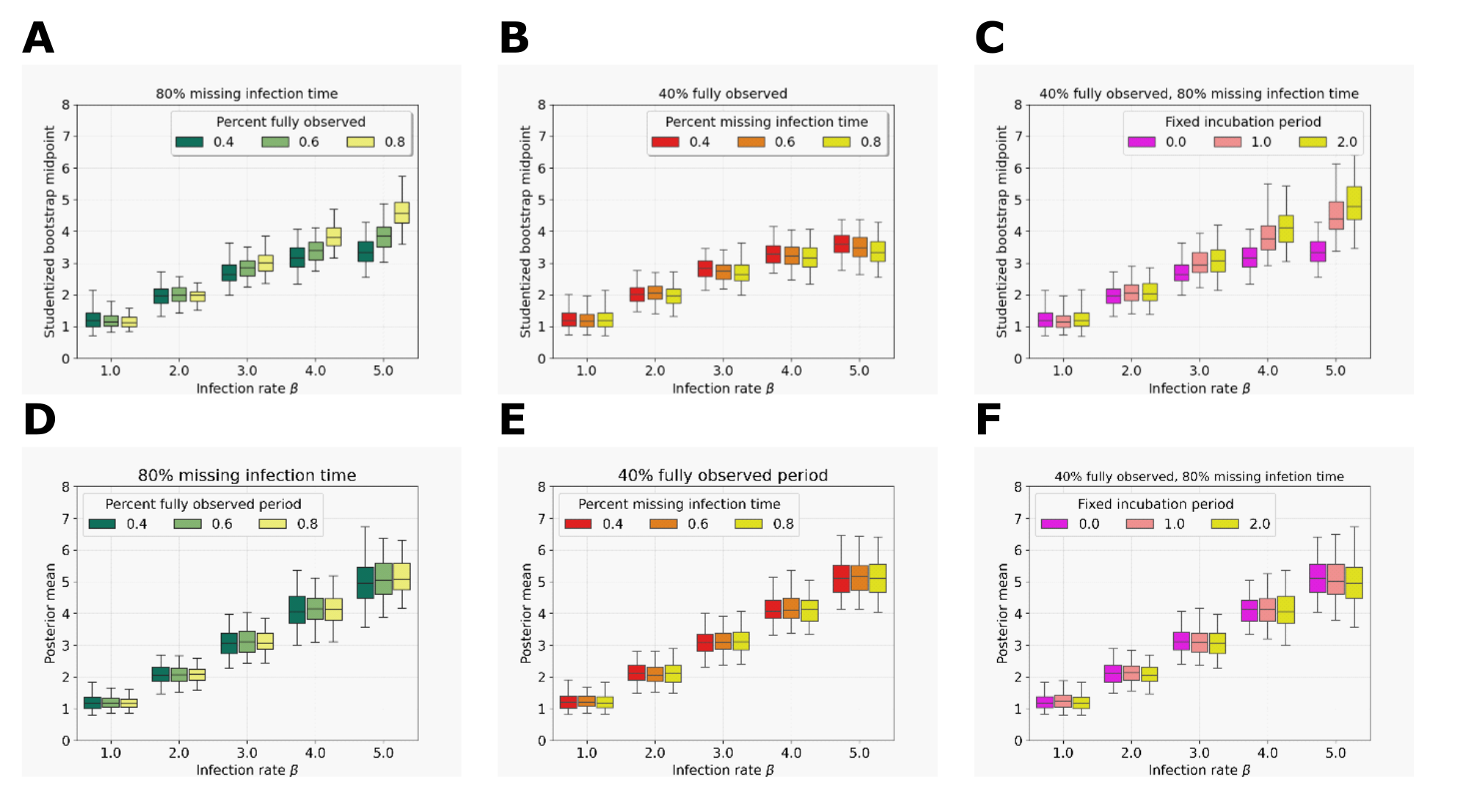}
\caption{Infection rate estimates in SEIR epidemic models. Box plots show the 2.5th, 25th, 50th, 75th, and 97.5th percentiles of infection rate estimates from A-C) the studentized bootstrap midpoints and D-F) the Bayesian posterior means. The posterior means were calculated from 400 samples after a burn-in period of 100 samples. The default settings were that the expected proportion $p_1$ of completely observed infectious periods was 0.40, the expected proportion $p_2$ of missing infection times was 0.80, and the fixed incubation period $\delta$ was 0. Along the columns, we varied one of these parameters (legend) and held the other two parameters fixed. The susceptible population size $N$ was 100, the minimum epidemic size $n$ was 20, the removal rate $\gamma$ was 1, and the Erlang shape parameter $m$ was 1.}
\label{fig:main}
\end{figure}

\begin{figure}[!htp]
\centering\includegraphics[scale=0.85]{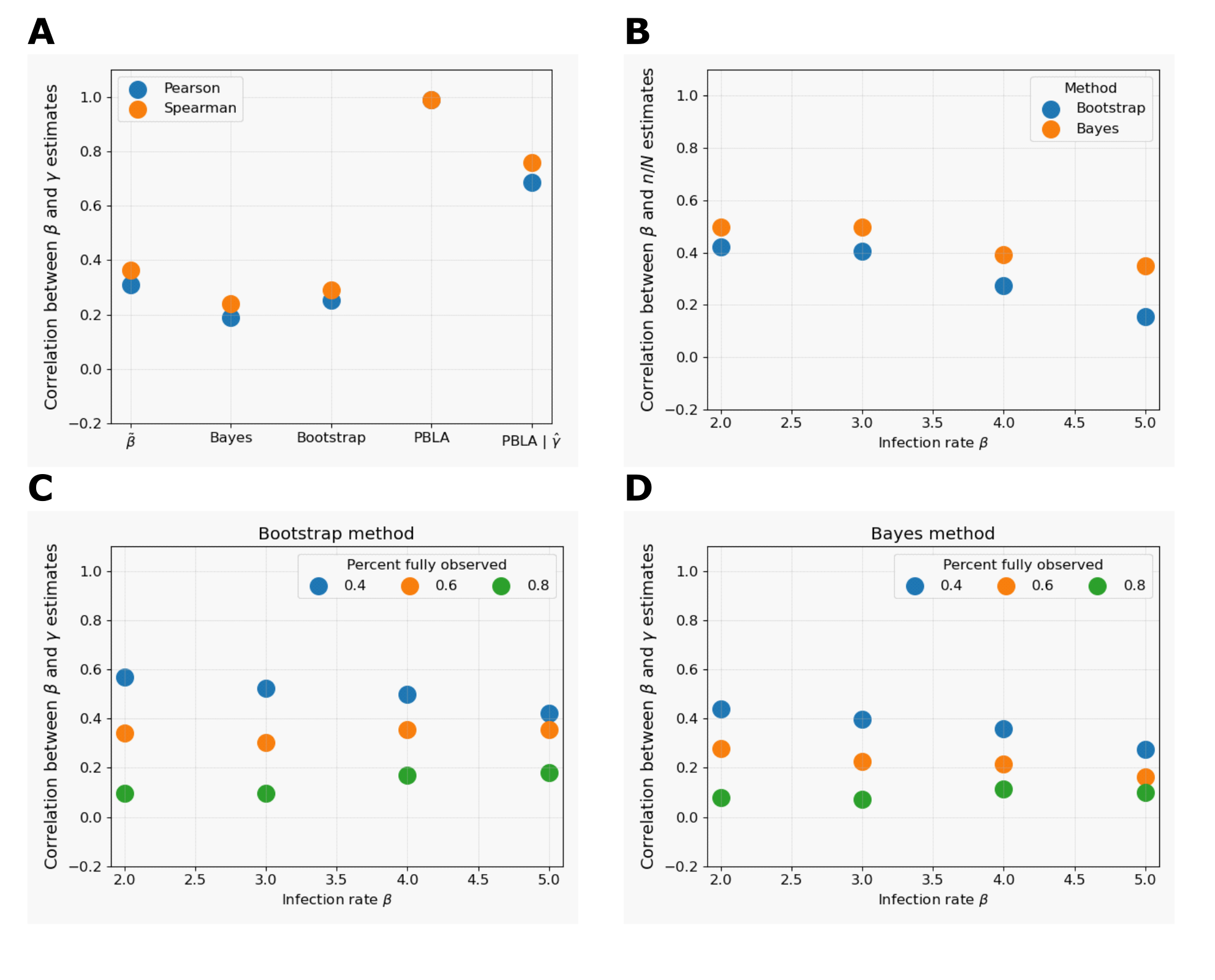}
\caption{Correlations between estimates and prevalences for different methods. A) The scatter plot shows the Pearson and Spearman (legend) correlations (y-axis) between $\beta$ and $\gamma$ estimates for the different methods (x-axis). B) The scatter plot shows the Pearson correlations between $\beta$ estimates and the prevalences $n/N$ for the bootstrap and Bayesian methods (legend). C-D) The scatter plot shows the Pearson correlations between $\beta$ and $\gamma$ estimates for the bootstrap and Bayesian methods (subplot titles) as we varied the expected proportion $p_1$ of fully observed infectious periods. All settings were as in Figure \ref{fig:main}.}
\label{fig:pearson}
\end{figure}

\begin{figure}[!htp]
\centering\includegraphics[scale=0.85]{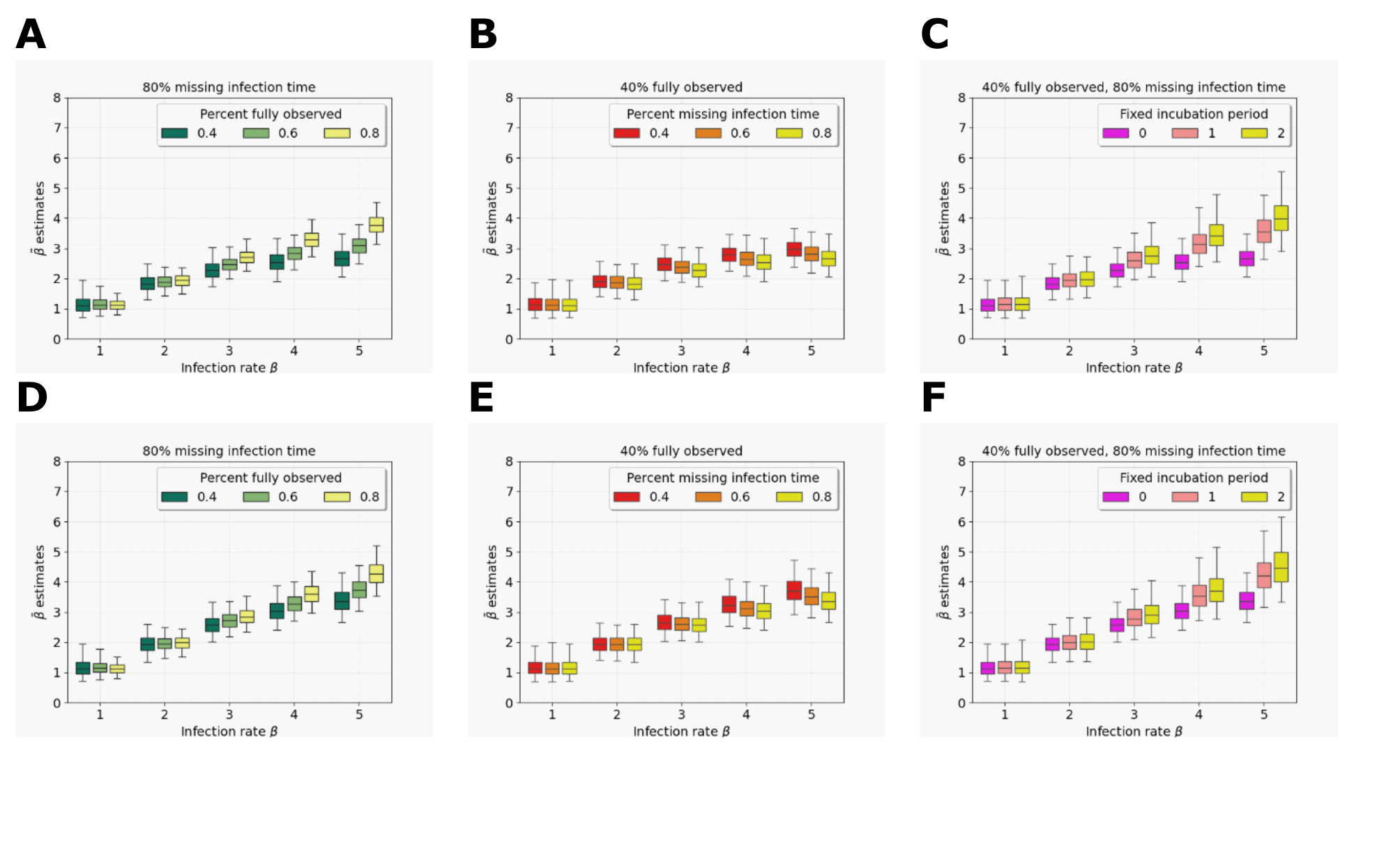}
\caption{Initial infection rate estimates in SEIR epidemic models. Box plots show the 2.5th, 25th, 50th, 75th, and 97.5th percentiles of infection rate estimates A-C) $\tilde{\beta}$ and D-F) $\bar{\beta}$ using the expectation-maximization approach. All settings were as in Figure \ref{fig:main}.
}
\label{fig:eminfection}
\end{figure}

\begin{figure}[!htp]
\centering\includegraphics[scale=0.85]{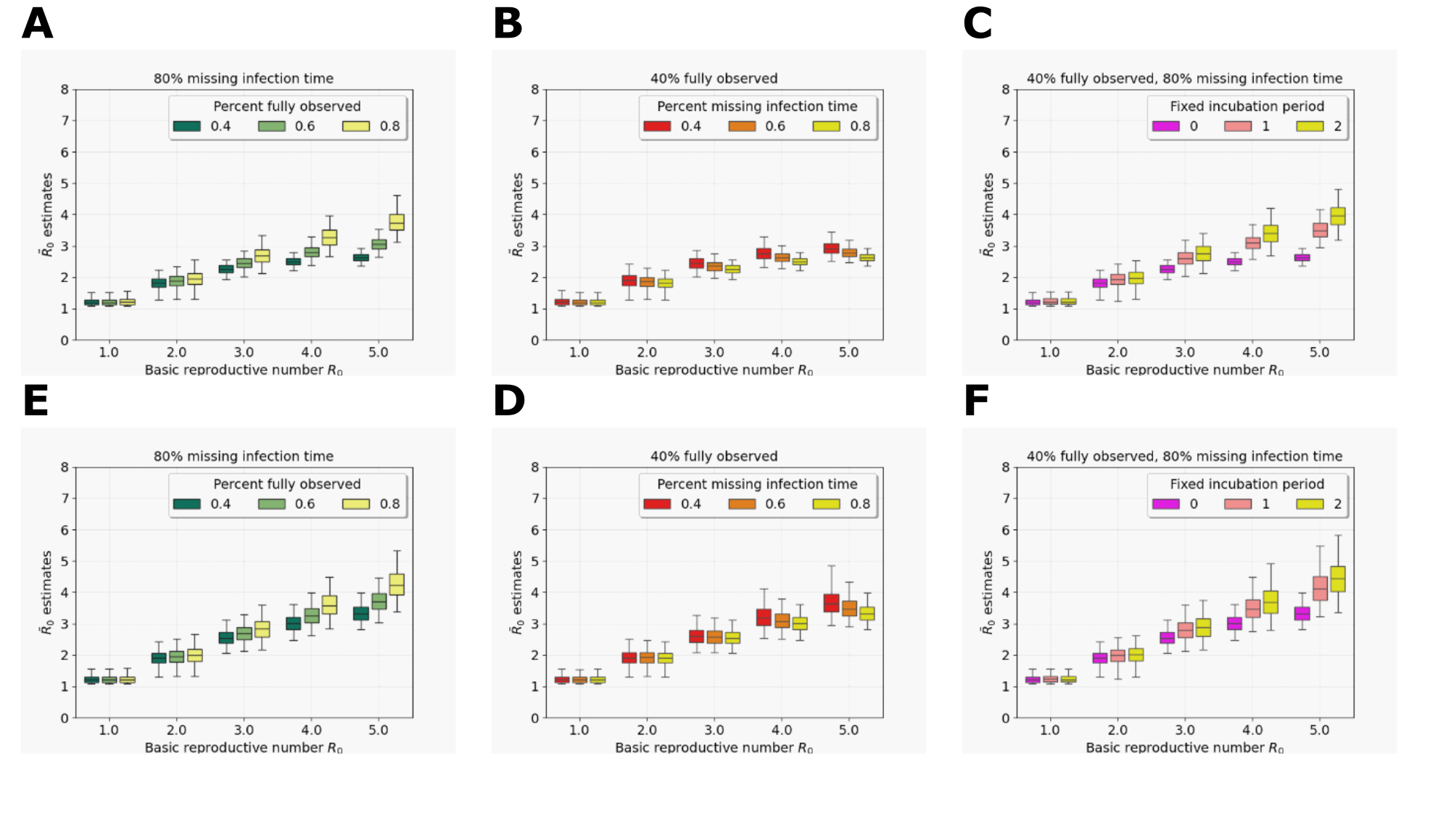}
\caption{Initial basic reproductive number estimates in SEIR epidemic models. Box plots show the 2.5th, 25th, 50th, 75th, and 97.5th percentiles of infection rate estimates A-C) $\tilde{R}_0$ and D-F) $\bar{R}_0$ using the expectation-maximization approach. All settings were as in Figure \ref{fig:main}.
}
\label{fig:embasicnumber}
\end{figure}

\begin{figure}[!htp]
\centering\includegraphics[scale=0.85]{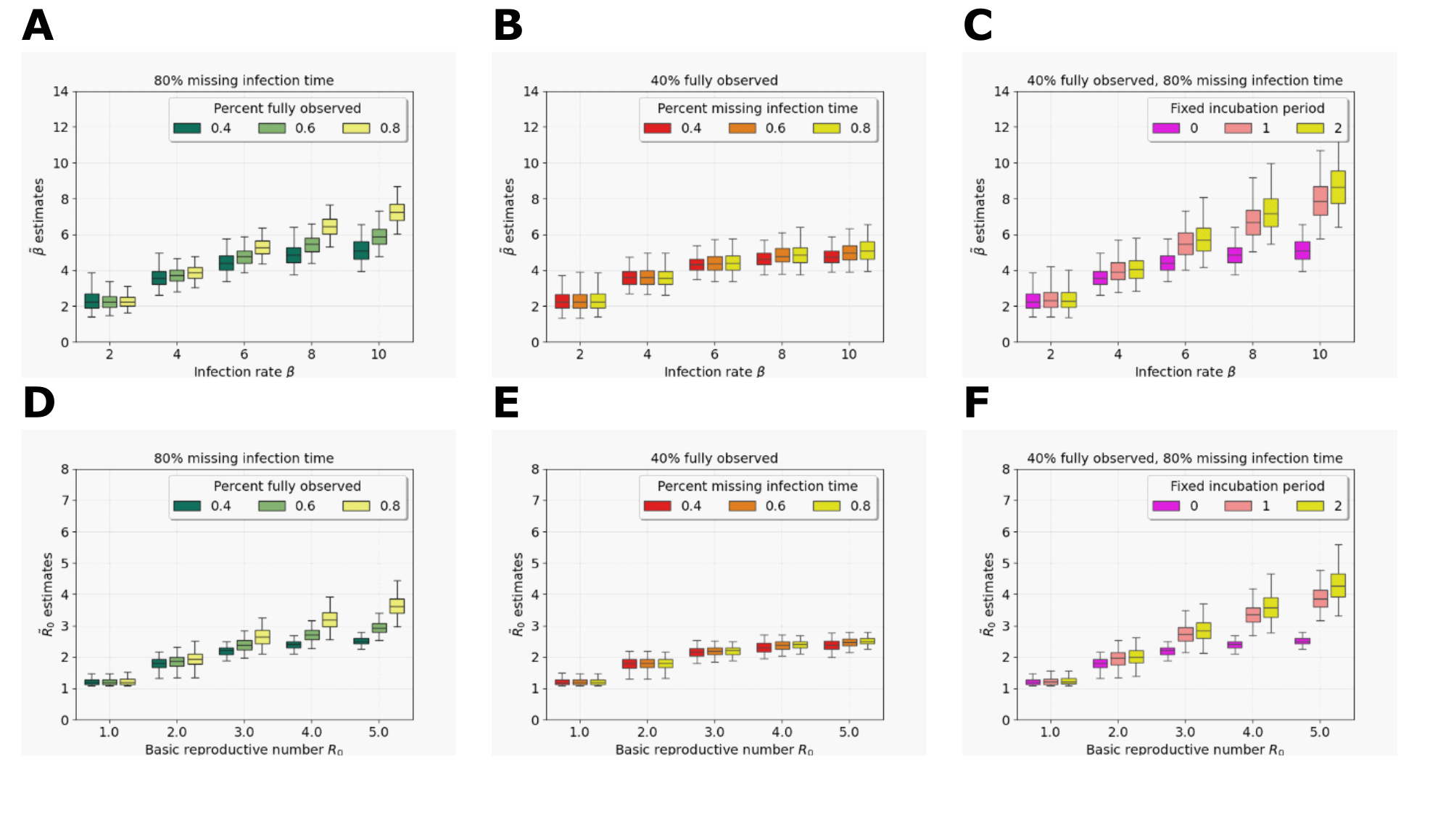}
\caption{Initial estimates in SEIR epidemic models with a larger removal rate. Box plots show the 2.5th, 25th, 50th, 75th, and 97.5th percentiles of A-C) infection rate estimates $\tilde{\beta}$ D-F) basic reproductive number estimates $\tilde{R}_0$ using the new conditional expectation method. The removal rate $\gamma$ is 2. All other settings were as in Figure \ref{fig:main}.}
\label{fig:bigremovalrate}
\end{figure}

\begin{figure}[!htp]
\centering\includegraphics[scale=0.85]{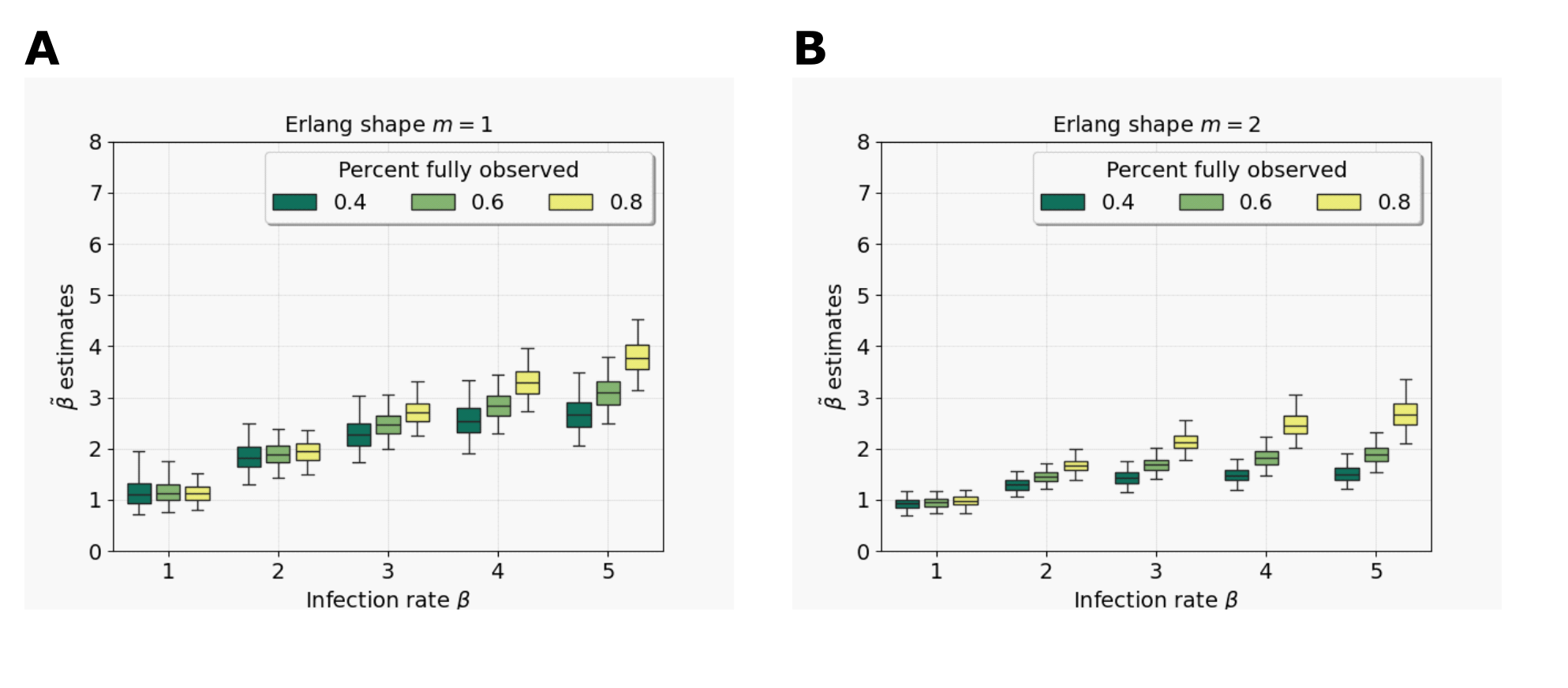}
\caption{Variability of infection rate estimates when infectious period distribution is misspecified. Box plots show the 2.5th, 25th, 50th, 75th, and 97.5th percentiles of infection rate estimates $\tilde{\beta}$. A) The Erlang shape parameter $m$ is 1 versus B) $m$ is 2. All other simulation settings were as in Figure \ref{fig:main}A.}
\label{fig:erlang}
\end{figure}

\begin{figure}[!htp]
\centering\includegraphics[scale=0.85]{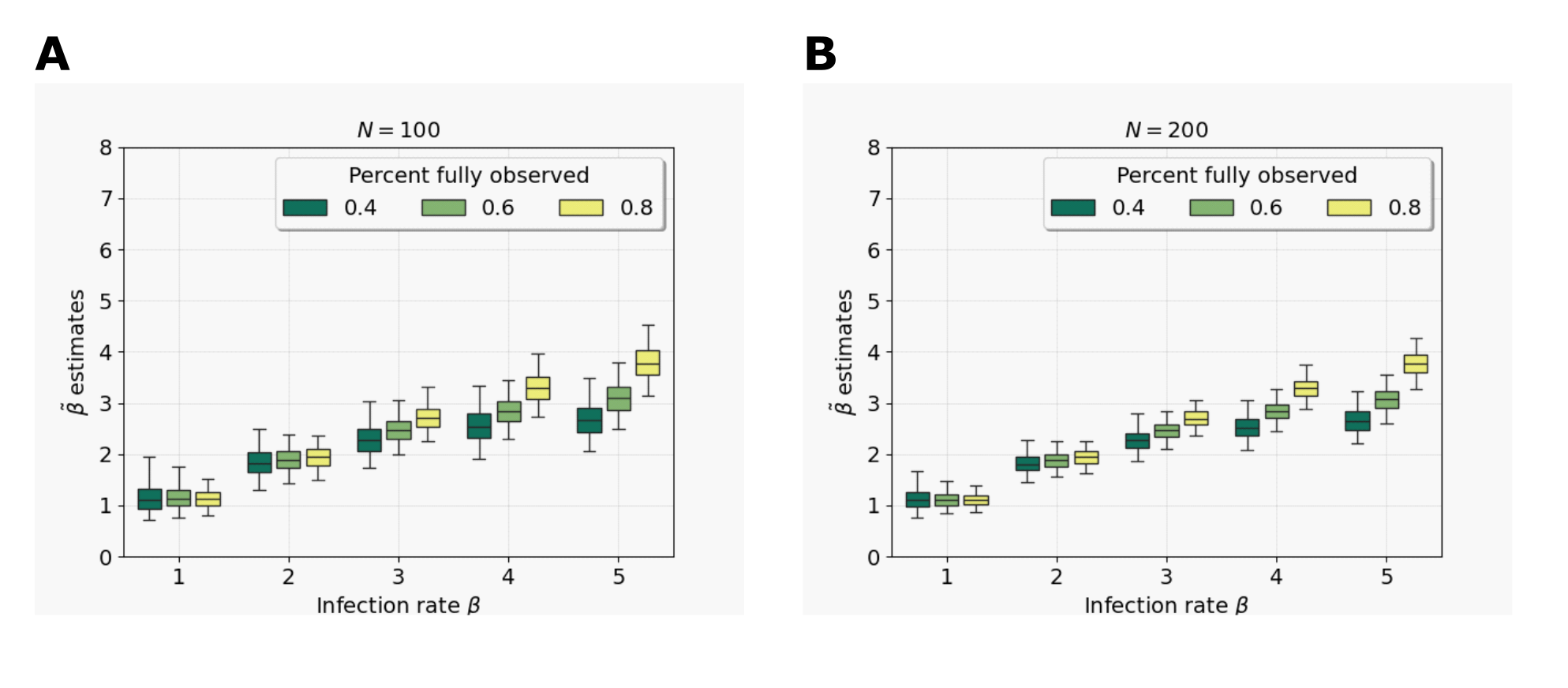}
\caption{Variability of infection rate estimates depends on the total susceptible population size. Box plots show the 2.5th, 25th, 50th, 75th, and 97.5th percentiles of infection rate estimates $\tilde{\beta}$. A) The total susceptible population size $N$ is 100 and the minimum epidemic size is 20 versus B) the total susceptible population size is 200 and the minimum epidemic size is 40. All other simulation settings were as in Figure \ref{fig:main}A.}
\label{fig:populationsize}
\end{figure}

\begin{figure}[!htp]
\centering\includegraphics[scale=0.85]{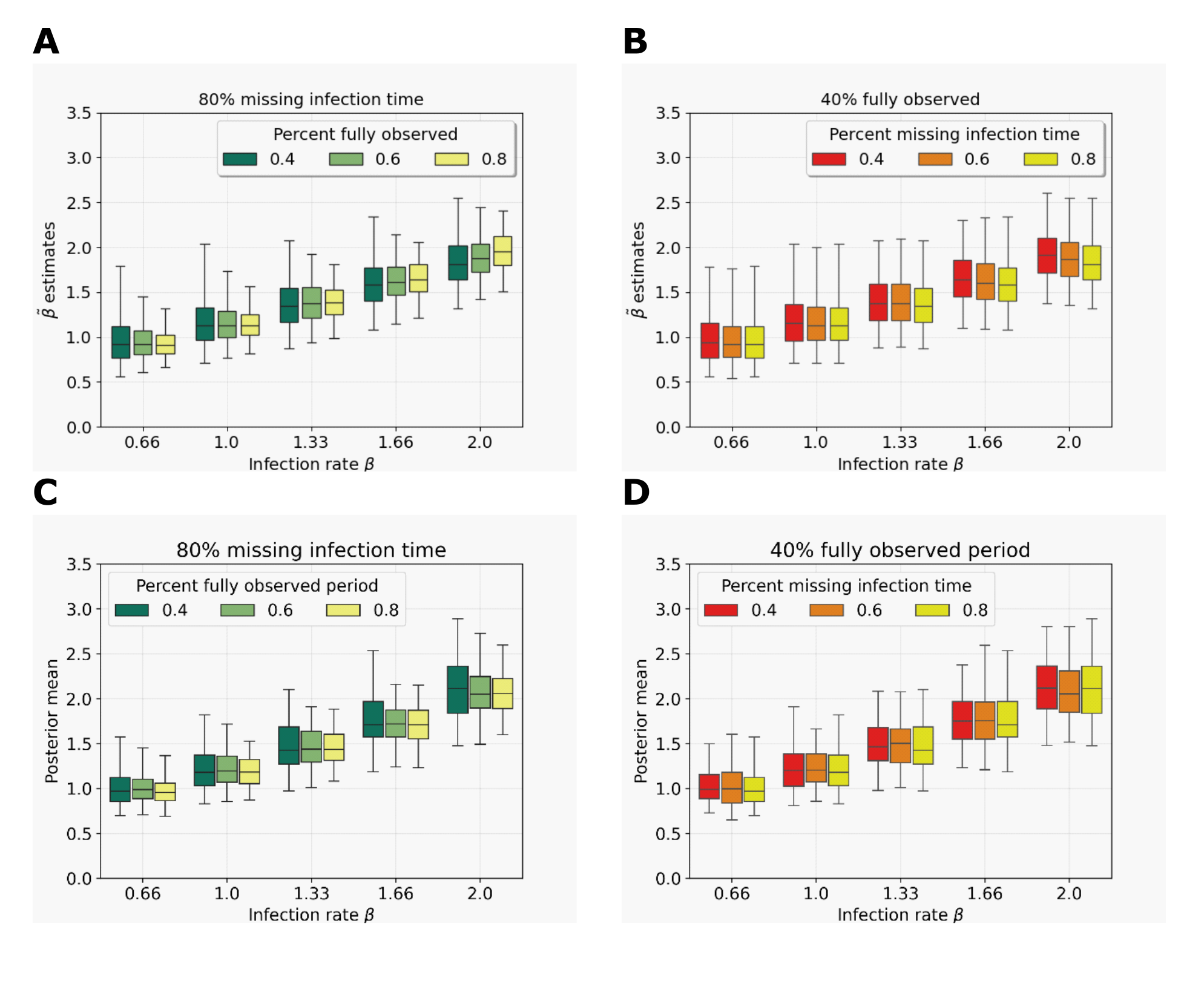}
\caption{Estimates of small infection rates in SIR epidemic models. Box plots show the 2.5th, 25th, 50th, 75th, and 97.5th percentiles of  A-B) infection rate estimates $\tilde{\beta}$ using the expectation-maximization approach and C-D) the Bayesian posterior means of $\beta$. The infection rates $\beta$ (x-axis) vary from 0.66 to 2. The fixed incubation period $\delta$ is 0. All other simulation and MCMC settings were as in Figure \ref{fig:main}.}
\label{fig:smallinfectionrate}
\end{figure}

\begin{figure}[!htp]
\centering\includegraphics[scale=0.85]{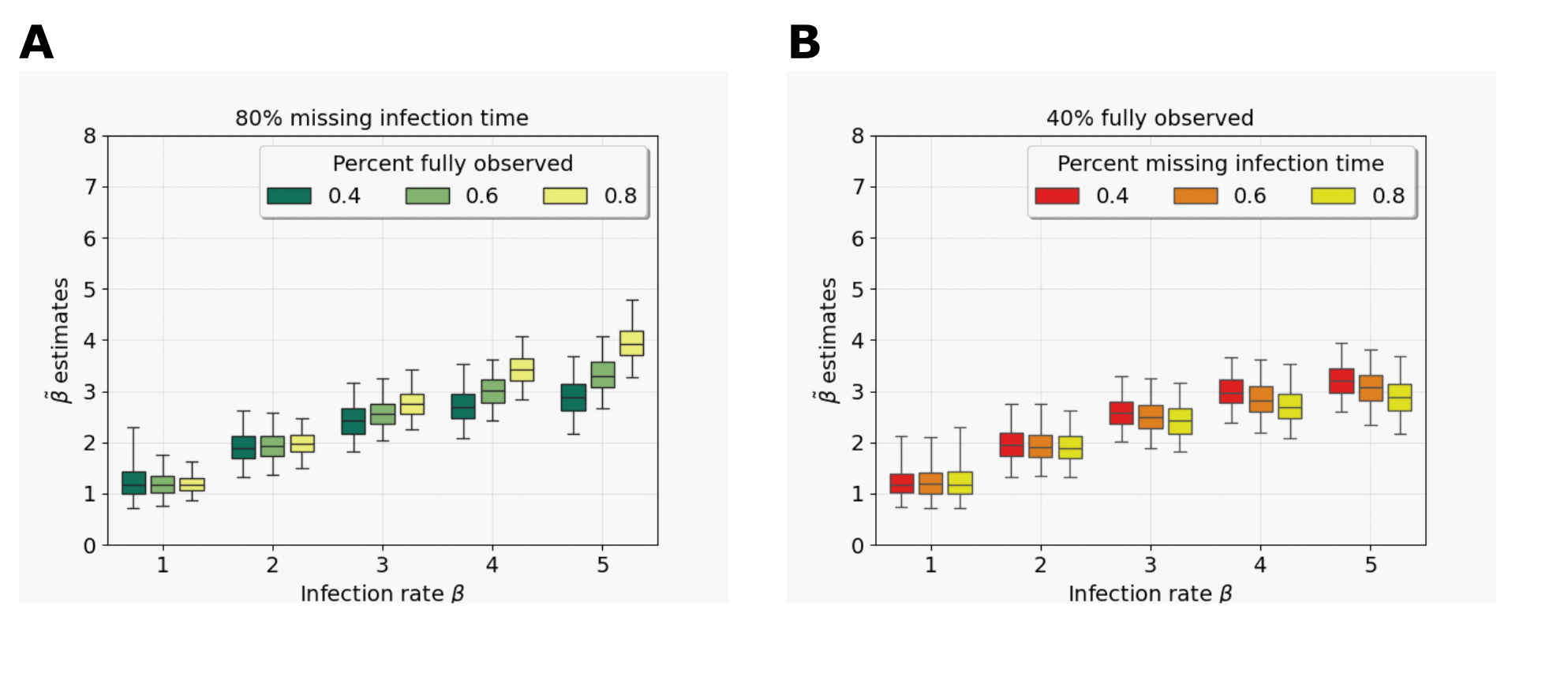}
\caption{Baseline infection rate estimates in spatial SIR epidemic models. Box plots show the 2.5th, 25th, 50th, 75th, and 97.5th percentiles of baseline infection rate estimates $\tilde{\beta}$. The spatial model and its distance kernel and the simulations are described in the main text. All other simulation settings were as in Figure \ref{fig:main}A-B.}
\label{fig:spatial}
\end{figure}

Based on the results in Section \ref{sec:initial}, we decided to partially correct the bias in the $\tilde{\beta}$ estimator with the bootstrap$-t$ procedure. That is, the midpoint of the studentized bootstrap interval (Algorithm \ref{algorithm3}) corrected some of the bias in the $\tilde{\beta}$ estimates (Figure \ref{fig:main}A-C). This technique reduced the biases in the $\tilde{\beta}$ estimates of $\beta \in \{3, 4,5\}$ between 0.5 and 1.0 points. In particular, the corrected $\tilde{\beta}$ estimates were now centered around the true $\beta=3$ for the stochastic SIR model ($\delta=0$), and the 2.5th and 97.5th percentiles of the corrected estimates always contained the true $\beta \in \{1,2,3,4\}$. The midpoint estimates were almost unbiased for $\beta \in \{4,5\}$ when $\delta \geq 1$. Running the bootstrap procedure is simple and automatic (no monitoring of MCMC diagnostics), but the posterior means from Algorithm \ref{algorithm4} remained superior estimators when $\delta < \gamma^{-1}$ (Figure \ref{fig:main}D-F).

\subsubsection{Behavior of the Bayesian posterior mean}

\begin{figure}[!htp]
\centering\includegraphics[scale=0.85]{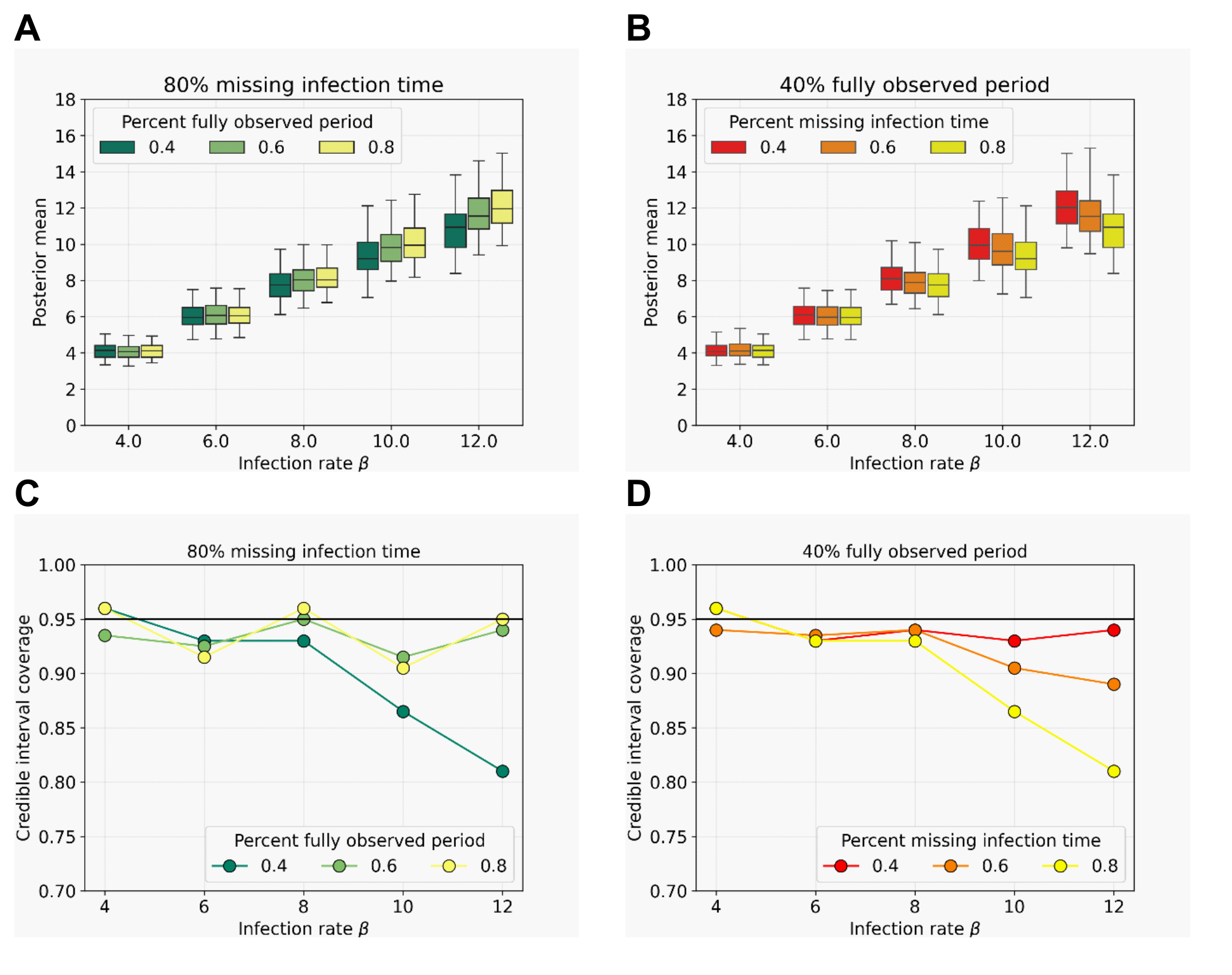}
\caption{Posterior means and credible interval coverages for large infection rates in SIR epidemic models. A-B) Box plots show the 2.5th, 25th, 50th, 75th, and 97.5th percentiles of  posterior means (y-axis) of the infection rate $\beta$ (x-axis). C-D) Line plots show the empirical coverage of $95\%$ credible intervals for $\beta$. The fixed incubation period $\delta$ is 0. All other simulation and MCMC settings were as in Figure \ref{fig:main}, which includes that $\gamma=1$.}
\label{fig:biginfectionrate}
\end{figure}

\begin{figure}[!htp]
\centering\includegraphics[scale=0.85]{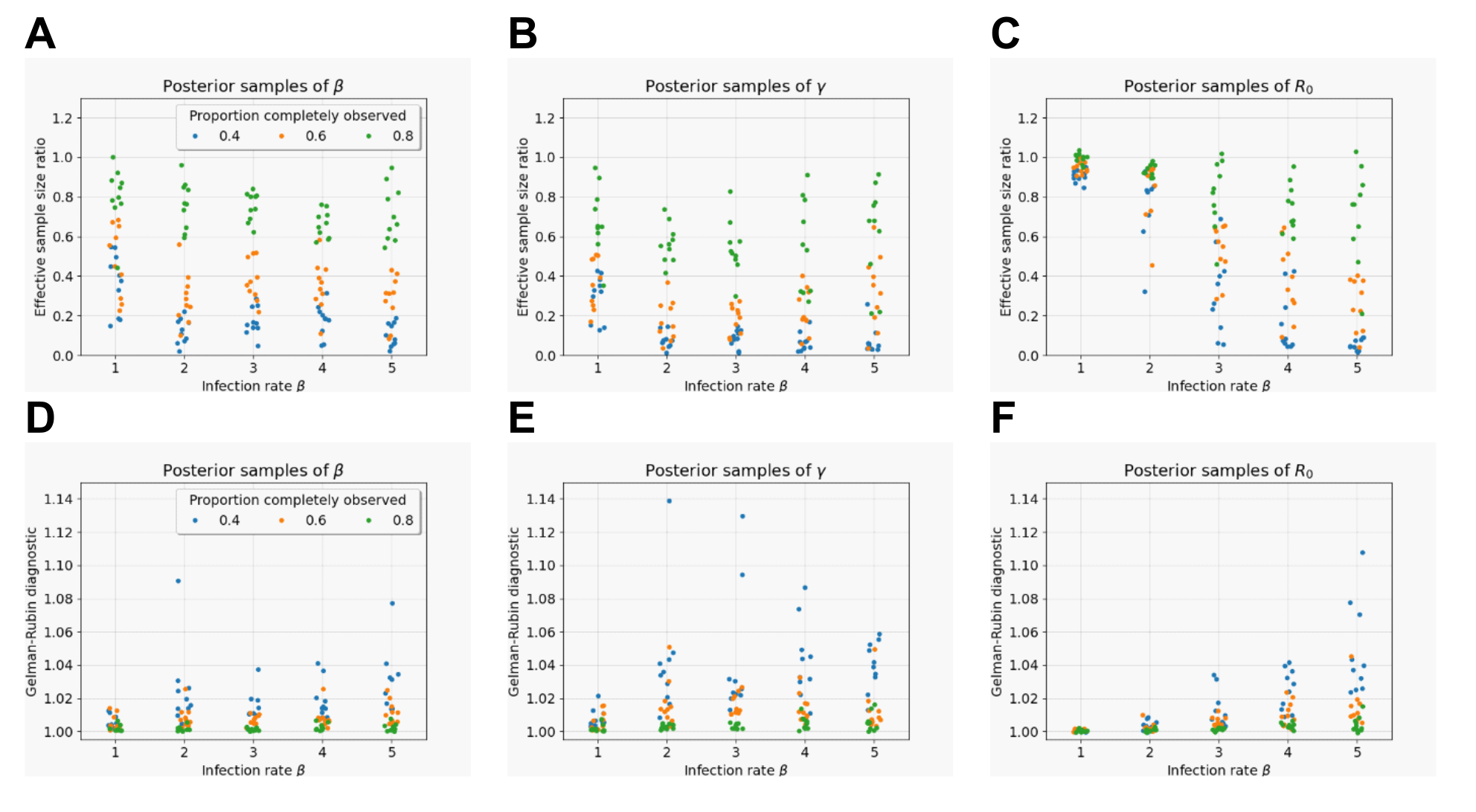}
\caption{Markov Chain Monte Carlo diagnostics for Bayesian SIR models. Strip plots show the A-C) effective sample size ratios and D-F) Gelman-Rubin statistic (y-axes) from posterior samples of $\beta$, $\gamma$, and $R_0$ (subplot titles). There are 10 simulations for each infection rate $\beta \in \{1,2,3,4,5\}$ (x-axis). We consider expected proportions of fully observed infectious periods $p_1 \in \{0.4,0.6,0.8\}$ (legend) and the expected proportion of those missing infection times $p_2 = 0.8$. All other settings were as in Figure \ref{fig:main}. We used 10 mixing chains. We discarded the first 100 posterior samples as burn-in and then calculated diagnostics from the remaining 400 posterior samples. The diagnostics are computed with the Python functions \texttt{ess} and \texttt{rhat} in the \texttt{arviz} package.}
\label{fig:partialmcmc}
\end{figure}

By and large, the posterior mean estimates from Algorithm \ref{algorithm4} were unbiased and not qualitatively different as we varied $p_1$, $p_2$, and $\delta$ (Figures \ref{fig:main}D-F and \ref{fig:smallinfectionrate}C-D). For very large $\beta \geq 8$ and $\gamma=1$, the posterior means were slight underestimates of the true $\beta$ (Figure \ref{fig:biginfectionrate}A-B). Autocorrelation makes successive MCMC draws partially redundant, so we checked the effective sample size (ESS) ratio to measure how much independent information was contained in chains from Algorithm \ref{algorithm4}. Figure \ref{fig:partialmcmc}A-C shows ESS ratios for $\beta$, $\gamma$, and $R_0$ that were mostly greater than 0.60 when $p_1=0.8$ but often less than 0.30 when $p_1=0.4$. The ESS ratios of $R_0$ decreased as the true $R_0$ increased above 3 whereas there was no clear trend for the ESS ratios of $\beta$ and $\gamma$. Figure \ref{fig:partialmcmc}D-F shows that the Gelman-Rubin diagnostics of $\beta, \gamma,$ and $R_0$ (from 10 mixing chains) were mostly less than 1.02 for $0.4 \leq p_1 \leq 0.8$ and decreased as a function of $p_1$. The mixing chains had mostly converged when some full infectious periods were observed, compared to the poor convergence we observed when no infection times $\mathbf{i}$ were observed (Figure \ref{fig:nonemcmc}).

\subsubsection{Interval estimates and coverage}

\begin{figure}[!htp]
\centering\includegraphics[scale=0.85]{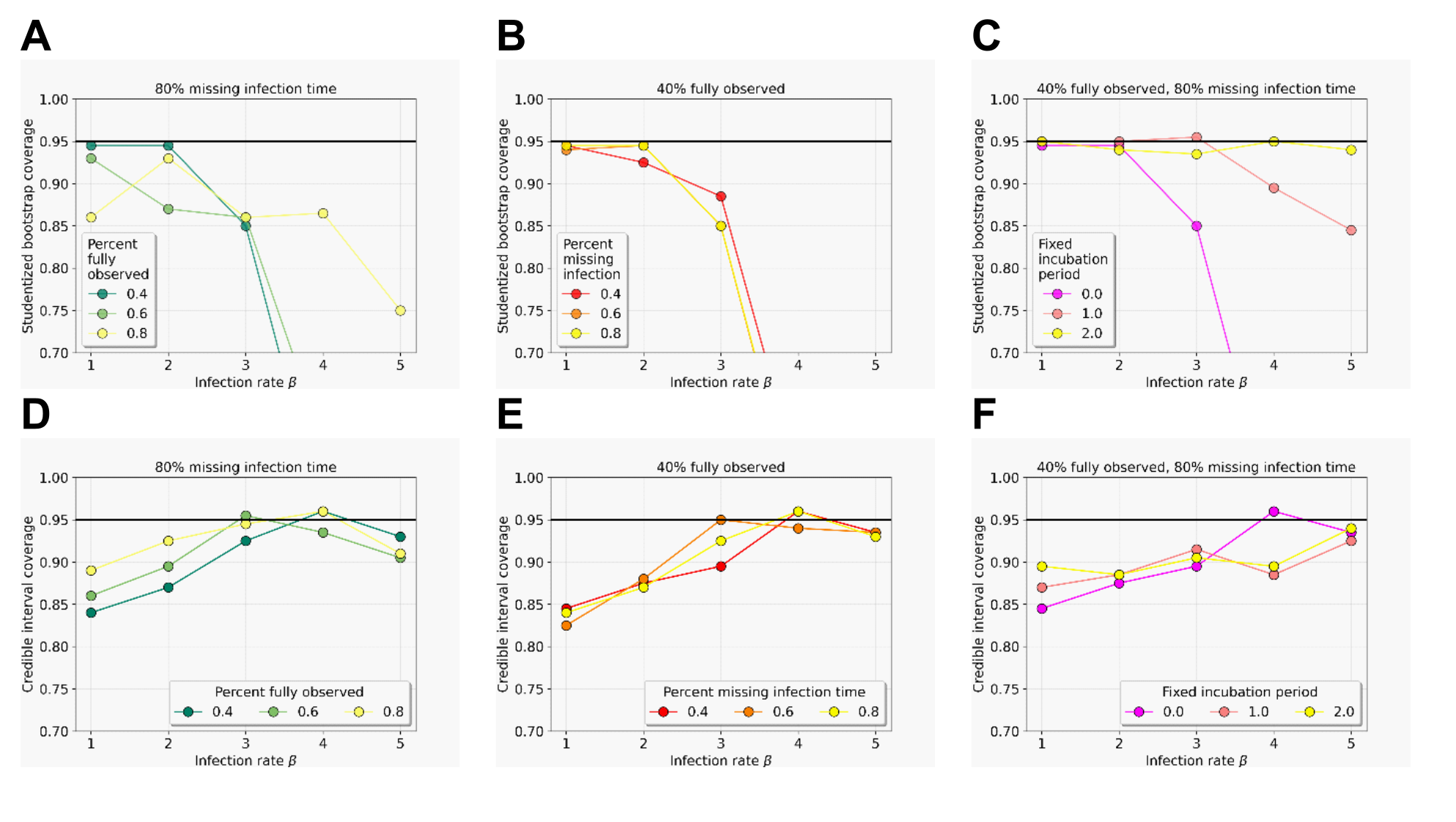}
\caption{Coverage of infection rate interval estimates in SEIR epidemic models. Line plots show the empirical coverage (y-axis) of A-C) studentized bootstrap and D-F) Bayesian credible intervals for the infection rate $\beta$ (x-axis). For the studentized bootstrap intervals, we estimated coverage from 1000 simulations. For the Bayesian credible intervals, we estimated coverage from 200 simulations. All simulation and MCMC settings were as in Figure \ref{fig:main}.}
\label{fig:coverage}
\end{figure}

\begin{figure}[!htp]
\centering\includegraphics[scale=0.85]{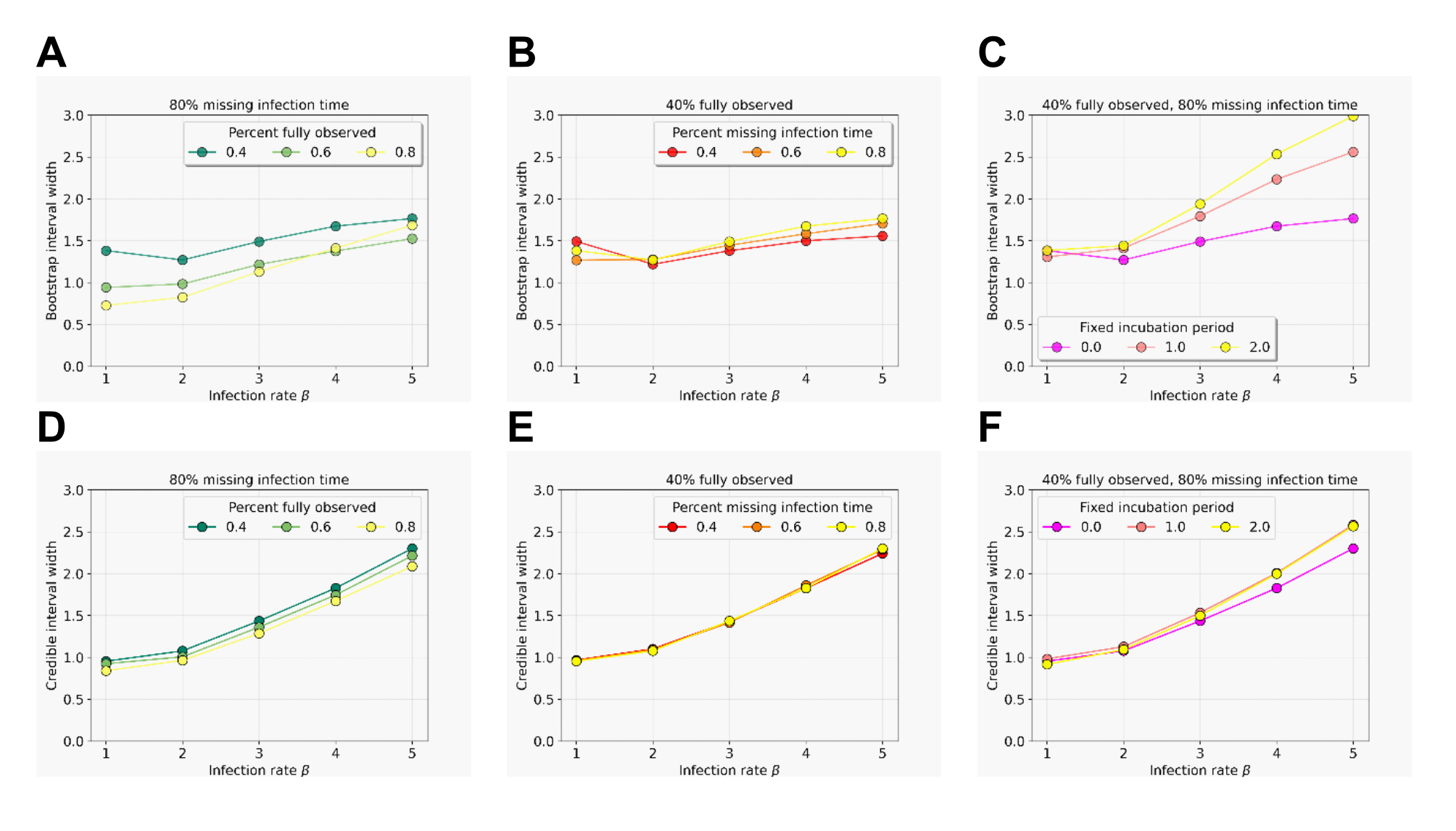}
\caption{Interval widths for infection rates in SIR epidemic models. Line plots show the $95\%$ studentized bootstrap and Bayesian credible interval widths for the infection rate $\beta$. All simulation and MCMC settings were as in Figure \ref{fig:main}.}
\label{fig:widths}
\end{figure}

Figure \ref{fig:coverage} shows the percentage of replicate simulations where the 95\% studentized bootstrap and Bayesian credible intervals contain the true infection rate $\beta \in \{1,2,3,4,5\}$, and Table \ref{Table1} reports the numerical results for $p_1 \in \{0.4,0.8\}, p_2=0.8,$ and $\delta=1$. For the studentized bootstrap intervals, the empirical coverages exceeded 0.85 for $1\leq \beta \leq 3$ and $0.4 \leq p_1 \leq 0.8$. The bootstrap coverages improved to more than 0.89 for $1 \leq \beta \leq 4$ when the fixed incubation period was $\delta \geq 1$. For Bayesian credible intervals, empirical coverage exceeded 0.85 for $3 < \beta \leq 10$ under all experimental conditions (Figure \ref{fig:biginfectionrate}C-D). Bayesian coverages often decreased to a few percentage points less than those of the studentized bootstrap intervals when $\beta$ decreased below 3. Figure \ref{fig:widths} shows the average widths of the studentized bootstrap and the Bayesian credible intervals. The bootstrap interval widths decreased when $p_1$ grew or $\delta$ shrunk, whereas the Bayesian credible interval widths barely changed when $p_1$ or $\delta$ did. Neither the bootstrap nor the credible interval widths changed when we varied $p_2$. In general, the bootstrap and credible interval widths for $\beta \in \{1,2,3,4,5\}$ were within 0.25 and at most 1.0 points of each other under all experimental conditions. From this finding, we argue that the decrease in coverage of the studentized bootstrap interval as $\beta$ rose above 3 (Figure \ref{fig:coverage}A-C) came from unaddressed bias (Figure \ref{fig:main}A-C) as opposed to insufficient interval size.

\section{Application to Hagelloch measles outbreak}\label{sec:measles}

\cite{Pfeilsticker1863-rz} collected rich information on an 1861 measles epidemic in the isolated village of Hagelloch, Germany, where nearly all susceptible children became infected. The spatiotemporal dataset contains individual-level features, including home location, age, sex, family, school class, and times of prodromal symptoms and the appearance of measles rash \citep{Oesterle1993-yk}. Previous work has explored models that include all of these features \citep{Groendyke2012-wr,Neal2004-ir}. We focused only on an SEIR model for school class-specific infection rates $(\beta_1, \beta_2,\beta_3)$ (Equation \ref{eq:betahatgroup}), as we were more interested in statistical inference when some of the infection and removal times were simulated as missing than other transmission dynamics. 

\begin{figure}[!htp]
\centering\includegraphics[scale=0.85]{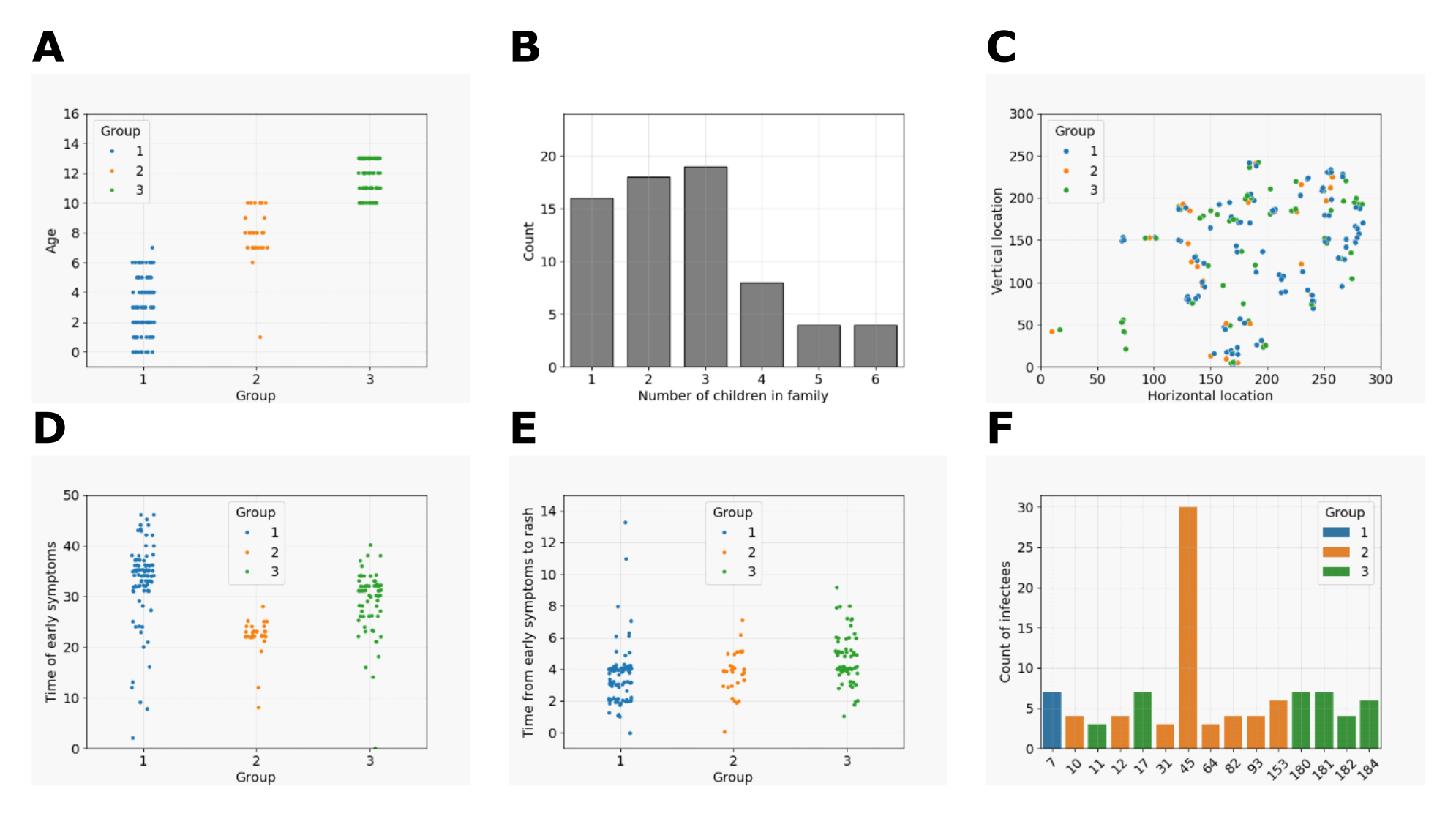}
\caption{Features of the Hagelloch dataset. A) The strip plot shows the ages (y-axis) of the children by group (x-axis). B) The histogram shows the count (y-axis) of children that each family had (x-axis). C) The scatter plot shows the location of each child's home, jittered by the Normal(0,3) random variable. D) The strip plot shows the jittered date of the prodromal symptoms (y-axis) for the children in each group (x-axis). E) The strip plot shows the jittered time difference between the prodromal symptoms and the rash appearance (y-axis) for the children in each group (x-axis). F) The histogram shows the count of infectees (y-axis) for the top 15 suspected infectors (x-axis), colored by group.}
\label{fig:hagelloch}
\end{figure}

We adopted the same assumptions as those of \cite{Neal2004-ir} about the total susceptible population. That is, there are 185 susceptible children, divided according to age into school classes of 90, 30, and 65 children (Figure \ref{fig:hagelloch}A). More than 50\% of the children belonged to families of 3 or more children, which is a transmission dynamic that we ignored (Figure \ref{fig:hagelloch}B). There is no obvious difference in the distribution of home locations between school classes (Figure \ref{fig:hagelloch}C). \cite{Neal2004-ir} showed that school class was an important feature in their best fitting model to the data on top of these two pieces of spatial information.

Based on \cite{Neal2004-ir}, we set the infection times at 1 day before the time of the first symptoms and the removal times at 3 days after the appearance of the measles rash. The time between initial symptoms and the appearance of the rash was typically between 2 and 6 days, irrespective of school class (Figure \ref{fig:hagelloch}E). Based on \cite{Dardis2012-ht}, we assumed a fixed incubation period of 10 days. In \cite{Neal2004-ir}, they found that reasonable changes in these parameters did not affect the qualitative conclusions of their study.

\begin{figure}[!htp]
\centering\includegraphics[scale=0.85]{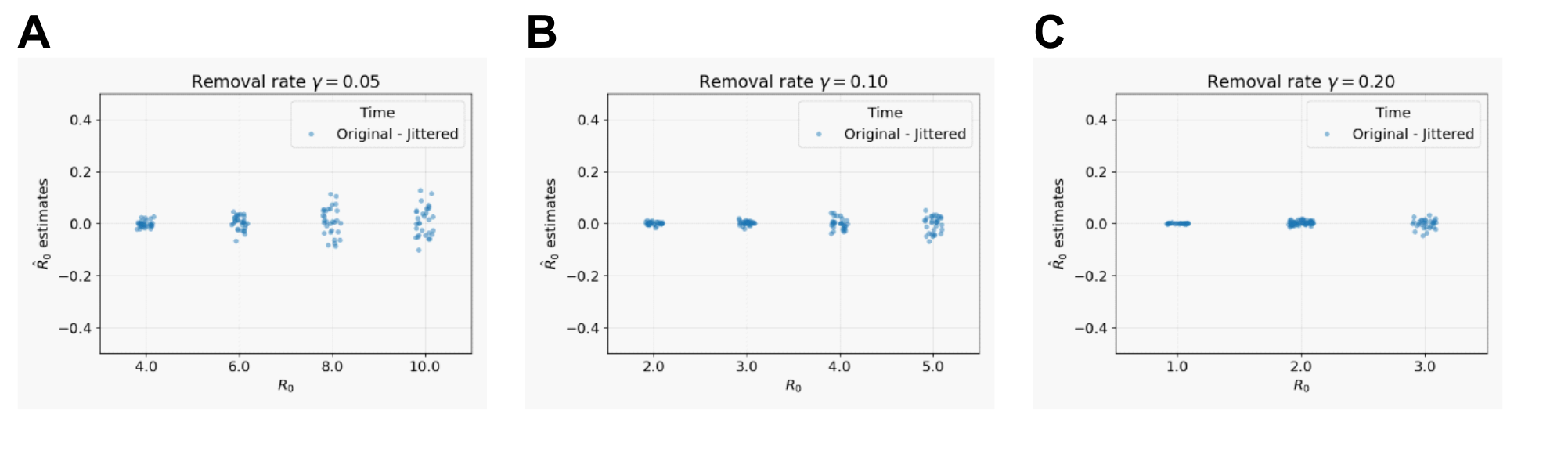}
\caption{Estimation from daily data. Strip plots show the difference in maximum likelihood estimates of the basic reproductive number $\hat{R}_0 = \hat{\beta} / \hat{\gamma}$ from exact infection and removal times versus from times discretized at a daily level. We rounded times to the nearest integer and then added white noise with mean 0 and standard deviation 0.33 (larger than in the Hagelloch study). For each combination of the infection rates $\beta \in \{0.2, 0.3, 0.4, 0.5\}$ and the removal rates $\gamma \in \{0.05, 0.10, 0.20\}$, we simulated 30 epidemics consisting of at least 20 infected individuals out of the 200 total susceptible individuals. The removal rates $\gamma$ are A) 0.05, B) 0.10, and C) 0.20, and they correspond to expected infectious periods of 20, 10, and 5 (days). The fixed lag $\delta$ is 10 (days). }
\label{fig:dailydata}
\end{figure}

The temporal information was reported as integer--valued dates. We do not have analytical formulae for $\E[\tau_{kj}]$ in discrete--time models \citep{Lekone2006-nm,O-Neill2005-ax}. To accommodate our continuous--time model, we added white noise with mean 0 and standard deviation 0.1 to these time values. Figure \ref{fig:dailydata} shows that adding white noise to daily data perturbed the estimates of $R_0 \in \{4,6,8,10\}$ by up to 0.2 points in simulated epidemics with model parameters similar to ours.

\begin{figure}[!htp]
\centering\includegraphics[scale=0.85]{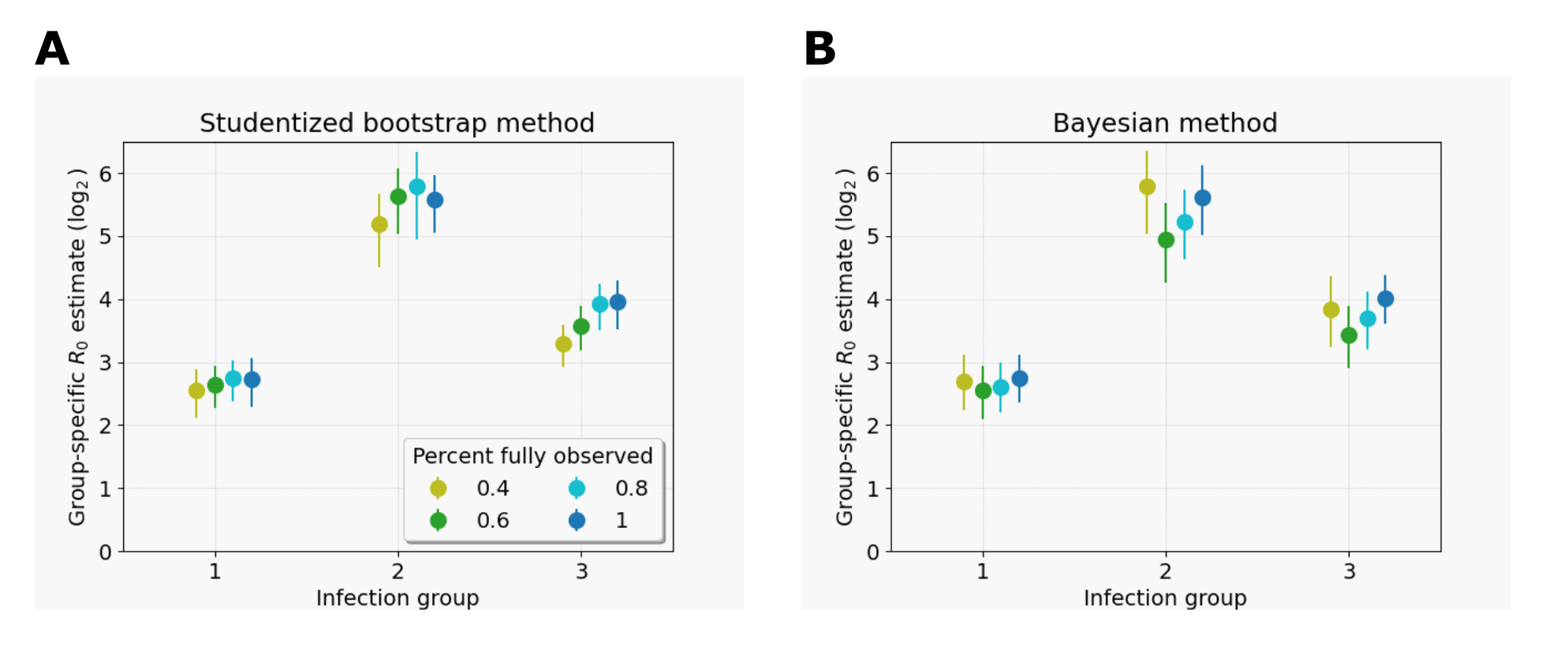}
\caption{Group-specific reproductive number estimates of the 1861 measles epidemic in Hagelloch, Germany. Error plots show the point estimates and 95\% confidence intervals of basic reproductive number estimates (y-axis) for each group (x-axis) from the A) studentized bootstrap or B) Bayesian methods. The y-axis is on the $\log_2$ scale. We varied the expected proportion $p_1$ of fully observed infectious periods (legend). The expected proportion $p_2$ of missing infection times was 0.8. The fixed incubation period was 10 days.}
\label{fig:measles}
\end{figure}

First, we report the school class--specific basic reproductive number estimates for the complete data. Figure \ref{fig:measles} shows that the 95\% confidence intervals for these group-specific parameters did not overlap (on the $\log_2$ scale). The studentized bootstrap and Bayesian credible intervals were nearly identical, as we expected for the complete data. Table \ref{Table2} provides the numerical estimates for these interval estimates. The upper and lower ends for the extremely high basic reproductive numbers of the second group ranged from 32 to 70. Figure \ref{fig:hagelloch}D shows that most children in the second class (aged 6 to 10) experienced symptoms within a couple of days of each other. The individual suspected of infecting 30 other children also belonged to this school class (Figure \ref{fig:hagelloch}F). The infection times of the individuals in the other two school classes were much more spread out and no individual was suspected of infecting more than 7 other children. Children between 0 and 6 years old had the lowest group-specific estimate. These individuals may not have met in common places as often as the older children in the other school classes. The Bayesian credible interval for the global removal rate $\gamma$ ranged from 0.108 to 0.145, which is consistent with an expected duration of the infectious period between 7 and 10 days.

We simulated missing data in the Hagelloch dataset with $p_1 \in \{0.4,0.6,0.8\}$ and $p_2=0.8$, and then we reanalyzed the dropout datasets with the partial data methods (Section \ref{sec:methods}). Figure \ref{fig:measles} and Table \ref{Table2} show that the group-specific estimates from the partial data methods were similar to those of the complete data methods. The midpoints of the bootstrap interval increased as we increased $p_1$, consistent with our simulation study. The bootstrap intervals were similar to the Bayesian credible intervals, as was the case in our simulation study when the fixed incubation period was close to or more than the expected duration of the infectious period. Importantly, we derived the same conclusion from the partial data that the group-specific interval estimates did not overlap.

These results do not necessarily mean that within school class transmission is the only important dynamic of the spread. We performed the nonparametric test of \cite{Aristotelous2025-fy}, which measures the spread within groups of ordered infection or removal times. Excluding a few obvious outliers (Figure \ref{fig:hagelloch}D), we calculated a $p-$value of 0.62 for the ordered times of prodromal symptoms. So, while our analysis indicates that there are differences between school classes that merit further investigation (Figure \ref{fig:measles}), household dynamics and global spatial spread may also be important in modeling \citep{Neal2004-ir}.

\section{Conclusion}
\label{sec:discussion}

Joint statistical inference of the infection and removal rates $\beta$ and $\gamma$ in SIR/SEIR models can be unreliable when infection times are entirely unobserved. Here, we present an estimation procedure that leverages a small calibration sample of fully observed infectious periods to accurately estimate $\gamma$, and thus stabilizes an imputation-based estimator of $\beta$. Crucially, we derive a theoretical result indicating that removal time--only methods may be unreliable when the basic reproductive number $R_0 > 2$. For Bayesian and frequentist estimators, the inclusion of the calibration sample substantially reduces bias and improves coverage.

In simulations and in a reanalysis of the Hagelloch measles outbreak under simulated missingness, we found that observing tens of complete infectious periods was sufficient to recover qualitatively stable transmission inferences, including school class--specific differences with non-overlapping interval estimates. These results suggest a pragmatic data collection strategy for emerging outbreaks: prioritize broad collection of removal times, but also collect a modest set of complete infectious periods to decouple estimation of $\gamma$ from $\beta$ and improve inference for $R_0$. Extending these ideas to richer feature--dependent transmission models and to principled model selection is a possible direction for future work.

\section{Software}
\label{sec:software}

Software in the form of R code is available at \url{https://github.com/sdtemple/peirrs} and \url{https://github.com/sdtemple/pblas}.



\section*{Acknowledgments}

S.D.T. acknowledges funding from the Eric and Wendy Schmidt AI in Science Postdoctoral Fellowship by Schmidt Sciences, LLC. This research was supported in part by the National Institute of General Medical Sciences of the NIH under award number R35GM151145.
The content is solely the responsibility of the authors and does not necessarily represent the official views of the NIH.

\noindent{\it Conflict of Interest}: None declared.

\clearpage


\section*{Algorithms}\label{sec:algorithms}

\begin{algorithm}[!ht]
\caption{The studentized SEIR bootstrap}
\label{algorithm3}
\begin{algorithmic}
\vskip5pt
\State \textbf{Input:} infection rate $\tilde{\beta}$, removal rate $\hat{\gamma}$, incubation period $\delta$, shape $m$, population size $N$, epidemic size $n$, expected fraction of fully observed periods $p_1$, expected fraction of missing infection times $p_2$, within percentage $\omega$, outer bootstrap size $B_{\mathrm{out}}$, inner bootstrap size $B_{\mathrm{in}}$, and level $1-\alpha$.
\State \textbf{Output:} a bootstrap$-t$ confidence interval for $\beta$ and $R_0$.
\vskip5pt
\For{$b_o=1,\ldots,B_{\mathrm{out}}$}
    \State \begin{enumerate}
    \item Run Algorithm \ref{algorithm2} with pairwise infection rates $\{\tilde{\beta}/N\}$ and the removal rate $\hat{\gamma}$ to generate $\{(r_j^*,i_j^*)\}_{1:n^*}$ until $(1-\omega)n \leq n^* \leq (1+\omega)n$.
    \item Select $x_1 \sim \text{Binomial}(n^*, p_1)$ infectious periods to have partially observed $(r_j^*,i_j^*)$.
    \item Select $x_2 \sim \text{Binomial}(x_1, p_2)$ and $n^* - x_1$ of those periods to have missing $i_j^*$ and $r_j^*$.
    \item Estimate $\tilde{\beta}^*$ and $R_0^*=\tilde{\beta}^*/\hat{\gamma}^*$ (Equation \ref{eq:betahattau}) from the updated $\{(r_j^*,i_j^*)\}_{1:n^*}$.
    \item Repeat steps 1 to 4 for $B_{\mathrm{in}}$ runs with $(\tilde{\beta}^*,\hat{\gamma}^*)$ to generate $\{(\tilde{\beta}^{**},\tilde{R}_0^{**})\}_{1:B_{\mathrm{in}}}$.
    \item Calculate $\widehat{\mathrm{se}}(\tilde{\beta}^*) \gets \mathrm{sd}(\tilde{\beta}^{**}_1,\dots,\tilde{\beta}^{**}_{\mathrm{B}_{in}})$ and $\widehat{\mathrm{se}}(\tilde{R}_0^*) \gets \mathrm{sd}(\tilde{R}^{**}_{0,1},\dots,\tilde{R}^{**}_{0,\mathrm{B}_{in}})$.
    \item Studentize $t_{\beta,b_o}^* \gets (\tilde{\beta}^*-\tilde{\beta}) / \widehat{\mathrm{se}}(\tilde{\beta}^*)$ and $t_{R_0,b_o}^* \gets (\tilde{R}_0^*-\tilde{R}_0) / \widehat{\mathrm{se}}(\tilde{R}_0^*)$.
    \end{enumerate}
\EndFor
\State Let $t^*_{\theta,\alpha/2}$ and $t^*_{\theta,1-\alpha/2}$ be the empirical $\alpha/2$ and $1-\alpha/2$ quantiles of some $\{t_{\theta,b_o}^{*}\}_{b_o=1}^B$.
\State Let $\widehat{\mathrm{se}}(\tilde{\theta})$ be a standard error of some estimate $\tilde{\theta}$ from 100 bootstraps
\State Form the bootstrap$-t$ intervals:
\State \hspace{1em}$\Big[\tilde{\beta} - t^*_{\beta,1-\alpha/2} \times \widehat{\mathrm{se}}\,(\tilde{\beta}),\ \tilde{\beta} - t^*_{\beta,\alpha/2} \times \widehat{\mathrm{se}}\,(\tilde{\beta})\Big]$,
\State \hspace{1em}$\Big[\tilde{R}_0 - t^*_{R_0,1-\alpha/2} \times \widehat{\mathrm{se}}\,(\tilde{R}_0),\ \tilde{R}_0 - t^*_{R_0,\alpha/2} \times \widehat{\mathrm{se}}\,(\tilde{R}_0)\Big]$.
\end{algorithmic}
\end{algorithm}

\clearpage

\begin{algorithm}[!ht]
\caption{A Metropolis-within-Gibbs sampler for the partially observed SEM}
\label{algorithm4}
\begin{algorithmic}
\vskip5pt
\State \textbf{Input:} times $\{(i_j,r_j)\}_{j=1:n}$, some of which are missing, population size $N$, shape $m$, priors $\pi(\beta_N) \sim\text{Gamma}(\xi_\beta,\zeta_\beta)$ and $\pi(\gamma)\sim\text{Gamma}(\xi_\gamma,\zeta_\gamma)$, iterations $T_1$, attempts $T_2$
\State \textbf{Output:} posterior samples $\{\beta^{(t)},\gamma^{(t)},\tilde{\mathbf{i}}^{(t)},\tilde{\mathbf{r}}^{(t)}\}_{t=1}^{T_1}$
\vskip5pt
\State Initialize $(\beta^{(0)},\gamma^{(0)})$ and augmented times $(\tilde{\mathbf{i}}^{(0)},\tilde{\mathbf{r}}^{(0)})$.
\For{$t=1,\ldots,T_1$}
    \State Draw $\beta^{(t)}_N \sim \pi(\beta_N \mid \tilde{\mathbf{i}}^{(t-1)},\tilde{\mathbf{r}}^{(t-1)})$ using Equation~\ref{eq:betagibbs}.
    \State Draw $\gamma^{(t)} \sim \pi(\gamma\mid N^{-1},\tilde{\mathbf{i}}^{(t-1)},\tilde{\mathbf{r}}^{(t-1)})$ using Equation~\ref{eq:gammagibbs}.

    \State Set $(\tilde{\mathbf{i}}^{(t)},\tilde{\mathbf{r}}^{(t)})\gets(\tilde{\mathbf{i}}^{(t-1)},\tilde{\mathbf{r}}^{(t-1)})$.

    \For{$T_2$ attempts}
    \State Randomly select individual $j$ with a missing endpoint
        \State Draw $\tilde{u}\sim\text{Erlang}(m,1)$.
        \If{$i_j$ is observed and $r_j$ is missing}
            \State Propose $\tilde{r}_j' \gets i_j + \gamma^{(t)}\cdot \tilde{u}$.
            \State Compute the Hastings ratio $H\big(\tilde{r}_j\to \tilde{r}_j'\big)$ using Equation~\ref{eq:hastingremoval}.
            \State Update $\tilde{r}_j^{(t)}\gets \tilde{r}_j'$ with probability $\min\{1,H\}$.
        \ElsIf{ $r_j$ is observed and $i_j$ is missing}
            \State Propose $\tilde{i}_j' \gets r_j - \gamma^{(t)}\cdot \tilde{u}$.
            \State Compute the Hastings ratio $H\big(\tilde{i}_j\to \tilde{i}_j'\big)$ using Equation~\ref{eq:hastinginfection}.
            \State Update $\tilde{i}_j^{(t)}\gets \tilde{i}_j'$ with probability $\min\{1,H\}$.
        \EndIf
    \EndFor
\EndFor
\end{algorithmic}
\end{algorithm}

\clearpage

\section*{Figures}\label{sec:figures}

\begin{figure}[!htp]
\centering\includegraphics[scale=0.4]{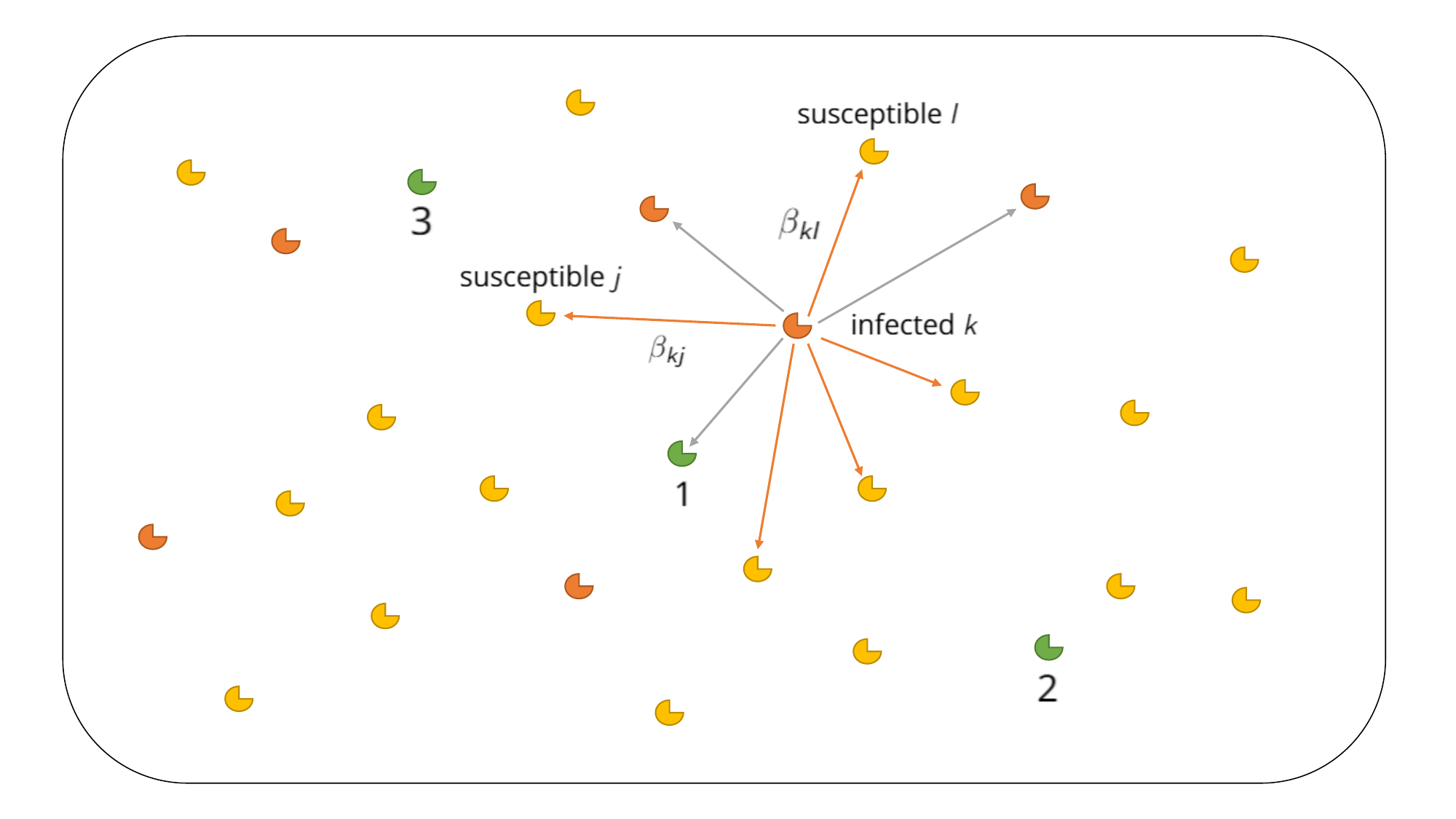}
\caption{Diagram of the stochastic SIR model. Infectious individuals $\mathcal{I}(t)$ (orange) attempt to infect susceptible individuals $\mathcal{S}(t)$ (yellow) at time $t$. The pair-specific rates $\beta_{kj}$ and $\beta_{kl}$ from infected individual $k$ to susceptible individuals $j$ and $l$, depicted as orange directed arrows. The three removed individuals $\mathcal{R}(t)$ (green) cannot be reinfected, depicted as grey arrows.}
\label{fig:cartoon}
\end{figure}

\clearpage

\clearpage

\clearpage

\newpage
\clearpage

\section*{Tables}\label{sec:tables}

\begin{table}[!htp]
\tblcaption{Glossary of mathematical notation.
\label{tab:glossary}}
{\tabcolsep=4.25pt
\begin{tabular}{@{}ll@{}}
\tblhead{
Parameters & Description}
$\beta_{kj}$ & infection rate the infected $k$ applies to the susceptible $j$ \\
$\gamma_j$ & removal rate of the infected $j$ \\
$R_0$ & basic reproductive number \\
$\tau_{kj}$ & time that the infected $k$ applies pressure to the susceptible $j$ \\
$r_j$ & the removal time of infected $j$ \\
$i_j$ & the infection time of infected $j$ \\
$e_j$ & the exposure time of infected $j$ \\
$N$ & initial population size of susceptible individuals \\
$n$ & total number of infected individuals \\
$p_1$ & percent of fully observed periods $\{r_j - i_j\}$ \\
$p_2$ & percent of missing infection times $\{i_j\}$ \\
$\hat{\gamma}$ & maximum likelihood estimator (of the removal rate) \\
$\tilde{\beta}, \bar{\beta}$ & plug--in estimators (of the infection rate) \\
$\breve{\beta}$ & posterior mean estimators (of the infection rate)
\lastline
\end{tabular}}
\end{table}

\clearpage

\begin{table}[!htp]
\tblcaption{Interval estimates in simulated SEIR models. We report the empirical coverages and widths of 95\% studentized bootstrap and Bayesian credible intervals for the frequentist estimator $\tilde{\beta}$ and the posterior mean $\breve{\beta}$, respectively. We varied the infection rates $\beta$ and expected proportion $p_1$ of fully observed periods from 0.4 to 0.8. The removal rate $\gamma$ was 1, the fixed incubation period was 1, and expected proportion $p_2$ of missing infection times was 0.8.
\label{Table1}}
{\tabcolsep=4.25pt
\begin{tabular}{@{}ccccccc@{}}
\tblhead{
$\beta$ & $p_1$ & $\tilde{\beta}$ coverage & $\tilde{\beta}$ width & $\breve{\beta}$ coverage & $\breve{\beta}$ width}
1 & 0.40 & 0.95 & 1.31 & 0.84 & 0.98 \\
1 & 0.80 & 0.88 & 0.71 & 0.90 & 0.83 \\
2 & 0.40 & 0.95 & 1.41 & 0.88 & 1.13 \\
2 & 0.80 & 0.91 & 0.91 & 0.95 & 0.99 \\
3 & 0.40 & 0.95 & 1.79 & 0.90 & 1.53 \\
3 & 0.80 & 0.94 & 1.31 & 0.93 & 1.33 \\
4 & 0.40 & 0.90 & 2.23 & 0.92 & 2.01 \\
4 & 0.80 & 0.91 & 1.72 & 0.94 & 1.76 \\
5 & 0.40 & 0.84 & 2.56 & 0.92 & 2.58 \\
5 & 0.80 & 0.88 & 2.09 & 0.94 & 2.18
\lastline
\end{tabular}}
\end{table}

\clearpage
\begin{table}[!htp]
\tblcaption{
Group-specific reproductive number estimates for the 1861 measles epidemic in Hagelloch, Germany. We report the 2.5th and 97.5th percentiles of the studentized bootstrap and Bayesian credible intervals, as well as their midpoints $\tilde{R}_0$ and the posterior means $\breve{R}_0$, respectively. We varied the percent $p_1$ of fully observed periods in dropout simulations. The fixed incubation period was 10 days, and the expected proportion $p_2$ of missing infection times was 0.8.
\label{Table2}}
{\tabcolsep=4.25pt
\begin{tabular}{@{}cccccccc@{}}
\tblhead{
Group & $p_1$ & $\tilde{R}_0$ & $\tilde{R}_{0,2.5}$ & $\tilde{R}_{0,97.5}$& $\breve{R}_0$ & $\breve{R}_{0,2.5}$ & $\breve{R}_{0,97.5}$}
1 & 0.40 & 5.86 & 4.31 & 7.40 & 6.49 & 4.72 & 8.66 \\
1 & 0.60 & 6.27 & 4.84 & 7.69 & 5.84 & 4.29 & 7.68 \\
1 & 0.80 & 6.68 & 5.20 & 8.16 & 6.10 & 4.61 & 7.94 \\
1 & 1.00 & 6.64 & 4.91 & 8.36 & 6.71 & 5.14 & 8.64 \\
2 & 0.40 & 36.75 & 22.78 & 50.72 & 55.19 & 32.85 & 81.77 \\
2 & 0.60 & 49.82 & 32.79 & 66.84 & 30.73 & 19.17 & 45.86 \\
2 & 0.80 & 55.53 & 30.72 & 80.34 & 37.36 & 24.78 & 53.25 \\
2 & 1.00 & 47.77 & 33.00 & 62.55 & 48.97 & 32.48 & 69.79 \\
3 & 0.40 & 9.81 & 7.60 & 12.01 & 14.25 & 9.41 & 20.63 \\
3 & 0.60 & 11.95 & 9.13 & 14.77 & 10.83 & 7.49 & 14.88 \\
3 & 0.80 & 15.11 & 11.27 & 18.95 & 12.99 & 9.27 & 17.33 \\
3 & 1.00 & 15.54 & 11.48 & 19.59 & 16.15 & 12.15 & 20.91
\lastline
\end{tabular}}
\end{table}

\clearpage

\appendix

\section{Multitype and distance-based models}\label{sec:hsem}
This section describes an extension to the model presented in the main text for 
studying multitype and/or distance-based infection rates.

Let $\beta_{kj} = \beta_{g}/N$ if individual $j$ belongs to some infection group $\mathcal{G}$. This group-specific infection rate depends on the susceptible individual $j$ and not the infector $k$. We define $N_g:= |\mathcal{G} \cap \{2,\dots,N\}|$ and $n_g:= |\mathcal{G} \cap \{2,\dots,n\}|$ as the number of individuals and the number of infected individuals in the infection group $\mathcal{G}$, respectively, beyond the initial infected individual. After modifying the log--likelihood (Equation \ref{eq:loglikelihood}) to accommodate group-specific infection rates, we take the partial derivative with respect to $\beta_g$.
\begin{equation}\label{eq:partialgroup}
    \begin{split}
        \frac{\partial{\ell(\cdot)}}{\partial{\beta_g}} &= \beta_g^{-1} \cdot n_g - N^{-1} \cdot \bigg( (N_g-n_g) \cdot \sum_{j=1}^n (r_j - i_j) + \sum_{j \in \mathcal{G}, \hs 2\leq j \leq n} \sum_{k\neq j}^n \tau_{kj} \bigg).
    \end{split}
\end{equation}
We have insisted that the group-specific infection rate $\beta_g$ depends on the susceptible individual $j$ but not the infector $k$ to pull $\beta_g$ out of the log-sum in the log-likelihood.
\begin{equation}
    \begin{split}
        \sum_{j=2}^n \log \bigg( \sum_{k\neq j}^n \beta_{kj} 1(i_k<i_j<r_k) \bigg) &= \sum_{j \in \mathcal{G}}^n \log \beta_g +  \log \bigg( \sum_{k\neq j}^n  1(i_k<i_j<r_k) \bigg) - \log N \\
        &+ \sum_{j \notin \mathcal{G}}^n \log \bigg( \sum_{k\neq j}^n \beta_{kj} 1(i_k<i_j<r_k) \bigg)
    \end{split}
\end{equation}
Next, the MLE of $\beta_g$ is found by setting the partial derivative to zero.
\begin{equation}\label{eq:betahatgroup}
    \begin{split}
        \hat{\beta}_g &= \frac{n_g}{\sum_{j \in \mathcal{G}, \hs 2\leq j \leq n} \sum_{k\neq j}^n \tau_{kj} + (N_g-n_g) \cdot \sum_{j=1}^n (r_j - i_j)} \times N
    \end{split}
\end{equation}
The numerator counts observed infections within the infection group $\mathcal{G}$ beyond the initial infected individual, while the denominator aggregates the total exposure (infectious pressure) experienced by the infection group, split amongst infected individuals and individuals that were never infected. If there is only one infection group, then Equation \ref{eq:betahatgroup} becomes Equation \ref{eq:betahat}.

For group-specific removal rates, we let $\gamma_f = \gamma_j$ if individual $j$ belongs to some removal group $\mathcal{F}$. The removal group $\mathcal{F}$ and infection group $\mathcal{G}$ need not contain the same individuals; in particular, $\mathcal{F}\cap\mathcal{G}$ may be a proper subset of $\mathcal{G}$. We define $n_f:= |\mathcal{F} \cap \{1,\dots,n\}|$ as the number of infected individuals in the removal group $\mathcal{F}$. The group-specific MLE of the removal rate $\gamma_f$ is
\begin{equation}\label{eq:gammahat_f}
    \begin{split}
        \hat{\gamma}_f &= \frac{m \cdot n_f}{\sum_{j \in \mathcal{F}} (r_j - i_j)}.
    \end{split}
\end{equation}

We may also model the infection rate $\beta_{kj} = \beta_0 / N \cdot h(\mathbf{x}_{k}, \mathbf{x}_j)$ as a baseline infection rate $\beta$ times some kernel function $h$ of the features $\mathbf{x}_k$ and $\mathbf{x}_j$ of individuals $k$ and $j$, respectively. For example, features $\mathbf{x}$ could be position on a 2- or 3-dimensional grid, and the kernel function could encode exponential decay based on the Euclidean distance. In these cases, the MLE of the baseline infection rate $\beta_0$ is
\begin{equation}\label{eq:betahatbase}
    \begin{split}
        \hat{\beta}_0 &= \frac{n - 1}{\sum_{j=2}^n \sum_{k\neq j}^n \tau_{kj} \cdot h(\mathbf{x}_k, \mathbf{x}_j) + \sum_{j=1}^n (r_j - i_j) \cdot \sum_{k=n+1}^N  h(\mathbf{x}_k, \mathbf{x}_j)} \times N
    \end{split}
\end{equation}
If we set $h(\mathbf{x}_k, \mathbf{x}_j)=1$ for all $(k,j)$, then Equation \ref{eq:betahatbase} becomes Equation \ref{eq:betahat}.

\section{Conditional expectations for $\tau_{kj}$}\label{sec:appendix}

This appendix collects derivations of the conditional expectations of the infective time pressure
\[
\tau_{kj} := r_k \wedge i_j - i_j \wedge i_k
\]
under partial endpoint observation. We first assume that there is no incubation period such that $e_j=i_j$. For a fixed incubation period of length $\delta$, the same formulas apply after shifting times as described in the main text.

From Poisson processes (PPs) \citep{Durrett1999-yz}, we make some derivations intuitively \citep{Stockdale2021-ez}, but, for others, we have to solve complicated integrals. \cite{Stockdale2021-ez} considered PPs going backward from $r_k$ and $r_j$, whereas we also consider PPs going forward from $i_k$ and $i_j$. We derived the expected values by hand and checked them with Mathematica to ensure correctness. Recall that when $i_j$ is observed, the expected values with $r_j$ observed versus missing are the same, so there are two fewer expected values to derive.

\subsection{Exponentially distributed infectious periods}

Throughout, infectious periods are independent and exponentially distributed unless otherwise stated: $X_\ell:=r_\ell-i_\ell\sim\expon(\gamma_\ell)$. The survival and cumulative distribution functions are denoted as $F_\gamma$ and $S_\gamma$.  

\begin{lemma}\label{lemma2}
    If $i_k$ and $i_j$ are observed and $r_k$ is missing, then
    $$\E[\tau_{kj}] = \begin{cases}
        0 & i_j < i_k \\
        S_{\gamma_k}(i_j - i_k) \cdot (i_j - i_k) & i_j > i_k \\
        + \frac{\exppeir{\gamma_k}{i_j}{i_k} (\gamma_k i_k - \gamma_k i_j - 1) + 1}{\gamma_k}.&
    \end{cases} 
    $$
    \begin{proof}
    First, if $i_j < i_k$, then $r_k \wedge i_j = 0$ and $i_k \wedge i_j = 0$. Second, if $i_j > i_k$, the probability that $r_k > i_j$ is $S_{\gamma_k}(i_j-i_k)$, in which case $\tau_{kj} = i_j-i_k$. The other additive term is the integral
    $$ \int_{i_k}^{i_j} (r_k-i_k) \cdot \gamma_k \exppeir{\gamma_k}{r_k}{i_k} dr_k, $$
    which has $i_j$ as the upper bound of $r_k$.
        
    \end{proof}
\end{lemma}

\begin{lemma}\label{lemma4}
    If $r_k, r_j,$ and $i_k$ are observed and $i_j$ is missing, then
    $$\E[\tau_{kj}] = \begin{cases}
        0 & r_j < i_k \\
        \frac{\exppeir{\gamma_j}{r_j}{i_k} - \gamma_j i_k + \gamma_j r_j - 1}{\gamma_j} & i_k < r_j < r_k \\
        S_{\gamma_j}(r_j-r_k) \cdot \frac{\exppeir{\gamma_j}{r_k}{i_k} - \gamma_j i_k + \gamma_j r_k - 1}{\gamma_j} & r_j > r_k \\
        + F_{\gamma_j}(r_j-r_k) \cdot (r_k - i_k).& 
    \end{cases} 
    $$
    \begin{proof}

    First, for $r_j< i_k$, then $i_j<i_k$ which means $\tau_{kj}=0$. Second, for $i_k < r_j < r_k$, we compute an integral where $i_k$ is the lower bound of $i_j$; otherwise $\tau_{kj}$ would be 0.
    $$ \int_{i_k}^{r_j} (i_j - i_k) \cdot \gamma_j \exp(-\gamma_j (r_j-i_j)) di_j.  $$
    Third, for $r_j > r_k$, we have the probability $F_{\gamma_j}(r_j-r_k)$ that a $\gamma_j-$PP renewal happens before $r_k$, resulting in $\tau_{kj} = r_k - i_k$. The other additive term is the probability $S_{\gamma_j}(r_j-r_k)$ that there is no renewal of the $\gamma_j-$PP multiplied by the integral
    $$ \int_{i_k}^{r_k} (i_j - i_k) \cdot \gamma_j \exp(-\gamma_j (r_k-i_j)) di_j.  $$
        
    \end{proof}
\end{lemma}

\begin{lemma}\label{lemma3}
    If $r_k$ and $i_j$ are observed and $i_k$ is missing, then
    $$\E[\tau_{kj}] = \begin{cases}
        S_{\gamma_k}(r_k-i_j) \cdot \gamma_{k}^{-1}  & i_j < r_k \\
        \gamma_k^{-1} & i_j > r_k.
    \end{cases} 
    $$
    \begin{proof}
    First, for $i_j < r_k$, if $i_j$ were also less than $i_k$, $\tau_{kj}$ would be 0. The probability that there is no renewal of the $\gamma_k-$PP between $r_k$ and $i_j$ is $S_{\gamma_k}(r_k-i_j)$, and then the expected waiting time of the $\gamma_k-$PP is $\gamma_k^{-1}$. Second, if $i_j > r_k$, $\E_{\gamma_k}[r_k - i_k] = \gamma_k^{-1}$.
        
    \end{proof}
\end{lemma}

\begin{lemma}\label{lemma1}
    If $r_k$ and $r_j$ are observed and $i_k$ and $i_j$ are missing, then
    $$\E[\tau_{kj}] = \begin{cases}
        S_{\gamma_k}(r_k-r_j)\gamma_j \gamma_k^{-1} \invpeir{\gamma_k}{\gamma_j} & r_j < r_k \\
        S_{\gamma_j}(r_j-r_k) \gamma_j \gamma_k^{-1} \invpeir{\gamma_k}{\gamma_j} & r_j > r_k \\
        + \gamma_k^{-1} F_{\gamma_j}(r_j - r_k).&
    \end{cases} 
    $$
    \begin{proof}
    First, let $r_j < r_k$. The term $S_{\gamma_k}(r_k-r_j)$ is the probability that there is no renewal of the $\gamma_k-$PP between $r_k - r_j$. The term $\gamma_j\invpeir{\gamma_k}{\gamma_j}$ is the probability that the $\gamma_j-$PP renews before the $\gamma_k-$PP, resulting in the infection time $i_k < i_j < r_k$. The term $\gamma_k^{-1}$ is the expected waiting time from $i_j$ until $i_k$.

    Second, let $r_j > r_k$. For the scenario where $i_j$ is before $r_k$, the term $S_{\gamma_j}(r_j-r_k)$ is the probability that there is no renewal of the $\gamma_j-$PP between $r_j - r_k$, and the term $\gamma_j \gamma_k^{-1} \invpeir{\gamma_k}{\gamma_j}$ has the same logic as above. The other scenario is that $i_j$ renews before $r_k$, which happens with probability $F_{\gamma_j}(r_j-r_k)$, and then the term $\gamma_k^{-1}$ is the expected waiting time from $r_k$ until $i_k$.   
    \end{proof}
\end{lemma}

The time $\tau_{kj} := r_k \wedge i_j - i_j \wedge i_k$ that the infected individual $k$ tries to infect the susceptible individual $j$ is easier to calculate when at least 2 of the 3 variables $r_k, i_j,$ and $i_k$ are observed. When only $r_j$ and $i_k$ are available, $r_j$ provides some indirect information about $i_j$. This case is the most cumbersome formula to derive of all the partial observation patterns.

\begin{lemma}\label{hardlemma}

    There is an analytical formula for the expected value $\E[\tau_{kj}]$ when $r_j$ and $i_k$ are observed and $r_k$ and $i_j$ are missing.

    \begin{proof}

    First, if $r_j < i_k$, then $\tau_{kj} = 0$. Second, if $r_j > i_k$, then the expected value splits into two additive terms. The additive terms come from integrating with the indicators $1(i_j > r_k)$ and $1(i_k < i_j < r_k)$. We compute these integrals through intermediary steps denoted $s_l$.

    \begin{flalign}
        \begin{split}
            &\E[(r_k-i_k) \cdot 1(i_j > r_k) \hs | \hs r_j > i_k] \\
            &= \int_{i_k}^{r_j} \gamma_k \exppeir{\gamma_j}{r_j}{i_j} \int_{i_k}^{i_j} (r_k - i_k) \gamma_j \exppeir{\gamma_k}{ r_k}{i_k} dr_k di_j \\
            &= \gamma_j \gamma_k \exp(-\gamma_j r_j) \exp(\gamma_k i_k) \int_{i_k}^{r_j} \exp(\gamma_j i_j) \int_{i_k}^{i_j} (r_k-i_k) \exp(-\gamma_k r_k) dr_k di_j \\
            &= \gamma_j \gamma_k \exp(-\gamma_j r_j) \exp(\gamma_k i_k) \times (s_2 - s_1).
        \end{split} &&
    \end{flalign}
\begin{flalign}
        \begin{split}
            &s_1 = i_k \int_{i_k}^{r_j} \exp(\gamma_j i_j) \int_{i_k}^{i_j} \exp(-\gamma_k r_k) dr_k di_j \\
            &= i_k \gamma_k^{-1} \int_{i_k}^{r_j} \exp(\gamma_j i_j) \bigg[ \exp(-\gamma_k i_k) - \exp(- \gamma_k i_j) \bigg] di_j \\
            &= i_k \gamma_k^{-1} \bigg[ \gamma_j^{-1} \exp(-\gamma_k i_k) \exp(\gamma_j i_j) \bigg|_{i_k}^{r_j} - s_3 \bigg].
        \end{split} &&
    \end{flalign}
    \begin{flalign}
        \begin{split}
            &s_3 = \int_{i_k}^{r_j} \exp((\gamma_j - \gamma_k) i_j) di_j = \begin{cases}
                (\gamma_j - \gamma_k)^{-1} \exp((\gamma_j - \gamma_k) i_j) \bigg|_{i_k}^{r_j} & \gamma_k \neq \gamma_j \\
                r_j - i_k & \gamma_j = \gamma_k
            \end{cases}
        \end{split} &&
    \end{flalign}
    \begin{flalign}
        \begin{split}
            s_2 &=  \int_{i_k}^{r_j} \exp(\gamma_j i_j) \int_{i_k}^{i_j} r_k \exp(-\gamma_k r_k) dr_k di_j \\
            &= \gamma_k^{-2} \int_{i_k}^{r_j} \exp(\gamma_j i_j) \bigg[ (1+\gamma_k i_k) \exp(-\gamma_k i_k) - (1+\gamma_k i_j) \exp(-\gamma_k i_j) \bigg] di_j  \\
            &= \gamma_k^{-2} (s_4 - s_5).
        \end{split} &&
    \end{flalign}
    \begin{flalign}
        \begin{split}
            s_4 &= (1+\gamma_k i_k) \exp(-\gamma_k i_k) \int_{i_k}^{r_j} \exp(\gamma_j i_j) di_j = \gamma_j^{-1} (1+\gamma_k i_k) \exp(-\gamma_k i_k) \exp(\gamma_j i_j) \bigg|_{i_k}^{r_j}.
        \end{split} &&
    \end{flalign}
    \begin{flalign}
        \begin{split}
            s_5 &= \int_{i_k}^{r_j} (1+\gamma_k i_j) \exp((\gamma_j-\gamma_k) i_j) di_j = s_6 + s_7.
        \end{split} &&
    \end{flalign}
    \begin{flalign}
        \begin{split}
            s_6 &= \int_{i_k}^{r_j} \exp((\gamma_j - \gamma_k) i_j) di_j = \begin{cases}
                (\gamma_j - \gamma_k)^{-1} \exp((\gamma_j - \gamma_k)i_j) \bigg|_{i_k}^{r_j} & \gamma_j \neq \gamma_k \\
                (r_j - i_k) & \gamma_j = \gamma_k.
            \end{cases} 
        \end{split}&&
    \end{flalign}
    \begin{flalign}
        \begin{split}
            s_7 &= \gamma_k \int_{i_k}^{r_j} 
 i_j \exp((\gamma_j - \gamma_k) i_j) di_j = \begin{cases}
      0.5 \cdot \gamma_k (r_j^2 - i_k^2) & \gamma_j = \gamma_k \\
     s_8 & \gamma_j \neq \gamma_k .
 \end{cases}
        \end{split}&&
    \end{flalign}
\begin{flalign}
    \begin{split}
        s_8 &= \gamma_k (\gamma_j - \gamma_k)^{-2}
        (1 - (\gamma_j - \gamma_k) i_j) \exp((\gamma_j - \gamma_k) i_j) \bigg|_{r_j}^{i_k}.
    \end{split}&&
\end{flalign}
\begin{flalign}
    \begin{split}
        \E[(i_j - i_k) \cdot 1(i_k < i_j < r_k \hs | \hs r_j > i_k)] = s_9 + s_{10}.
    \end{split}&&
\end{flalign}
\begin{flalign}
    \begin{split}
        s_9 &= \gamma_k \gamma_j \int_{i_k}^{r_j} \exp(-\gamma_j (r_j - i_j)) \int_{r_j}^{\infty} (i_j - i_k) \exp(-\gamma_k (r_k - i_k)) dr_k di_j \\
        &= \gamma_k \gamma_j \exp(-\gamma_j r_j) \exp(\gamma_k i_k) (s_{11} - s_{12}).
    \end{split}&&
\end{flalign}
\begin{flalign}
    \begin{split}
        s_{11} &= \int_{i_k}^{r_j} i_j \exp(\gamma_j i_j) \int_{r_j}^\infty \exp(-\gamma_k r_k) dr_k di_j \\
        &= \gamma_k^{-1} \exp(-\gamma_k r_j) \int_{i_k}^{r_j} i_j \exp(\gamma_j i_j) di_j \\
        &= \gamma_k^{-1} \gamma_{j}^{-2} \exp(-\gamma_k r_j )(1 - \gamma_j i_j) \exp(\gamma_j i_j) \bigg|^{i_k}_{r_j}.
    \end{split}&&
\end{flalign}
\begin{flalign}
    \begin{split}
        s_{12} &= i_k \int_{i_k}^{r_j} \exp(\gamma_j i_j) \int_{r_j}^\infty \exp(-\gamma_k r_k) dr_k di_j \\
        &= i_k \gamma_k^{-1} \exp(-\gamma_k r_j) \int_{i_k}^{r_j} \exp(\gamma_j i_j) di_j \\
        &= i_k \gamma_k^{-1} \gamma_j^{-1} \exp(-\gamma_k r_j) \exp(\gamma_j i_j) \bigg|_{i_k}^{r_j}.
    \end{split}&&
\end{flalign}
\begin{flalign}
    \begin{split}
        s_{10} &= \gamma_k \gamma_j \int_{i_k}^{r_j} \exp(-\gamma_j (r_j - i_j)) \int_{i_j}^{r_j} (i_j - i_k) \exp(-\gamma_k (r_k - i_k)) dr_k di_j \\
        &= \gamma_k \gamma_j \exp(-\gamma_j r_j) \exp(\gamma_k i_k) (s_{13} - s_{14}).
    \end{split}&&
\end{flalign}
\begin{flalign}
    \begin{split}
        s_{13} &= \int_{i_k}^{r_j} i_j \exp(\gamma_j i_j) \int_{i_j}^{r_j} \exp(-\gamma_k r_k) dr_k di_j \\
        &= \gamma_k^{-1} \int_{i_k}^{r_j} i_j \exp(\gamma_j i_j) \bigg[ \exp(-\gamma_k i_j) - \exp(- \gamma_k r_j) \bigg] di_j \\
        &= \gamma_k^{-1} (s_{15} - s_{16}).
    \end{split}&&
\end{flalign}
\begin{flalign}
    \begin{split}
        s_{15} &= \int_{i_k}^{r_j} i_j \exp((\gamma_j-\gamma_k) i_j) di_j = \begin{cases}
            s_{17} & \gamma_k \neq \gamma_j \\
            0.5 (r_j^2 - i_k^2) & \gamma_k = \gamma_j.
        \end{cases}
    \end{split}&&
\end{flalign}
\begin{flalign}
    \begin{split}
        s_{17} &= (\gamma_j - \gamma_k)^{-2}
        (1 - (\gamma_j - \gamma_k) i_j) \exp((\gamma_j - \gamma_k) i_j) \bigg|_{r_j}^{i_k}.
    \end{split}&&
\end{flalign}
\begin{flalign}
    \begin{split}
        s_{16} &= \exp(-\gamma_k r_j) \int_{i_k}^{r_j} i_j \exp(\gamma_j i_j) di_j = \gamma_j^{-2} (1 - \gamma_j i_j)\exp(-\gamma_k r_j) \exp(\gamma_j i_j) \bigg|^{i_k}_{r_j}.
    \end{split}&&
\end{flalign}
\begin{flalign}
\begin{split}
       s_{14} &= i_k \int_{i_k}^{r_j} \exp(\gamma_j i_j) \int_{i_j}^{r_j} \exp(-\gamma_k r_k) dr_k di_j \\
    &= i_k \gamma_k^{-1} \int_{i_k}^{r_j} \exp(\gamma_j i_j) \bigg[ \exp(-\gamma_k i_j) - \exp(- \gamma_k r_j) \bigg] di_j \\ 
    &= i_k \gamma_k^{-1} (s_{18} - s_{19}).
\end{split}&&
\end{flalign}
\begin{flalign}
    \begin{split}
        s_{18} &= \int_{i_k}^{r_j} \exp((\gamma_j - \gamma_k) i_j) di_j = \begin{cases}
            r_j - i_k & \gamma_k = \gamma_j \\
             (\gamma_j - \gamma_k)^{-1} \exp((\gamma_j - \gamma_k)i_j)\bigg|_{i_k}^{r_j} & \gamma_k \neq \gamma_j .
        \end{cases}
    \end{split}&&
\end{flalign}
\begin{flalign}
    \begin{split}
        s_{19} &= \exp(-\gamma_k r_j) \int_{i_k}^{r_j} \exp(\gamma_j i_j)  = \gamma_j^{-1} \exp(-\gamma_k r_j) \exp(\gamma_j i_j) \bigg|_{i_k}^{r_j}.
    \end{split}&&
\end{flalign}
        
    \end{proof}
    
\end{lemma}

\subsection{Erlang distributed infectious periods}

We suspect that there are also analytical solutions when the infectious periods are Erlang distributed ($m \in \mathbb{N}$ and $m > 2$), following arguments like those in \cite{Stockdale2019-kd}, but direct and even symbolic integration would be difficult, error--prone, and not particularly interesting. We offer examples of how to extend Lemmas \ref{lemma3} and \ref{lemma1} to include Erlang distributions. These are the simpler cases that can be derived from probabilistic arguments about PPs as opposed to direct integration.

Now, let the infectious periods be independent and Erlang distributed. The survival and cumulative distribution functions are denoted as $F_{\gamma,m}$ and $S_{\gamma,m}$. Let $P_{\gamma \cdot w}(l)$ be the probability mass function of the number $l$ for a Poisson($\gamma \cdot w$) random variable, where $w$ is the width of an interval, and denote $c_{kj} = \gamma_k / (\gamma_k+\gamma_j).$

\begin{lemma}\label{lemma5}
     If $r_k$ and $r_j$ are observed, $i_k$ and $i_j$ are missing, and the Erlang shape is $m$, then
    $$\E[\tau_{kj}] = \begin{cases}
        \sum_{l_1=0}^{m-1} P_{\gamma_k(r_k-r_j)}(l_1) \times \sum_{l_2=0}^{m-l_1-1} \binom{m+l_2 - 1}{l_2} c_{kj}^{l_2} (1-c_{kj})^{m} \times (m - l_1 - l_2) \gamma_{k}^{-1} & r_j < r_k \\
        \sum_{l_1=0}^{m-1} P_{\gamma_j(r_j-r_k)}(l_1) \times \sum_{l_2=0}^{m-1} \binom{m - l_1 - 1 + l_2}{l_2} c_{kj}^{l_2} (1-c_{kj})^{m - l_1} \times (m - l_2) \gamma_{k}^{-1} & r_j > r_k \\
        + F_{\gamma_j,m}(r_j-r_k) \cdot m \gamma_k^{-1}.&
    \end{cases} 
    $$
    \begin{proof}
    First, let $r_j < r_k$. The probabilistic argument is that there are (i) $l_1$ renewals of the $\gamma_k-$PP between $r_k$ and $r_j$, (ii) then $l_2 + m - 1$ renewals of $\gamma_k-$ and $\gamma_j-$PPs, $l_2$ of them being from the $\gamma_k-$PP, (iii) and finally the last renewal of the $\gamma_j-$PP. We multiply this probability by the expected waiting time of the remaining $m - l_1 - l_2$ renewals of the $\gamma_k-$PP.
    \begin{equation}
        \sum_{l_1=0}^{m-1} P_{\gamma_k(r_k-r_j)}(l_1) \times \sum_{l_2=0}^{m-l_1-1} \binom{m+l_2 - 1}{l_2} c_{kj}^{l_2} (1-c_{kj})^{m} \times (m - l_1 - l_2) \gamma_{k}^{-1}.
    \end{equation}

    Second, let $r_j > r_k$. The probabilistic argument is that there are (i) $l_1$ renewals of the $\gamma_j-$PP between $r_j$ and $r_k$, (ii) then $l_2 + m - l_1 - 1$ renewals of $\gamma_k-$ and $\gamma_j-$PPs, $l_2$ of them being from the $\gamma_k-$PP, (iii) and finally the last renewal of the $\gamma_j-$PP. We multiply this probability by the expected waiting time of the remaining $m - l_2$ renewals of the $\gamma_k-$PP.

    \begin{equation}
        \sum_{l_1=0}^{m-1} P_{\gamma_j(r_j-r_k)}(l_1) \times \sum_{l_2=0}^{m-1} \binom{m - l_1 - 1 + l_2}{l_2} c_{kj}^{l_2} (1-c_{kj})^{m - l_1} \times (m - l_2) \gamma_{k}^{-1}.
    \end{equation}

    The other additive term $F_{\gamma_j,m} \cdot m \gamma_k^{-1}$ is the probability that there are $m$ renewals of the $\gamma_j-$PP between $r_j$ and $r_k$ multiplied by the expected waiting time of $m$ renewals of the $\gamma_k-$PP.
    
    \end{proof}
\end{lemma}

\begin{lemma}\label{lemma6}
    If $r_k$ and $i_j$ are observed, $i_k$ is missing, and the Erlang shape is $m$, then
    $$\E[\tau_{kj}] = \begin{cases}
        \sum_{l=0}^{m-1} \binom{m}{l} P_{\gamma_k(r_k-i_j)}(l) 
        \cdot (m-l) \gamma_{k}^{-1}  & i_j < r_k \\
        m \gamma_k^{-1} & i_j > r_k.
    \end{cases} 
    $$
    \begin{proof}
    First, for $i_j < r_k$, we must sum over the probability of $l$ renewals of the $\gamma_k-$PP before $i_j$ times the expected waiting time of $m - l$ renewals after $i_j$. Second, if $i_j > r_k$, $\E_{\gamma_k}[r_k - i_k] = m \gamma_k^{-1}$.
        
    \end{proof}
\end{lemma}

Some but not all terms in Lemmas \ref{lemma2} and \ref{lemma4} are also easy to calculate for $\mathrm{Erlang}(m,\gamma)$ infectious periods. In Lemma \ref{lemma2}, for the case where $i_j > i_k$, the additive term $S_{\gamma_k}\cdot (i_j-i_k)$ is replaced with $S_{\gamma_k,m}\cdot (i_j-i_k)$. In Lemma \ref{lemma4}, for the case where $r_j > r_k$, the additive term $F_{\gamma_j}(r_j-r_k) \cdot (r_k - i_k)$ is replaced with $F_{\gamma_j,m}(r_j-r_k) \cdot (r_k - i_k)$

\clearpage

\section{Hastings ratios}\label{sec:hastings}

Let $\tilde{\mathbf{r}}$ and $\tilde{\mathbf{i}}$ be the observed plus the augmented removal and infection times, respectively. That is, we may have some observed removal times $\{r_j\}$ and some augmented removal times $\{\tilde{r}_k\}$ in $\tilde{\mathbf{r}}$, and likewise for the infection times $\tilde{\mathbf{i}}$. We use a marginal target density $\pi(\tilde{i}_j, \tilde{\mathbf{i}}_{-j}, \tilde{\mathbf{r}} \mid \gamma )$ or $\pi(\tilde{r}_j, \tilde{\mathbf{i}}, \tilde{\mathbf{r}}_{-j} \mid \gamma)$ that integrates the pairwise infection rate $\beta_N$ out, so as to minimize any potential posterior correlations between the augmented time point $\tilde{r}_j$ or $\tilde{i}_j$ and $\beta_N$.

\begin{lemma} 
Given the proposal density $f_\gamma$ and prior $\pi(\beta \cdot N^{-1}) \sim \text{Gamma}(\xi_\beta, \zeta_\beta)$, the acceptance probability for a proposed infection time $\tilde{i}^\prime_j$ versus the current infection time $\tilde{i}_j$ is
\begin{equation}\label{eq:hastinginfection}
    \begin{split}
        H(\tilde{i}_j \to \tilde{i}_j^\prime) =\frac{\pi(\tilde{i}^\prime_j, \tilde{\mathbf{i}}_{-j}, \tilde{\mathbf{r}} | \gamma )\cdot f_\gamma(r_j-\tilde{i}_j)}{\pi(\tilde{\mathbf{i}}, \tilde{\mathbf{r}} | \gamma) \cdot f_\gamma(r_j-\tilde{i}^\prime_j)} = \frac{C(\tilde{\mathbf{i}}^\prime,\tilde{\mathbf{r}}) \cdot (\zeta_\beta + B(\tilde{\mathbf{i}},\tilde{\mathbf{r}}))^{\xi_\beta + n - 1}}{C(\tilde{\mathbf{i}},\tilde{\mathbf{r}}) \cdot (\zeta_\beta + B(\tilde{\mathbf{i}}^\prime,\tilde{\mathbf{r}}))^{\xi_\beta + n - 1}},
    \end{split}
\end{equation}
where $C(\mathbf{i},\mathbf{r})=\prod_{j=2}^n \sum_{k \neq j}^n 1(i_k < i_j < r_k)$. The acceptance probability for a proposed removal time $\tilde{r}^\prime_j$ versus the current removal time $\tilde{r}_j$ is
\begin{equation}\label{eq:hastingremoval}
    \begin{split}
        H(\tilde{r}_j \to \tilde{r}_j^\prime) = \frac{\pi(\tilde{r}^\prime_j, \tilde{\mathbf{i}}, \tilde{\mathbf{r}}_{-j} | \gamma) \cdot f_\gamma(\tilde{r}_j-i_j)}{\pi(\tilde{\mathbf{i}}, \tilde{\mathbf{r}} | \gamma) \cdot f_\gamma(\tilde{r}^\prime_j-i_j)} = \frac{C(\tilde{\mathbf{i}},\tilde{\mathbf{r}}^\prime) \cdot (\zeta_\beta + B(\tilde{\mathbf{i}},\tilde{\mathbf{r}}))^{\xi_\beta + n - 1}}{C(\tilde{\mathbf{i}},\tilde{\mathbf{r}}) \cdot (\zeta_\beta + B(\tilde{\mathbf{i}},\tilde{\mathbf{r}}^\prime))^{\xi_\beta + n - 1}}.
    \end{split}
\end{equation}

\begin{proof}
    First, we decompose the probability $\pi(\mathbf{r},\mathbf{i}|\beta,\gamma)$ of the data given the parameters (Equation \ref{eq:likelihood}) into three parts as in \cite{Stockdale2019-kd}.
    \begin{equation}\label{eq:l1}
        L_1(\mathbf{r},\mathbf{i}) = \prod_{j=2}^n \sum_{k \neq j}^n \beta_N \cdot 1(i_k < i_j < r_k) = \beta^{n-1}_N C(\mathbf{r},\mathbf{i}),
    \end{equation}
    \begin{equation}\label{eq:l2}
        L_2(\mathbf{r},\mathbf{i}) = \exp\bigg( - \beta_N \sum_{j=1}^n \sum_{k=1}^N \tau_{kj} \bigg) = \exp(- \beta_N \cdot B(\mathbf{r},\mathbf{i})),
    \end{equation}
    \begin{equation}\label{eq:l3}
        L_3(\mathbf{r},\mathbf{i}) = \prod_{j=1}^n f_\gamma(r_j-i_j).
    \end{equation}
Next, we simplify a target density that does not depend on $\beta$ into constituent parts.
\begin{equation}\label{eq:nasty}
\begin{split}
    \pi(i_j, \mathbf{i}_{-j}, \mathbf{r} | \gamma) &= \int \pi(i_j, \mathbf{i}_{-j}, \mathbf{r}, \beta_N | \gamma) d\beta_N \\
    &= \pi(\gamma) \int \pi(\mathbf{i},\mathbf{r}|\beta,\gamma) \pi(\beta_N) d\beta_n \\
    &= \pi(\gamma) \cdot L_3(\mathbf{r},\mathbf{i}) \cdot C(\mathbf{r},\mathbf{i}) \int \beta^{n-1}_N \exp(-\beta_N \cdot B(\mathbf{r},\mathbf{i})) \pi(\beta_N) d\beta \\
    &=  \pi(\gamma) \cdot L_3(\mathbf{r},\mathbf{i}) \cdot C(\mathbf{r},\mathbf{i}) \times \frac{\zeta_\beta^{\xi_\beta} \Gamma(\xi_\beta + n -1)}{ (\zeta_\beta + B(\mathbf{r},\mathbf{i}))^{\xi_\beta+n-1} \Gamma(\xi_\beta)}.
    \end{split}
\end{equation}
The terms in Equation \ref{eq:nasty} that do not involve the infection time $i_j$ cancel out in the Hastings ratio, as well as the terms $\Gamma(\xi_\beta)$ and $\zeta_\beta^{\xi_\beta}$ from the prior $\pi(\beta_N)$ on the pairwise infection rate $\beta_N $. The $L_3$ term cancels out with the choice of proposal density $f_\gamma$. The target density $\pi(r_j, \mathbf{i}, \mathbf{r}_{-j} | \gamma)$ is the same as $\pi(i_j, \mathbf{i}_{-j}, \mathbf{r} | \gamma)$, so the Hastings ratio is the same as well.

\end{proof}    
\end{lemma}

\beginsupplement

\clearpage

\section{Supplemental algorithms}\label{sec:suppalgs}

\begin{algorithm}[!ht]
\caption{Simulating a stochastic SIR model}
\label{algorithm1}
\begin{algorithmic}
\vskip5pt
\State \textbf{Input:} infection rate $\beta$, removal rate $\gamma$, population size $N$
\State \textbf{Output:} removal and infection times $\{(r_j,i_j)\}_{1:n}$ for epidemic size $n$
\State Initialize $t\gets 0$, $\mathcal{S}(0)=\{2,\ldots,N\}$, $\mathcal{I}(0)=\{1\}$, $\mathcal{R}(0)=\emptyset$.
\State Compute $S(0) = |\mathcal{S}(0)| = N-1$, $I(0)= |\mathcal{I}(0)| = 1$, and $R(0) = |\mathcal{R}(0)| = 0$.
\State Initialize $\mathbf{r}$ and $\mathbf{i}$ as $\{ \infty: j \in \{1,\dots, N\}\}$ and $i_0 \gets 0$.
\While{$I(t) > 0$}
    \State Compute $\lambda_{SI}(t) = \beta/N \cdot I(t) \cdot S(t)$ and $\lambda_{IR}(t) = \gamma \cdot I(t)$.
    \State Draw $\Delta t \sim \text{Exponential}(\lambda_{SI}(t) + \lambda_{IR}(t))$. Update $t \gets t + \Delta t$.
    \If{$u \sim \text{Bernoulli}(q(t))$}
        \State Decrement $S(t) \gets S(t) - 1$ and increment $I(t) \gets I(t) + 1$.
        \State Randomly select a susceptible individual $j \in \mathcal{S}(t)$ and move $j$ to $\mathcal{I}(t)$.
        \State Update the infection time $i_j \gets t$
    \Else
        \State Decrement $I(t) \gets I(t) - 1$ and increment $R(t) \gets R(t) + 1$.
        \State Randomly select an infected individual $j \in \mathcal{I}(t)$ and move it to $\mathcal{R}(t)$.
        \State Update the removal time $r_j \gets t$
    \EndIf
\EndWhile
\State Delete all $i_j$ and $r_j$ that are $\infty$.
\State Return $\{(r_j,i_j)\}_{1:n}$ where $n$ is the epidemic size.
\end{algorithmic}
\end{algorithm}

\clearpage

\begin{algorithm}[!ht]
\caption{Simulating a stochastic heterogeneous SEIR model}
\label{algorithm2}
\begin{algorithmic}
\State \textbf{Input:} infection rates $\{\beta_{kj}\}$, removal rate $\gamma$, incubation period $\delta$, shape $m$, population size $N$
\State \textbf{Output:} removal and infection times $\{(r_j,i_j)\}_{1:n}$ for epidemic size $n$
\State Initialize $t\gets 0$, $\mathcal{S}(0)=\{2,\ldots,N\}$, $\mathcal{E}(0)=\emptyset$, $\mathcal{I}(0)=\{1\}$, $\mathcal{R}(0)=\emptyset$.
\State Initialize $\mathbf{r}$, $\mathbf{i}$, $\mathbf{e}$ as $\{ \infty: j \in \{1,\dots, N\}\}$, $\boldsymbol{\mu} = \{0: j \in \{1,\dots,N\}\}$, and $e_0, i_0 \gets 0$.
\While{$\mathcal{I}(t)\neq\emptyset$}
    \State Compute $\lambda_{SE}(t)= \sum_{k\in\mathcal{I}(t)}\sum_{j\in\mathcal{S}(t)}\beta_{kj}$ and $\lambda_{IR}(t)=\sum_{j\in\mathcal{I}(t)}\gamma_j$.
    \State Draw $ \Delta t \sim\text{Exponential}(\lambda_{SE}(t)+\lambda_{IR}(t))$ and set $t_{\mathrm{ctmc}}\gets t+\Delta t$.
    \State Set $t_{\mathrm{prog}}\gets \min\{i_j: j\in\mathcal{E}(t)\}$ (and $t_{\mathrm{prog}}\gets\infty$ if $\mathcal{E}(t)=\emptyset$).
    \If{$t_{\mathrm{prog}} < t_{\mathrm{ctmc}}$}
        \State Set $t\gets t_{\mathrm{prog}}$ and let $j\gets\arg\min\{i_j: j\in\mathcal{E}(t)\}$.
        \State Move $j$ from $\mathcal{E}(t)$ to $\mathcal{I}(t)$.
    \Else
        \State Set $t\gets t_{\mathrm{ctmc}}$.
        \If{$u\sim\text{Bernoulli}\big(\lambda_{SE}(t)/(\lambda_{SE}(t)+\lambda_{IR}(t))\big)$}
            \State Draw $j$ among $\mathcal{S}(t)$ from a categorical distribution with Equation \ref{eq:draw_het_infect} probabilities.
            \State Move $j$ from $\mathcal{S}(t)$ to $\mathcal{E}(t)$ and update $e_j\gets t$ and $i_j \gets t + \delta$.
        \Else
            \State Draw $j$ among $\mathcal{I}(t)$ from a categorical distribution with Equation \ref{eq:draw_het_remove} probabilities.
            \State Set $\mu_j \gets \mu_j+1$. If $\mu_j=m$, move $j$ from $\mathcal{I}(t)$ to $\mathcal{R}(t)$ and update $r_j \gets t$.
        \EndIf
    \EndIf
\EndWhile
\State Delete all $e_j, i_j,$ and $r_j$ that are $\infty$.
\State Return $\{(r_j,i_j,e_j)\}_{1:n}$ where $n$ is the epidemic size.
\end{algorithmic}
\end{algorithm}

\clearpage

\section*{Supplemental figures}\label{sec:suppfigures}

\clearpage

\clearpage

\clearpage

\clearpage

\clearpage

\clearpage

\clearpage

\clearpage

\clearpage

\clearpage

\clearpage

\clearpage

\clearpage

\clearpage

\clearpage

\clearpage

\clearpage

\clearpage

\clearpage
\makeatletter
\def\ps@headings{%
  \let\@oddhead\@empty
  \let\@evenhead\@empty
  \let\@oddfoot\@empty
  \let\@evenfoot\@empty}
\pagestyle{headings}
\makeatother

\makeatletter
\def\@shorttitle{} 
\makeatother


\bibliographystyle{biorefs}
\bibliography{peirrs-2601}

\end{document}